\documentclass[prc,tightenlines,floatfix,
nofootinbib,eqsecnum,superscriptaddress,twocolumn]{revtex4-2}

\usepackage[T1]{fontenc}		

\usepackage{amsmath,amsfonts,amssymb,amstext,mathrsfs}
\usepackage{mathpazo}

\usepackage[dvips]{graphicx}
\usepackage{epsf,float}
\usepackage{revsymb}

\usepackage{dcolumn}
\usepackage{braket}
\usepackage{color,xcolor}
\usepackage{graphicx}
\usepackage{subfigure}
\usepackage{multirow}
\usepackage{tabularx}
\usepackage{pstricks}
\usepackage[section]{placeins}
\usepackage{booktabs}
\usepackage{array}

\usepackage{hyperref}

\usepackage{lineno}

\newcommand{\Reg}{\mathbb{R}}

\newcommand{\bp}{\mbox{\boldmath $p$}}
\newcommand{\bk}{\mbox{\boldmath $k$}}
\newcommand{\bq}{\mbox{\boldmath $q$}}

\newcommand{\btau}{\mbox{\boldmath $\tau$}}
\newcommand{\bPhi}{\mbox{\boldmath $\Phi$}}

\renewcommand\slash[1]{\not \! #1}

\usepackage[normalem]{ulem}  

\begin{document}


\title{Exclusive $\eta$ production in proton-proton collisions at energies available at \\the GSI Facility for Antiproton and Ion Research}

\vspace{0.6cm}

\author{Piotr Lebiedowicz}
\email{Piotr.Lebiedowicz@ifj.edu.pl}
\affiliation{Institute of Nuclear Physics Polish Academy of Sciences, Radzikowskiego 152, PL-31342 Krak{\'o}w, Poland}

\begin{abstract}
The cross sections for the $p p \to p p \eta$ reaction are evaluated at energies relevant for the HADES, PANDA, and SIS100 experiments at GSI-FAIR. 
Consideration includes the $\eta$-bremsstrahlung mechanism involving intermediate proton exchange via the $\pi^{0}$, $\eta$, $\rho^{0}$, and $\omega$ exchanges and the mechanism involving 
the nucleon resonances $N(1535)$, $N(1650)$, $N(1710)$, and $N(1880)$ excited 
via the pseudoscalar- and/or vector-meson exchanges, depending on the model.
The role of the $\omega \omega$- and $\rho^{0} \rho^{0}$-fusion processes with the reggeized vector-meson exchanges is also discussed.
The calculation is done in an effective Lagrangian approach.
To determine the parameters of the model, the $\gamma p \to \eta p$ and $\pi^{-} p \to \eta n$ reactions are studied and the results are then compared with the Crystal Ball and CLAS data.
For the $p p \to p p \eta$ reaction, the model results are compared with the energy dependence of the cross section measured far from the threshold and with the differential distributions
$d\sigma/d\cos{\theta_{\eta}}$ and $d\sigma/dp_{\eta}$
measured by the DISTO Collaboration.
The comparison shows that the $N(1535)$ resonance is the dominant contribution and that other contributions are also important due to interference effects.
Assuming that the $\rho$ exchange is the dominant resonant excitation process, the model is able to describe the available data, e.g., it reproduces the shape of the angular distributions measured by the DISTO Collaboration.
Predictions are given for the HADES experiment at center-of-mass energy of 
$\sqrt{s} = 3.46$ GeV, which can be verified in the near future, and for planned PANDA and SIS100 experiments at higher energies.
\end{abstract}


\maketitle

\section{Introduction}
\label{sec:intro}

The production of light pseudoscalar $\eta$ and $\eta'(958)$ mesons 
with quantum numbers $I^{G}(J^{PC}) = 0^{+}(0^{-+})$ 
is very interesting. 
In particular, the production mechanism of $\eta$ has been discussed in a number of papers. 
However, it still requires further theoretical and experimental investigation.
For a nice review of physics with the $\eta$ and $\eta'$ mesons see \cite{Gan:2020aco}.
The exclusive production of the $\eta$ meson 
near the kinematical threshold 
in nucleon-nucleon collisions
was discussed in
\cite{Germond:1990zd,Laget:1990xc,Vetter:1991sr,Batinic:1996me,
Gedalin:1997fa,Santra:1998jf,Calen:1998kx,Faldt:2001uz,
Baru:2002rs,Nakayama:2002mu,Nakayama:2003jn,Fix:2003gs,
Deloff:2003te,Ceci:2004cn,
Czyzykiewicz:2006jb,Moskal:2008pi,
Kaptari:2007ss,Shyam:2007iz,Nakayama:2008tg,Klaja:2010vy,Lu:2015pva}
and in \cite{Krusche:2014ava}.

Most calculations of $\eta$ production in nucleon-nucleon collisions indicate a dominant role of resonances, in particular the $N(1535)$ resonance formed by the exchange of virtual mesons.
Several phenomenological models, although based on different assumptions about the excitation mechanism of nucleon resonances, describe the $pp \to pp \eta$ near-threshold data quite well.
For example, in \cite{Germond:1990zd,Laget:1990xc,Gedalin:1997fa,Santra:1998jf} the $\rho$ exchange is found to contribute much more to $N(1535)$ than the $\pi$ or $\eta$ exchanges.
The importance of the $\omega$ exchange was emphasized in \cite{Vetter:1991sr}, while some importance of the two-$\pi$ exchange (simulated by a $\sigma$ exchange) was found in \cite{Batinic:1996me}.
The one-meson exchange model developed by F{\"a}ldt and Wilkin \cite{Faldt:2001uz}, which assumes a dominant excitation of the $N(1535)$ resonance by the exchange of the $\rho$ meson with destructive $\rho / \pi$ interference, reproduces the near-threshold data.
There it was found that the contribution of the $\rho$ exchange is larger than that of the $\pi$ exchange. 
An alternative model discussed in \cite{Baru:2002rs} explains that the relative importance of the different meson exchange (rescattering) contributions is strongly influenced by the nucleon-nucleon initial state interaction (ISI) effect. It was shown that the $\pi$ and $\rho$ exchanges have almost the same strength and give comparable cross sections before the ISI effect is included. After including the ISI effect, the $\pi$ exchange plays the dominant role. However, the contributions of the other terms ($\rho$, $\eta$, $\sigma$) and the direct term are still significant due to interference effects with the $\pi$ term. 
In the model presented in \cite{Ceci:2004cn}, 
the $\eta$ exchange contribution is comparable to 
the leading $\pi$ exchange term, 
and due to the phenomenological final state interaction 
there is no need to introduce other meson exchanges.
In recent calculations \cite{Kaptari:2007ss,Shyam:2007iz,Nakayama:2008tg},
it was shown that a good agreement with the near-threshold data can also be obtained by excitation of the nucleon resonances by different exchange mesons, including nucleonic and mesonic currents and possible interference effects. The interference between several amplitudes influences the behavior of differential cross-section distributions, which is a way to distinguish different models by comparison with experiment. 
In \cite{Ceci:2004cn,Shyam:2007iz,Kaptari:2007ss} it was suggested that the initial ($NN$ ISI) and final state ($NN$ and $\eta N$ FSI) interactions are essential to obtain the shape and size of the total cross section. In \cite{Nakayama:2002mu,Shyam:2007iz} it was found that terms corresponding to the excitation of the $N(1535)$ resonance via the $\pi$-exchange process dominate the cross section.
In \cite{Nakayama:2008tg} Nakayama and co-authors
developed the model based on a combined analysis of $\eta$ meson 
hadro- and photo-production off nucleons,
and discussed two scenarios for the $NN \to NN \eta$ reactions 
that differ in the dynamical content.
In the first scenario, the $S_{11}(1535+1650)$ term dominates over the $D_{13}(1520+1700)$ resonance contribution to the cross section, while in the second case the $D_{13}$ resonance excitation is the dominant production mechanism, especially at lower energies.
The quality of the near-threshold experimental data 
does not allow one to unambiguously distinguish 
different dynamical contributions.
It still remains to be verified whether 
the dominance of the $D_{13}$ contribution discussed 
in \cite{Nakayama:2008tg} is indeed true.
The predictions available in the literature are still not conclusive. 
It is clear that high-precision data on the $p p \to p p \eta$ reaction are needed to draw firm conclusions about its dynamics.

The present study estimates the total and differential cross sections for the $p p \to p p \eta$ reaction at higher energies.
The emphasis is on the production mechanism of the $\eta$ meson in the kinematic region where existing data are scarce and of relatively low accuracy.
The results presented in this work will be compared with data obtained by the DISTO Collaboration 
\cite{Balestra:2004kg}
and the HADES Collaboration 
\cite{Teilab:2010gv,Teilab_thesis,HADES:2012aa}.
Recently, efforts have been made to complement and extend 
cross-section database, 
in particular at the GSI facilities in Darmstadt (Germany).
In February 2022, proton-proton reactions
at 4.5~GeV beam kinetic energy (center-of-mass energy 
$\sqrt{s} = 3.46$~GeV) were measured
by the HADES Collaboration; see e.g. \cite{Trelinski_talk,Trelinski_Ciepal}.
They may provide new and valuable information to test theoretical models for the exclusive production of light mesons,
and notably in a kinematic region that has been poorly studied so far.
In Ref.~\cite{Lebiedowicz:2021gub}, 
the exclusive production of $f_{1}(1285)$ meson 
at energies relevant for the HADES and PANDA experiments 
at GSI-FAIR was discussed.
There, the vector-meson fusion mechanism was investigated.
It is worth mentioning, that at higher energies
the diffractive mechanism
described by Reggeon and Pomeron exchanges, 
as well as the fusion mechanism of corresponding
Regge exchanges, begin to play an important role.
Additional measurements, including those in the energy range of SIS100 at GSI \cite{SIS100} ($\sqrt{s} \approx 8$~GeV), 
are needed to explore the size and significance 
of individual contributions and would therefore be very welcome.

Diffractive exclusive production of 
$\eta$ and $\eta'$ mesons 
in proton-proton collisions at high energies
was discussed in \cite{Kochelev:1999wv,Kochelev:2000wm}, 
more recently in \cite{Lebiedowicz:2013ika,Lebiedowicz:2025num}
within the tensor-Pomeron approach~\cite{Ewerz:2013kda},
and in \cite{Anderson:2014jia,Anderson:2016zon} 
in the Sakai-Sugimoto model.
Two-gluon component of the $\eta$ and $\eta'$ mesons
to leading-twist accuracy was discussed in \cite{Kroll:2002nt}.
For a study of gluon dynamics in low-energy QCD and a discussion of the connection of light pseudoscalar mesons to anomalous glue, interested readers are referred to \cite{Bass:2018xmz}.
Hopefully, new experimental data for the $pp \to pp \eta$ reaction will be accurate enough to learn more about short-range hadron dynamics.

For a review of $\eta$ and $\eta'$ photo- and hadro-production
on nucleons and on nuclei see e.g. \cite{Krusche:2014ava}.
The $\eta$ 
photoproduction $\gamma p \to \eta p$ was measured 
by several experimental groups,
e.g.,
MIT \cite{Bellenger:1968zz}, DESY \cite{Braunschweig:1970jb},
Dewire \textit{et al.} \cite{Dewire:1972kk},
CLAS \cite{Dugger:2002ft,Williams:2009yj}, 
CBELSA/TAPS \cite{Crede:2009zzb}, and
MAMI \cite{McNicoll:2010qk}.
In Refs.~\cite{Crede:2009zzb,McNicoll:2010qk} the $\eta$ mesons 
were observed in their neutral decay modes.
Recently, the CLAS Collaboration \cite{Hu:2020ecf} 
measured the photoproduction of $\eta$ mesons 
in their dominant charged decay mode 
$\eta \to \pi^{+} \pi^{-} \pi^{0}$.
Several approaches were developed to describe the experimental data available at that time; see, e.g.,
\cite{Chiang:2002vq}, 
\cite{Nakayama:2008tg},
a chiral quark model \cite{Zhong:2011ti},
JPAC \cite{Nys:2016vjz},
$\eta$-MAID~2018 \cite{Kashevarov:2017vyl,Tiator:2018heh}.
There, the $t$-channel $\rho$ and $\omega$ exchange
(either Regge trajectories or meson exchanges)
is the dominant reaction mechanism 
for the small-$t$ behavior of the cross section, 
i.e., in the forward scattering region.
The CLAS data on $\eta$ photoproduction \cite{Hu:2020ecf}
give access to the energy range beyond 
the nucleon resonance region and thus allow a comparison 
with the Regge-based model predictions.

The paper is organized as follows. 
In Sec.~\ref{sec:formalism}, the model for the $\eta$ production
in proton-proton collisions is briefly described. 
Section~\ref{sec:results} presents the numerical results of the calculations
for the total and differential cross sections
of the $pp \to pp \eta$ reaction.
The numerical results for three models 
are presented and discussed
for energies far from the threshold relevant for upcoming and planned measurements.
Section~\ref{sec:conclusions} summarizes my conclusions.
In Appendixes~\ref{sec:appendixA}~and~\ref{sec:appendixB}, 
the $\pi^{-} p \to \eta n$
and $\gamma p \to \eta p$ reactions
which are helpful in determining and validating
some model parameters used for the $pp \to pp \eta$ reaction
are discussed.
Appendix~\ref{sec:appendixC} focuses on
$\rho N N^{*}$ coupling constants.

\section{Formalism}
\label{sec:formalism}

The exclusive production of the pseudoscalar $\eta$ meson in proton-proton collisions is the subject of this study:
\begin{eqnarray}
p(p_{a},\lambda_{a}) + p(p_{b},\lambda_{b}) \to
p(p_{1},\lambda_{1}) + p(p_{2},\lambda_{2})  + \eta(k) \,, \nonumber\\
\label{2to3_reaction} 
\end{eqnarray}
where $p_{a,b}$, $p_{1,2}$, and $\lambda_{a,b}$, 
$\lambda_{1,2} = \pm \frac{1}{2}$
denote the four-momenta and helicities of the nucleons, respectively, 
and $k$ denotes the four-momentum of the $\eta$ meson.
The~cross section is as follows:
\begin{widetext}
\begin{eqnarray}
\sigma({pp \to pp \eta}) &=&
\frac{1}{2}
\frac{1}{2\sqrt{s(s-4 m_{p}^{2})}}\,
\int 
\frac{d^{3}k}{(2 \pi)^{3} \,2 k^{0}}
\frac{d^{3}p_{1}}{(2 \pi)^{3} \,2 p_{1}^{0}}
\frac{d^{3}p_{2}}{(2 \pi)^{3} \,2 p_{2}^{0}}
(2 \pi)^{4} \delta^{(4)}(p_{1}+p_{2}+k-p_{a}-p_{b})
\nonumber \\
&&\times 
\frac{1}{4}
\sum_{\rm spins}
|{\cal M}_{pp \to pp \eta}|^{2}\,,
\label{xs_2to3}
\end{eqnarray}
including a statistics factor $\frac{1}{2}$
due to identical particles appear in the final state.
\end{widetext}

The amplitude reads
\begin{equation}
{\cal M}_{pp \to pp \eta} = {\cal M}_{pp \to pp \eta}(p_{1},p_{2})
- {\cal M}_{pp \to pp \eta}(p_{2},p_{1})\,.
\label{xs_2to3_amp}
\end{equation}
The relative minus sign here
is due to the Fermi statistics, 
which requires the amplitude to be antisymmetric
under interchange of the two final protons.

In the calculations, the complete amplitude 
${\cal M}_{pp \to pp \eta}$ is considered as a sum due to the bremsstrahlung (BS) and the $VV$ fusion processes 
($VV = \rho^{0}\rho^{0}$, $\omega\omega$) 
contributing to $pp \to pp \eta$:
\begin{equation}
{\cal M}_{pp \to pp \eta} = 
{\cal M}_{pp \to pp \eta}^{(\rm BS)} + 
{\cal M}_{pp \to pp \eta}^{(\rm VV)}\,.
\label{amp_total}
\end{equation}
Details will be given when discussing the production mechanisms below. 

\subsection{The $\eta$-bremsstrahlung mechanism}
\label{sec:nucleon_resonances}

The $\eta$-production mechanism
via an intermediate nucleon and nucleon resonances
is shown in Fig.~\ref{fig:diagrams_Nresonances}.

\begin{widetext}

\begin{figure}[!ht]
(a)\includegraphics[width=5.4cm]{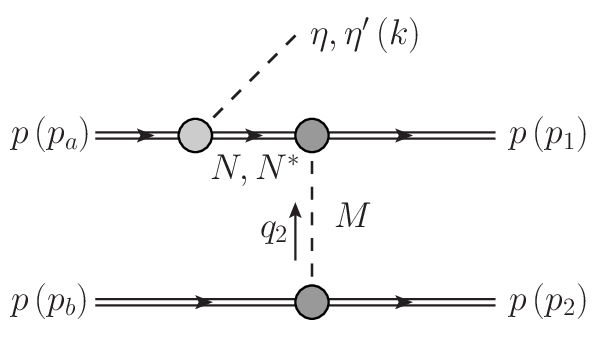}
\quad (b)\includegraphics[width=5.9cm]{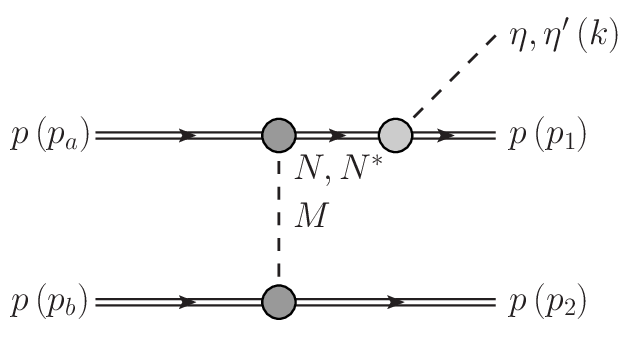}\\
(c)\includegraphics[width=5.4cm]{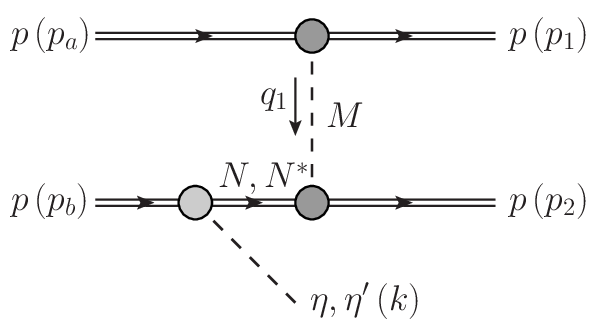}
\quad (d)\includegraphics[width=5.9cm]{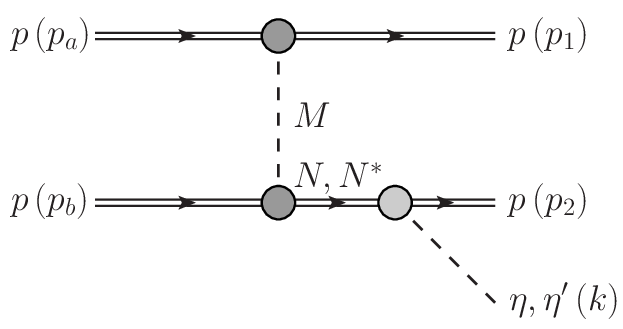}
\caption{The diagrams for the $\eta$ (or $\eta'$) production 
with an intermediate nucleon ($N$)
or a nucleon resonance ($N^{*}$).
$M$ stands for a single meson exchange.
The diagrams (a), (c) and (b), (d) are usually called
pre- and post-emission diagrams, respectively.
There are also the diagrams corresponding to the interchange
of the two final protons 
$p(p_{1}) \leftrightarrow p(p_{2})$.
These are not shown here.}
\label{fig:diagrams_Nresonances}
\end{figure}

The generic amplitude 
for the $pp \to pp \eta$ reaction reads
\begin{align}
{\cal M}_{pp \to pp \eta}^{(\rm BS)} = 
\sum_{i=N, N^{*}}\sum_{j=M}
\left( {\cal M}^{(a) ij}
+
{\cal M}^{(b) ij} 
+
{\cal M}^{(c) ij}
+
{\cal M}^{(d) ij} \right)
\,.
\label{sum_amplitudes_bremss}
\end{align}

\end{widetext}

The amplitudes for the diagrams
corresponding to the exchange of protons
$p(p_{1}) \leftrightarrow p(p_{2})$ 
are also taken into account with appropriate kinematics and
with a negative sign.

For the $\eta$-bremsstrahlung mechanism with the intermediate
proton all set of the diagrams is shown in
Fig.~\ref{fig:diagrams_Nresonances}
with the exchange of virtual mesons
$M = \pi^{0}, \eta, \rho^{0}, \omega$.
In the present analysis,
the processes
with the $\sigma$, $\eta'$, and $a_{0}$ exchanges were neglected
due to small or/and rather poorly known 
relevant $MNN$ and $MNN^{*}$ coupling constants.
In addition, these exchanges are expected to be small
relative to the $\pi^{0}$ exchange because of the heavier mass occurring in the meson propagator.
The analysis includes
the mechanism involving the $N^{*}$ resonances
excited through the exchange of a single meson,
limiting one to the so-called post-emission diagrams,
i.e., the diagrams (b) and (d) of Fig.~\ref{fig:diagrams_Nresonances}.
In these diagrams, the $\eta$ meson is formed through 
the decay of the $N^{*}$ resonance.
Pre-emission diagrams give only a negligible contribution because of kinematic reasons.
This is a reasonable simplification, since the analysis of the $\pi^{-} p \to \eta n$ and $\gamma p \to \eta p$ processes indicates
that the $u$-channel $N^{*}$-exchange contributions
are very small;
see the discussion in Appendixes~\ref{sec:appendixA} and \ref{sec:appendixB}.

As an example, the amplitude for the $pp \to pp \eta$ reaction
with excitation of the intermediate spin-1/2 nucleon resonance 
after the $\pi^{0}$-meson exchange
[the post-emission diagram shown in Fig.~\ref{fig:diagrams_Nresonances}~(b)]
is written as
\begin{widetext}
\begin{eqnarray}
{\cal M}^{(b) N^{*}_{1/2} \pi^{0}}_{\lambda_{a} \lambda_{b} \to \lambda_{1} \lambda_{2} \eta}(s,t_{2})
&=& (-i)
\bar{u}(p_{1}, \lambda_{1}) 
i\Gamma^{(\eta N N^{*}_{1/2})}(p_{1}, p_{1f})
iP^{(N^{*}_{1/2})}(p_{1f}^{2})
i\Gamma^{(\pi N N^{*}_{1/2})}(p_{1f},p_{a})
u(p_{a}, \lambda_{a})\nonumber \\
&& \times 
i\Delta^{(\pi)}(q_{2}) \,
\bar{u}(p_{2}, \lambda_{2}) 
i\Gamma^{(\pi N N)}(p_{2}, p_{b})
u(p_{b}, \lambda_{b}) \,,
\label{amplitude_2b}
\end{eqnarray}
\end{widetext}
where $p_{1f} = p_{a} + q_{2} = p_{1} + k$,
$q_{2} = p_{b} - p_{2}$, $t_{2} = q_{2}^{2}$.
The effective vertices $\pi NN$, $\eta NN$, 
$\pi NN^{*}$, $\eta NN^{*}$ are
obtained from the interaction Lagrangians given in
Appendix~\ref{sec:appendixA}.
The absolute value of the coupling constants $MNN^{*}$ can be determined by using the partial decay widths of the resonances 
with the compilation of Particle Data Group (PDG) data; 
see Table~\ref{tab:table_par2} in Appendix~\ref{sec:appendixA}.

Each vertex in (\ref{amplitude_2b})
incorporates a phenomenological cutoff function
\begin{eqnarray}
F_{\pi NN^{*}}(q^{2},p_{N}^{2},p_{N^{*}}^{2}) = 
F_{\pi}(q^{2}) F_{B}(p_{N}^{2}) F_{B}(p_{N^{*}}^{2}) \,,
\label{ff_resonance}
\end{eqnarray}
where $q$ denotes the four-momentum 
of the intermediate pion $\pi^{0}$, while $p_{N}$ and $p_{N^{*}}$
are the four-momenta of the two baryons.
For the form factor $F_{B}(p^{2})$
with $B = N, N^{*}_{1/2}$ one proceeds as in (\ref{ff_baryon}),
\begin{eqnarray}
F_{\pi}(q^{2}) =
\frac{\Lambda_{\pi NN^{*}}^{2}-m_{\pi}^{2}}{\Lambda_{\pi NN^{*}}^{2}-q^{2}} \,.
\label{ff_meson}
\end{eqnarray}
In the calculations the cutoff parameter $\Lambda_{\pi NN^{*}} = 1.2$~GeV
for each $N^{*}$ resonance is used.
Furthermore, in the $\pi NN$ vertex,
the monopole form factor is taken:
\begin{eqnarray}
F_{\pi NN}(q^{2}) =
\frac{\Lambda_{\pi NN}^{2}-m_{\pi}^{2}}{\Lambda_{\pi NN}^{2}-q^{2}} \,.
\label{ff_MNN}
\end{eqnarray}
Here $\Lambda_{\pi NN} = 1.0$~GeV.
Analogously, the $\eta$-exchange amplitude is derived,
but with the parameters
$\Lambda_{\eta NN} = 1.0$~GeV and $\Lambda_{\eta NN^{*}} = 1.2$~GeV.

The amplitude with the $\rho^{0}$-meson exchange,
corresponding to
Fig.~\ref{fig:diagrams_Nresonances}~(b),
is obtained from (\ref{amplitude_2b}) 
by making the replacements:
\begin{eqnarray}
\Delta^{(\pi)}(q_{2}) &\to & 
\Delta^{(\rho)}_{\mu \nu}(q_{2})\,, \nonumber\\
\Gamma^{(\pi N N^{*}_{1/2})}(p_{1f},p_{a}) &\to &
\Gamma^{(\rho N N^{*}_{1/2}) \mu}(p_{1f},p_{a})\,, \nonumber\\
\Gamma^{(\pi N N)}(p_{2}, p_{b}) &\to &
\Gamma^{(\rho N N) \nu}(p_{2}, p_{b})\,.
\label{rho_replacement}
\end{eqnarray}
The vector-meson-proton vertex is given 
in (\ref{vertex_VNN}) and
the effective coupling for $\rho N N^{*}_{1/2}$
is discuss in Appendix~\ref{sec:appendixC}.

In the calculation of the amplitudes
the propagators for various baryons and mesons are needed.
For nucleon and nucleon resonances see
Eqs.~(\ref{propagator_N})--(\ref{propagator_Nresonance_3half}).
For pseudoscalar mesons one has
$\Delta^{(\pi, \eta)}(q) = 
(q^{2} - m_{\pi, \eta}^{2})^{-1}$.
The propagators for $\rho$ and $\omega$ mesons 
are given by Eq.~(\ref{vec_mes_prop}).
The amplitudes with vector-meson exchanges were multiplied by a purely phenomenological suppression function in order to ensure a correct energy dependence:
\begin{eqnarray}
f_{\rm sup}(s) = \exp \left(- \frac{s - s_{\rm thr}}{\Lambda_{\rm sup}^{2}} \right)
\label{suppression_function}
\end{eqnarray}
with $s_{\rm thr} = (2 m_p+m_{\eta})^2$ 
and $\Lambda_{\rm sup} = 4$~GeV (unless stated otherwise).

Before presenting the calculated results 
for the $pp \to pp \eta$ reaction, 
it is worth discuss the possible role
of the initial state interaction (ISI) 
and final state interaction (FSI) between the nucleons,
i.e., the effects before and after the $\eta$ creation.
The ISI and/or FSI effects were discussed
in more detail in \cite{Sibirtsev:1998yv,Hanhart:1998rn,Nakayama:1998zv,Baru:2002rs,Ceci:2004cn,Kaptari:2004sd,Shyam:2007iz,Kaptari:2007ss}.
The $NN$ FSI effect is significant for small excess energies
$Q_{\rm exc} = \sqrt{s} - \sqrt{s_{\rm thr}}$.
In the present case, for the $\eta$ production 
at $\sqrt{s} \geqslant 3.46$~GeV
one has $Q_{\rm exc} > 625$~MeV
and the $NN$ FSI effect can be safely neglected.
As mentioned in \cite{Baru:2002rs} for the $\eta$ production 
at near-threshold energies, 
for the model calculation without $NN$ ISI the dominant contributions 
come from $\pi$ and $\rho$ exchanges, both being of comparable magnitude.
There, however, the ISI reduces the contribution of $\rho$ exchange 
much more than that of $\pi$ exchange, 
leaving $\pi$ exchange as the only dominant production mechanism. 
The author of Ref.~\cite{Shyam:2007iz} predicted that, close to the threshold, 
the full ($NN$ and $\eta N$) FSI effect can have similar strength 
as $NN$ ISI, but increase the cross section.
It is known that ISI also plays an important role in the HADES and PANDA energy range;
see, e.g., Refs.~\cite{Shyam:2014dia,Shyam:2017gqp}.
Previous studies have shown that both ISI and FSI effects are not universal, 
but depend, among other factors, on the considered reaction
and the concrete meson production mechanism.
Further study on the consistent treatment of these effects in the considered approach, 
including their potential impact on the final results 
(both on the energy dependence of the cross section 
and on the shape of the distributions)
is postponed for future work.

\subsection{The $VV$-fusion mechanism}
\label{sec:VV_fusion}

\begin{figure}[!ht]
\includegraphics[width=5.2cm]{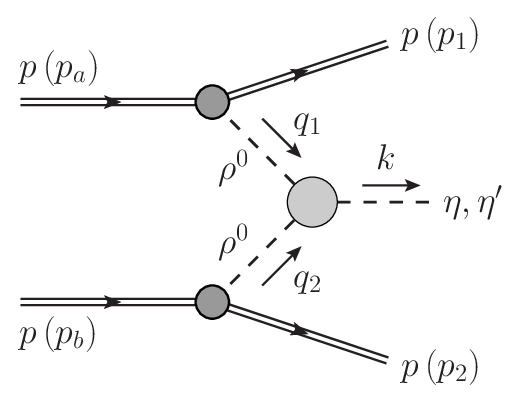}
\includegraphics[width=5.2cm]{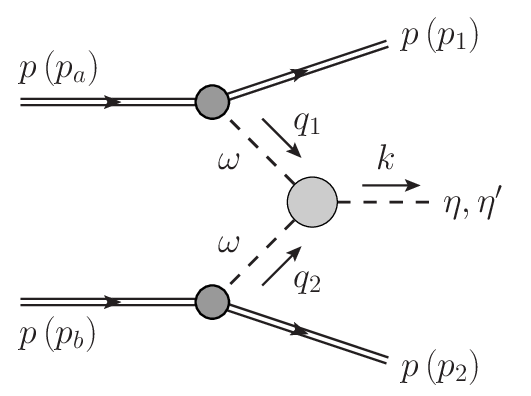}
\caption{The $\rho\rho$- and $\omega\omega$-fusion mechanism
for $\eta$ production in proton-proton collisions.
In addition there are the diagrams 
corresponding to the interchange 
$p(p_{1}) \leftrightarrow p(p_{2})$.}
\label{fig:diagrams}
\end{figure}

The $VV$-fusion mechanism 
($VV$ stands for $\rho^{0}\rho^{0}$ or $\omega\omega$)
is represented by the diagrams in Fig.~\ref{fig:diagrams}.
The kinematic variables 
are
\begin{eqnarray}
&&q_1 = p_{a} - p_{1}, \quad q_2 = p_{b} - p_{2}, \quad k = q_{1} + q_{2}, \nonumber \\
&&t_1 = q_{1}^{2}, \quad t_2 = q_{2}^{2}, \quad m_{\eta}^{2} = k^{2}, \nonumber \\
&&s = (p_{a} + p_{b})^{2} = (p_{1} + p_{2} + k)^{2}, \nonumber \\
&&    s_{1} = (p_{a} + q_{2})^{2} = (p_{1} + k)^{2}, \nonumber \\
&&    s_{2} = (p_{b} + q_{1})^{2} = (p_{2} + k)^{2}\,.
\label{2to3_kinematic}
\end{eqnarray}
For the kinematics see, e.g., Appendix~D of \cite{Lebiedowicz:2013ika}.

The $VV$-fusion amplitude ($VV = \rho^{0}\rho^{0}$, $\omega\omega$) 
is
\begin{eqnarray}
{\cal M}_{pp \to pp \eta}^{(\rm VV)} = 
{\cal M}_{pp \to pp \eta}^{(\rho \rho \; {\rm fusion})}
+
{\cal M}_{pp \to pp \eta}^{(\omega \omega \; {\rm fusion})}
\,,
\label{sum_amplitudes}
\end{eqnarray}
where one has\footnote{The last term in (\ref{amplitude_VV}),
corresponding to the exchange of protons
$p(p_{1}, \lambda_{1}) \leftrightarrow p(p_{2}, \lambda_{2})$, 
is calculated in a similar way as the first term
but with appropriate kinematics and with a negative sign.}
\begin{eqnarray}
{\cal M}^{(V V \; {\rm fusion})}_{\lambda_{a} \lambda_{b} \to \lambda_{1} \lambda_{2} \eta}
&=& (-i)
\bar{u}(p_{1}, \lambda_{1}) 
i\Gamma^{(V pp)}_{\mu_{1}}(p_{1},p_{a}) 
u(p_{a}, \lambda_{a}) \nonumber \\
&& \times 
i\tilde{\Delta}^{(V)\, \mu_{1} \nu_{1}}(s_{1},t_{1}) \,
i\Gamma^{(VV \eta)}_{\nu_{1} \nu_{2}}(q_{1},q_{2})\nonumber \\
&& \times 
i\tilde{\Delta}^{(V)\, \nu_{2}\mu_{2} }(s_{2},t_{2}) \nonumber \\
&& \times 
\bar{u}(p_{2}, \lambda_{2}) 
i\Gamma^{(V pp)}_{\mu_{2}}(p_{2},p_{b}) 
u(p_{b}, \lambda_{b}) 
\nonumber \\
&&
- [p_{1},\lambda_{1} \leftrightarrow p_{2},\lambda_{2}]
\,.
\label{amplitude_VV}
\end{eqnarray}
Here, omitting the Lorentz indices, $\Gamma^{(VV \eta)}$ and
$\Gamma^{(V pp)}$
are the $VV \eta$ and $V pp$ vertex functions, respectively,
and $\tilde{\Delta}^{(V)}$ is the propagator
for the reggeized vector meson $V$.
I now discuss all these quantities in turn.

The $V V \eta$ vertices
are derived from an effective Lagrangians
\cite{Nakayama:1999jb,Kaptari:2007ss} 
\begin{eqnarray}
&&{\cal L}_{\rho \rho \eta}
= \frac{g_{\rho \rho \eta}}{2m_{\rho}}\, 
\varepsilon_{\mu \nu \alpha \beta}\,
(\partial^{\mu} \bPhi_{\rho}^{\nu}
 \partial^{\alpha} \bPhi_{\rho}^{\beta})\,
\Phi_{\eta}\,, \nonumber \\
&&{\cal L}_{\omega \omega \eta}
= \frac{g_{\omega \omega \eta}}{2m_{\omega}}\, 
\varepsilon_{\mu \nu \alpha \beta}\,
(\partial^{\mu} \Phi_{\omega}^{\nu}\,
 \partial^{\alpha} \Phi_{\omega}^{\beta})\,
\Phi_{\eta}\,,
\label{VVPS_Lagrangian}
\end{eqnarray}
where 
$g_{V V \eta}$ are dimensionless coupling constants,
$\Phi$ are the meson fields
(a bold face letter stands for an isovector).
For the Levi-Civita symbol the convention 
$\varepsilon_{0123} = 1$ is used.
The $V V \eta$ vertex, including form factor,
with $q_{1}$, $\mu$ and $q_{2}$, $\nu$
the momenta and vector indices of the incoming $V$ mesons,
is then given by
\begin{eqnarray}
i\Gamma_{\mu \nu}^{(VV \eta)}(q_{1},q_{2}) 
= i \frac{g_{V V \eta}}{2 m_{V}}\, 
\varepsilon_{\mu \nu \alpha \beta} \,
q_{1}^{\alpha} q_{2}^{\beta}\,
F^{(VV \eta)}(q_{1}^{2},q_{2}^{2},k^{2})\,.\nonumber \\
\label{VVM}
\end{eqnarray}
%
%
%
According to the procedure for determining the coupling constant $g_{V V \eta}$ 
from the radiative meson decays
$V \to \eta \gamma$ 
in conjunction with the VMD assumption
(see Appendix~\ref{sec:appendixB}), 
the form factor $F^{(VV \eta)}$ is normalized to unity
when one vector meson is on-mass shell 
and the other one becomes massless, i.e.,
$F^{(VV \eta)}(0,m_{V}^{2},m_{\eta}^{2}) = 1$.
One can use
\begin{eqnarray}
F^{(VV \eta)}(t_{1},t_{2},m_{\eta}^{2}) = 
\frac{\Lambda_{V}^{2}}{\Lambda_{V}^{2}-t_{1}} 
\frac{\Lambda_{V}^{2}-m_{V}^{2}}{\Lambda_{V}^{2}-t_{2}}
\label{F_VV}
\end{eqnarray}
%
and $\Lambda_{V} = \Lambda_{V,\,{\rm mon}} = 1.2$~GeV
motivated by analysis of the $\gamma p \to \eta p$ reaction
in Appendix~\ref{sec:appendixB}.

The vector-meson-proton vertex is
\begin{eqnarray}
i\Gamma_{\mu}^{(V pp)}(p',p) &=& 
-i g_{V pp}\, F_{V NN}(t)\nonumber \\
&& \times 
\left[\gamma_{\mu} - i\frac{\kappa_{V}}{2m_{p}} \sigma_{\mu \nu} (p-p')^{\nu} 
\right]\quad
\label{vertex_VNN}
\end{eqnarray}
where $\kappa_{V} = f_{V NN}/g_{V NN}$ is
the tensor-to-vector coupling ratio
with
the following values for coupling constants:
\begin{eqnarray}
g_{\rho pp} = 3.0\,, \quad
\kappa_{\rho} = 6.1\,, \quad
g_{\omega pp} = 9.0\,, \quad
\kappa_{\omega} = 0\,; \qquad
\label{Vpp_couplings}
\end{eqnarray}
see also the discussion in \cite{Lebiedowicz:2021gub}.
The form factor $F_{VNN}(t)$ in (\ref{vertex_VNN}),
describing the $t$ dependence of the $V$-proton coupling,
can be parametrized as
\begin{eqnarray}
F_{VNN}(t) = 
\frac{\Lambda_{VNN}^{2}-m_{V}^{2}}{\Lambda_{VNN}^{2}-t}\,,
\label{F_V_M}
\end{eqnarray}
where $\Lambda_{VNN} > m_{V}$ and $t < 0$.
Here, $\Lambda_{VNN} = 1.4$~GeV was used
for both $\rho^{0}$- and $\omega$-proton coupling.\footnote{From the Bonn potential model \cite{Machleidt:1987hj}
$\Lambda_{\rho NN} = 1.4$~GeV and 
$\Lambda_{\omega NN} = 1.5$~GeV
are required for a fit to $NN$ scattering data;
see Table~4 of~\cite{Machleidt:1987hj}.}

The standard form of the vector-meson propagator 
is given, e.g., in (3.2) of \cite{Ewerz:2013kda}:
\begin{eqnarray}
i\Delta_{\mu \nu}^{(V)}(q) &=&
i \left(-g_{\mu \nu} + \frac{q_{\mu} q_{\nu}}{q^{2} + i \epsilon} \right) 
\Delta_{T}^{(V)}(q^{2}) \nonumber \\
&& 
- i \frac{q_{\mu} q_{\nu}}{q^{2} + i \epsilon}
\Delta_{L}^{(V)}(q^{2})\,.
\label{vec_mes_prop}
\end{eqnarray}
Here $\Delta_{T}^{(V)}(t) = (t - m_{V}^{2})^{-1}$.
With the relations 
for the $VV \eta$ vertex (\ref{VVM}),
\begin{eqnarray}
\Gamma_{\mu \nu}^{(VV \eta)}(q_{1},q_{2}) \,
q_{1}^{\mu} = 0\,, \quad 
\Gamma_{\mu \nu}^{(VV \eta)}(q_{1},q_{2})\,
q_{2}^{\nu} = 0\,, \quad
\label{relations_for_VVM_vertex}
\end{eqnarray}
the terms $\propto q_{\mu}q_{\nu}$
in (\ref{vec_mes_prop})
do not contribute in (\ref{amplitude_VV}).
According to the arguments presented in \cite{Lebiedowicz:2021gub}
the reggeized vector meson propagator 
$\tilde{\Delta}_{\mu \nu}^{(V)}(s_{i},t_{i})$ is used
in (\ref{amplitude_VV}).
It is obtained from (\ref{vec_mes_prop}) with the replacement
%
\begin{eqnarray}
\Delta_{T}^{(V)}(t_{i}) &\to& 
\tilde{\Delta}_{T}^{(V)}(s_{i},t_{i})\,,
\nonumber \\
\tilde{\Delta}_{T}^{(V)}(s_{i},t_{i}) &=&
\Delta_{T}^{(V)}(t_{i})
\left( \exp (i \phi(s_{i}))\,\frac{s_{i}}{s_{\rm thr}} 
\right)^{\alpha_{V}(t_{i})-1}\,, \nonumber \\
\label{reggeization_2}
\end{eqnarray}
where 
\begin{eqnarray}
&&\phi(s_{i}) =
\frac{\pi}{2}\exp\left(\frac{s_{\rm thr}-s_{i}}{s_{{\rm thr}}}\right)-\frac{\pi}{2}\,,
\label{reggeization_aux}\\
&&s_{\rm thr} = (m_p+m_{\eta})^2\,,
\label{sthr}
\end{eqnarray}
and the non-linear form of the vector-Regge trajectory,
the so-called ``square-root'' trajectory \cite{Brisudova:1999ut},
\begin{eqnarray}
\alpha_{V}(t) = 
\alpha_{V}(0) + 
\gamma \left( \sqrt{T_{V}} - \sqrt{T_{V} - t}  \right)\,.
\label{trajectory_nonlinear}
\end{eqnarray}
%
The parameters are fixed to be 
$\alpha_{\rho}(0) = 0.55$,
$\gamma = 3.65$~GeV$^{-1}$, $\sqrt{T_{\rho}} = 2.46$~GeV;
see Table~I of \cite{Brisudova:1999ut}.
%
%
The intercept of the $\rho$ trajectory
is rather well known $\alpha_{\rho}(0) \approx 0.55$;
see \cite{Brisudova:1999ut} and references therein.
Since $m_{\omega} > m_{\rho}$, the intercepts 
of $\omega$ and $\rho$ trajectories should satisfy 
$\alpha_{\omega}(0) < \alpha_{\rho}(0)$.
Note that $\alpha_{V}(0) \lesssim 0.5$
in Refs.~\cite{Nys:2016vjz,Kashevarov:2017vyl}.
Therefore, for the $\omega$ trajectory,
$\alpha_{\omega}(0) = 0.5$, and 
$\sqrt{T_{\omega}} = \sqrt{T_{\rho}}$
in (\ref{trajectory_nonlinear}), are used.

Other fusion mechanisms,
such as $\omega \phi (\phi \omega) \to \eta$
and $\phi \phi \to \eta$,
can be neglected 
due to much smaller coupling constants and
the heavier mass of $\phi$ occurring in the propagator;
see the discussion 
in \cite{Nakayama:1999jb,Nakayama:1999jx}.
There can be also 
the $a_{0} \pi^{0} (\pi^{0} a_{0}) \to \eta$ fusion processes discussed in \cite{Nakayama:2002mu}.
In the present work,
these contributions are neglected due to the uncertainty in the off-shell form factor and rather poorly known coupling constants.
Note that in \cite{Nakayama:1999jb,Nakayama:2002mu}
these processes were estimated to be small contributions.

\section{Results}
\label{sec:results}

This section presents the numerical results of the calculations 
and the comparison of these with existing experimental data.
It should be stressed that the results presented here for the $pp \to pp \eta$ reaction may not represent optimal solutions within the present approach. The results should be considered as examples of the possible features that can be obtained within a rather simplified model. 
The model parameters are obtained by fitting the model results 
to the available data for the $\gamma p \to \eta p$ and $\pi^{-} p \to \eta n$ reactions and by matching the experimental radiative decay widths, as explained in Appendixes~\ref{sec:appendixA}, \ref{sec:appendixB},
and \ref{sec:appendixC}.

In the following, theoretical predictions for three models are presented.
They differ in the way the resonances are excited (the number and type of exchanged mesons) and may differ in the type of couplings and their parameter values.
Model~1 and model~2 assume the dominance of the $\rho$ exchange (there is no exchange of pseudoscalar mesons). 
Model~1 uses the pseudoscalar (PS) type $g_{\eta NN^{*}}$ couplings (\ref{piNR_PS_Lagrangian}), (\ref{etaNR_PS_Lagrangian}) for all included $N^{*}$ resonances, i.e., for $N(1535)$, $N(1655)$, $N(1710)$ and $N(1880)$. 
Model~2 is the same as model~1 except that for the even parity resonances $N(1710)$, and $N(1880)$ 
the pseudovector (PV) type $g_{\eta NN^{*}}$ couplings (\ref{piNR_PV_Lagrangian}), (\ref{etaNR_PV_Lagrangian})
were used.
In both model~1 and model~2 the $g_{\eta NN^{*}}$ coupling constants were taken from Table~\ref{tab:table_par2} in Appendix~\ref{sec:appendixA}, assuming the upper value for $N(1710)$. 
For model~1, the $\rho N N^{*}$ coupling constants are as follows:
$g_{\rho N N(1535)} = 5.0$,
$g_{\rho N N(1650)} = 1.5$,
$g_{\rho N N(1710)} = 0.4$,
$g_{\rho N N(1880)} = 1.5$,
and a universal value of $\tilde{\Lambda}_{\rho} = 1.2$~GeV for all resonances.
Model~2, on the other hand, uses a smaller value of $g_{\rho N N(1535)} = 4.5$ 
and leaves the other parameters unchanged. 
In model~3, in addition to the $\rho$ exchange, the pseudoscalar meson exchanges $\pi^{0}$ and $\eta$ are also taken into account.
This model uses PS-type couplings (as in model~1) for both $g_{\pi NN^{*}}$ and $g_{\eta NN^{*}}$ for all $N^{*}$ resonances, 
but with the lower values for $g_{\pi N N(1710)}$ and $g_{\eta N N(1710)}$; see Table~\ref{tab:table_par2} of Appendix~\ref{sec:appendixA}.
Again, the same values were used for the $\rho N N^{*}$
couplings as in the model~1.
The important difference here is the change
of $\Lambda_{\rm sup}$ parameter that occurs in (\ref{suppression_function}), from 4 to 2~GeV.
It should be stressed again that, for all the models considered, the calculations for the nucleon resonances 
$N(1535)$, $N(1650)$, $N(1710)$, and $N(1880)$ were performed using the coupling constants $g_{\pi NN^{*}}$ and $g_{\eta NN^{*}}$ from Table~\ref{tab:table_par2} of Appendix~\ref{sec:appendixA} 
and $g_{\rho NN^{*}}$ from the analysis in Appendix~\ref{sec:appendixC}.

Figure~\ref{fig:sig_tot_W} shows results of integrated cross sections for the exclusive $\eta$ meson production 
as a function of center-of-mass (c.m.) energy $\sqrt{s}$ 
together with experimental data. 
Model results include the $\eta$-bremsstrahlung mechanism with the intermediate nucleon ($N$), the nucleon resonance contributions, and the $VV$-fusion contributions. The coherent sum of all contributions, which of course includes the interference term between them, is denoted by the black solid line (``total, model 1''). The individual contributions included in the amplitude are also shown. One can see that the $N(1535)$ component dominates.
The results calculated for the $N^{*}$ contributions 
excited by the $\rho$ exchange 
reproduce energy dependence of the measured $pp \to pp \eta$ cross section.
\begin{figure}[!ht]
\includegraphics[width=8.6cm]{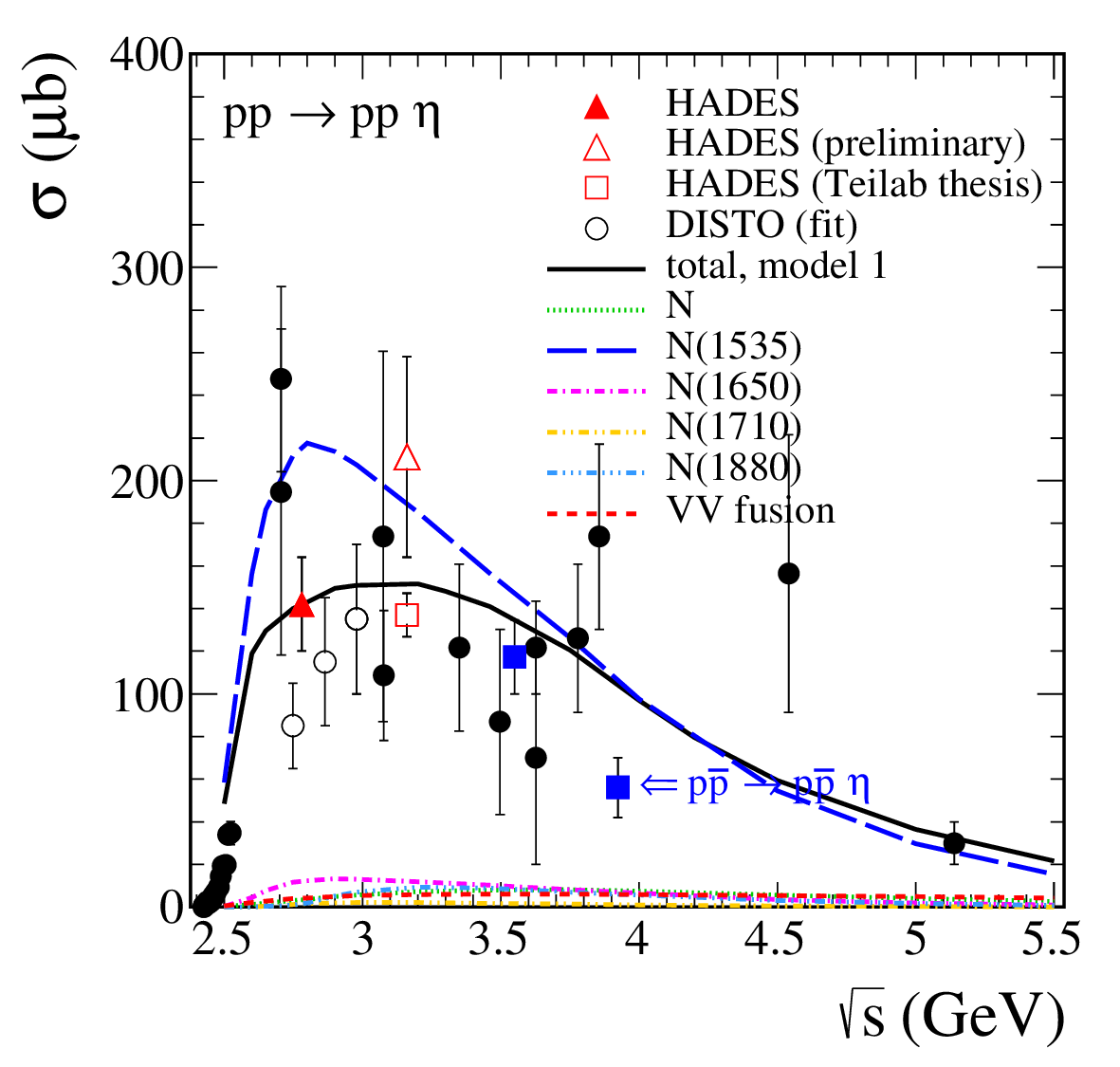}
\caption{Total cross section for the $p p \to p p \eta$ reaction as a function of collision energy $\sqrt{s}$. Shown are the complete model result (``total, model 1'') and the individual contributions included in the calculations together with the compilation of experimental data. The data points from the figure legend are referenced in the text.
There are also shown two experimental points
corresponding to $p \bar{p} \to p \bar{p} \eta$.}
\label{fig:sig_tot_W}
\end{figure}

It is worthwhile to review, even briefly, the experimental data
for the $pp \to pp \eta$ reaction,
which can be found in the literature.
The experimental data in Fig.~\ref{fig:sig_tot_W}
displayed as solid circles ($\bullet$)
are from the HEPData repository \cite{HEPData}, 
\cite{Baldini:1988ti},
and \cite{Pickup:1962zz,Alexander:1967zz,LeGuyader:1972ur,
Chiavassa:1994kf,
Calen:1996mn,Hibou:1998de,Moskal:2009bd}.
Some experimental results 
were corrected for the branching ratio
${\cal B}(\eta \to \pi^+ \pi^- \pi^0) = 23.02 \%$ from \cite{ParticleDataGroup:2024cfk}.
Figure~\ref{fig:sig_tot_W} presents also 
two experimental points
({\color{blue} $\blacksquare$})
which correspond to the $p \bar{p}$ interactions 
\cite{Atherton:1974ma,VanApeldoorn:1978jd}. 
Clearly, the bubble chamber experimental data  are not precise enough to provide a quantitative constraint on the theoretical model.
The DISTO Collaboration \cite{Balestra:2004kg} 
measured the $pp \to pp \eta$ reaction 
in the $\eta \to \pi^{+}\pi^{-}\pi^{0}$ decay channel 
at three c.m. energies $\sqrt{s} = 2.748, 2.865$, 
and 2.978~GeV 
(at proton beam kinetic energies $E_{\rm kin} = 2.15, 2.50$, 
and 2.85~GeV, respectively).
The data points marked as ($\circ$) in Fig.~\ref{fig:sig_tot_W}
are taken from \cite{Balestra:2004kg}
and have been estimated by the DISTO Collaboration; 
see \cite{DISTO:2000dfs} for a description of the method used.
It is important to mention that the DISTO Collaboration 
does not provide values 
for the total cross section for the $pp \to pp \eta$ reaction,
but only an absolute normalization of their data
(based on fits of data with $\sqrt{s}$
ranging from threshold up to 3.4~GeV
and without taking into account both results of Pickup \textit{et al.}
\cite{Pickup:1962zz} at $\sqrt{s} = 2.7$~GeV).
%
%
%
%
The data point marked as
({\color{red}{$\blacktriangle$}})
was obtained by the HADES Collaboration \cite{HADES:2012aa}. 
In this experiment, 
the exclusive $\eta$ production in $pp$ collisions
was studied in hadronic 
($\eta \to \pi^{+}\pi^{-}(\pi^{0})$)
and leptonic ($\eta \to \gamma e^{+} e^{-}$) channels
at $E_{\rm kin} = 2.2$~GeV ($\sqrt{s} = 2.78$~GeV).
The extracted cross section of $\sigma = 142 \pm 22$~$\mu$b
by HADES Collaboration has a larger value than those obtained by DISTO Collaboration in a similar energy range.
In Ref.~\cite{Teilab:2010gv} the preliminary 
total production cross section $\sigma = 211 \pm 47$~$\mu$b 
for $\sqrt{s} = 3.16$~GeV ($E_{\rm kin} = 3.5$~GeV) was determined. 
The result $\sigma = 136.9 \pm 0.9\, (\rm{stat})\pm 10.1 \,(\rm{syst})$~$\mu$b from Teilab's thesis \cite{Teilab_thesis} 
seems to be inconsistent with the result of \cite{Teilab:2010gv}.
Note the very small error in the estimation from \cite{Teilab_thesis}.
The HADES experiment at $\sqrt{s} = 3.46$~GeV 
($E_{\rm kin} = 4.5$~GeV) will provide further important 
information on the exclusive $\eta$ production; 
see \cite{Trelinski_talk,Trelinski_Ciepal} for some details on the analysis.


The integrated cross sections for the $pp \to pp \eta$ reaction 
from $\sqrt{s} = 2.748$ to 8.0~GeV for three models are given in Table~\ref{tab:table1}. 
For model 1 the cross sections of the individual contributions are also shown. 
The complete results for models 2 and 3 
are summarized at the end of the table. 
Presented theoretical results for different models
can be used as a prediction for future experiments
on the $pp \to pp \eta$ reaction.
\begin{widetext}

\begin{table}[!ht]
\centering
\caption{
Cross sections (in $\mu$b) for the $pp \to pp \eta$ reaction
calculated for $\sqrt{s}$ from 2.748 to 8.0~GeV.
Results for the individual contributions
and their coherent sum  (``Total, model 1'') for model 1 are given.
The last two columns contain the complete results for models 2 and 3.}
\label{tab:table1}
\begin{tabular}{l|r|r|r|r|r|r|r||r||r}
\hline
\hline
&
\multicolumn{9}{c}{$\sigma$ ($\mu$b)}
\\
\hline
$\sqrt{s}$ (GeV) & Total, model 1 & $N$ & $N(1535)$ & $N(1650)$ & $N(1710)$ & $N(1880)$ & $VV$ fusion & Total, model 2 
& Total, model 3\\
\hline
2.748   
& 139.0  & 3.1  & 212.1 & 11.6  & 1.3  & 0.9   & 3.9  
& 111.9 
& 174.6\\
2.978   
& 149.1  & 5.7   & 208.1 & 13.0  & 2.1  & 7.1   & 5.3  
& 129.4 
& 143.9\\
3.46    
& 136.9  & 8.0  & 156.5 & 10.2  & 1.7  & 8.8   & 6.1  
& 116.2 
& 80.8\\
5.0     
& 37.1   & 3.9   & 34.4  & 2.3   & 0.3  & 1.9   & 4.8  
& 31.9 
& 21.1\\
8.0     
& 3.3   & 0.7   & 0.3   & 0.02  & $<0.003$ & 0.01 & 2.5  
& 3.3  
& 5.4\\
\hline
\hline
\end{tabular}
\end{table}

\end{widetext}

From Fig.~\ref{fig:sig_tot_W} and 
Table~\ref{tab:table1}
one can see that the $VV$-fusion contributions 
are much lower than the experimental data.
The cross section from the $VV \to \eta$ fusion quickly rises 
from the reaction threshold 
$\sqrt{s_{\rm thr}} = 2 m_p + m_{\eta}$,
it reaches a maximum about $\sqrt{s} \approx 3.46$~GeV, 
and then it starts to decrease towards higher energies.
It should be mentioned that without the reggeization effect, 
calculated 
according to (\ref{reggeization_2})--(\ref{trajectory_nonlinear}),
the cross section would continuously grow from the threshold.
The reduction of cross section due to the reggeization
is about a factor of 10 at $\sqrt{s} = 3.46$~GeV.
Both the HADES and PANDA experiments have a good opportunity to verify this sizable effect of the model.
The $VV$-fusion contribution has different characteristics than the bremsstrahlung processes, 
and as can be seen from Table~\ref{tab:table1},
it turns out to be more significant at higher energies.

For the $\eta$-bremsstrahlung mechanism
with the intermediate proton exchange
(contribution denoted by $N$ in Table~\ref{tab:table1})
different type of processes were included in the calculations,
i.e., the pre- and post-emission diagrams of Fig.~\ref{fig:diagrams_Nresonances} w
ith the $\pi^{0}$, $\eta$, $\rho^{0}$, and $\omega$ exchanges.
For model~1, the total cross section 
for $\sqrt{s} = 3.46$~GeV
is $\sigma_{N, {\rm total}} = 8.02$~$\mu$b.
It is instructive to give the relative contribution 
to the cross section for each exchange.
The results are as follows:
$\sigma_{N, \pi^{0}} = 1.47$~$\mu$b,
$\sigma_{N, \eta} = 0.9$~nb,
$\sigma_{N, \rho^{0}} = 2.08$~$\mu$b, and
$\sigma_{N, \omega} = 2.43$~$\mu$b.
Clearly, the larger total cross-section value 
indicates interference effect between different exchanges in amplitude.
It is noteworthy that the predominant $N$ contribution 
to the cross-section comes from pre-emission diagrams (a) and (c) of Fig.~\ref{fig:diagrams_Nresonances} 
than from post-emission diagrams (b) and (d) of Fig.~\ref{fig:diagrams_Nresonances}.
There is also interference between them.
For the pre-emission diagrams
one finds $\sigma_{N, {\rm pre}} = 8.52$~$\mu$b, while
for the post-emission diagrams
one finds only $\sigma_{N, {\rm post}} = 0.26$~$\mu$b.
One can see from Table~\ref{tab:table1} 
that the $N$ contribution calculated within model~1
is important even going to higher collision energies.
For models 2 and 3, the $N$ contribution is less significant 
(see the results
of Figs.~\ref{fig:brems_1} and \ref{fig:brems_4}
below)
because there for $\rho$ and $\omega$ exchange contributions
the more restricting parameter 
$\Lambda_{\rm sup} = 2$~GeV 
in (\ref{suppression_function})
was assumed.
These predictions represent my educated guesses 
for the cross sections, the validity of which 
is to be determined by comparison to future data.

In Figs.~\ref{fig:brems_1}-\ref{fig:brems_3}, results 
obtained in models 1 and 2 are presented.
Figures~\ref{fig:brems_1} and \ref{fig:brems_2} 
show the distributions in $\cos\theta_{\eta}$, $p_{\eta}$ and $M_{p\eta}$, $M_{pp}$, respectively.
As explained above, several mechanisms 
for exclusive $\eta$ meson production were included.
Their individual contributions 
and the coherent sum (details in legend and description of Fig.~\ref{fig:brems_1})
for the two models 1 and 2 are presented.
One can see that the contribution of $N(1535)$ 
(through $\rho$ exchange) is dominant
and that there is a large interference of different 
components in the amplitude.
The shape of complete theoretical predictions 
of $\cos\theta_{\eta}$ and $p_{\eta}$ seems to be in good agreement with the DISTO data \cite{Balestra:2004kg}.
The experimental data points were normalized to the
values of the total cross sections from \cite{Balestra:2004kg}
and include both statistical and the relative systematical
uncertainties, but do not include additional global
systematic uncertainties of the absolute scale of 22\% and 33\%
for $\sqrt{s} = 2.748$ and 3.46~GeV, respectively.

Figure~\ref{fig:brems_3} shows the distribution
in Feynman-$x$ variable for the $\eta$ meson 
(defined as $x_{F, \eta} = 2 p_{z, \eta} / \sqrt{s}$ with the longitudinal momentum of the meson in the c.m. system)
and the distributions in meson transverse momentum
$p_{t, \eta}$ and proton transverse momentum $p_{t, p}$
for $\sqrt{s} = 3.46$~GeV.
From these and previous results, one can see that 
it is difficult to separate the $VV$-fusion contribution
(indicated by the red short-dashed line) from others. 
However, this contribution plays an increasingly important role at higher energies.
On the other hand, the contribution of the non-resonant bremsstrahlung (indicated by the green dotted line) 
is significant in the region of $\cos\theta_{\eta} \approx \pm 1$, 
$p_{\eta} \approx 1$~GeV, and for $|x_{F, \eta}| \approx 0.55$.
The nucleonic contribution is also characterized by smaller values of the $pp$ invariant mass 
and the $\eta$ and $p$ transverse momenta.
Experimental observation of an enhancement in these kinematic regions can be used to determine the magnitude of the nucleonic term, at least the strength of the $\eta NN$ coupling.

Figures~\ref{fig:brems_4} and \ref{fig:brems_5} represent
the differential cross sections for model~3.
It can be seen from Fig.~\ref{fig:brems_4} that with increasing energy, the contribution of $N^{*}$ resonances excited via $\pi^{0}$ exchange becomes more significant than the $N^{*}$ term excited via $\rho^{0}$ exchange. 
The predictions of model 3 with the admixture of pseudoscalar meson exchanges
describe the DISTO angular distributions less accurately 
than models 1 and 2 (see Fig.~\ref{fig:brems_1})
as far as the shape of the distributions is concerned.
The importance of the $N(1710)$ and $N(1880)$ resonances 
is more visible in model 3
[see two distinct resonant bumps after $N(1535)$ in $d\sigma/dM_{p\eta}$].
Such prediction of the model (resonant features) may be verified 
by examining the distributions in the proton transverse momentum 
and in the $\eta p$ invariant mass.
In particular, compared to $\rho$ exchange processes, 
excited states by pseudoscalar meson exchanges
have the $d\sigma/dp_{t, p}$ distribution 
shifted towards smaller values.

Figure~\ref{fig:brems_6} shows the two-dimensional
differential cross sections 
in the $M_{p \eta}^{(1)}$-$M_{p \eta}^{(2)}$ plane (left panels)
and in the $M_{p \eta}^{(1)}$-$M_{pp}$ plane (right panels)
calculated for model~1 (top panels) 
and for model~3 (bottom panels).
The resonance structure is clearly visible in both models
with dominance of $N(1535)$.
Model~3 highlights the possible role of heavier states 
$N(1710)$ and $N(1880)$
due to the inclusion of pseudoscalar meson excitations.
These resonance patterns provide crucial information about the reaction mechanism and can be used as benchmarks 
for experimental analysis.

At higher energies other fusion mechanisms may be important.
As discussed in \cite{Lebiedowicz:2013ika,Lebiedowicz:2025num},
the Pomeron-Pomeron-fusion mechanism 
and the subleading Reggeon exchanges
(Pomeron-$f_{2 \Reg}$, $f_{2 \Reg}$-Pomeron, 
$f_{2 \Reg}$-$f_{2 \Reg}$) 
seem to dominate at WA102 energy of $\sqrt{s} = 29.1$~GeV.
The inclusion of secondary Reggeons in the calculation 
seems to be particularly important to explain the available data from experiments at intermediate energies.
It can be expected that these processes will give 
a rather small contribution up to $\sqrt{s} \approx 5$~GeV.
To quantify the diffractive mechanism in the intermediate energy range and to understand the role of subleading exchanges, future measurements of exclusive meson production 
from the FAIR accelerator (SIS100) would be necessary.

\begin{widetext}

\begin{figure}[!ht]
\includegraphics[width=6.8cm]{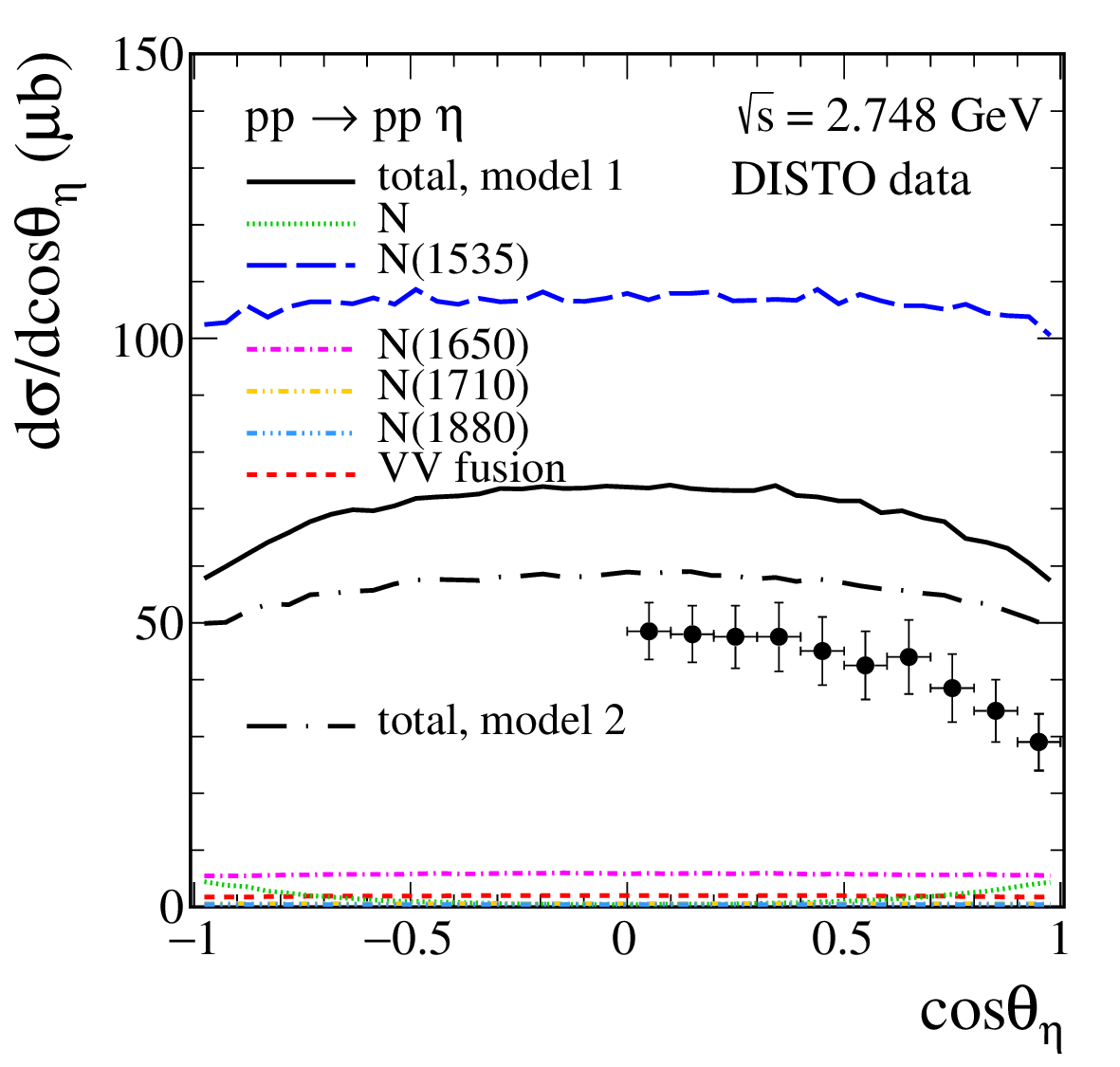}
\includegraphics[width=6.8cm]{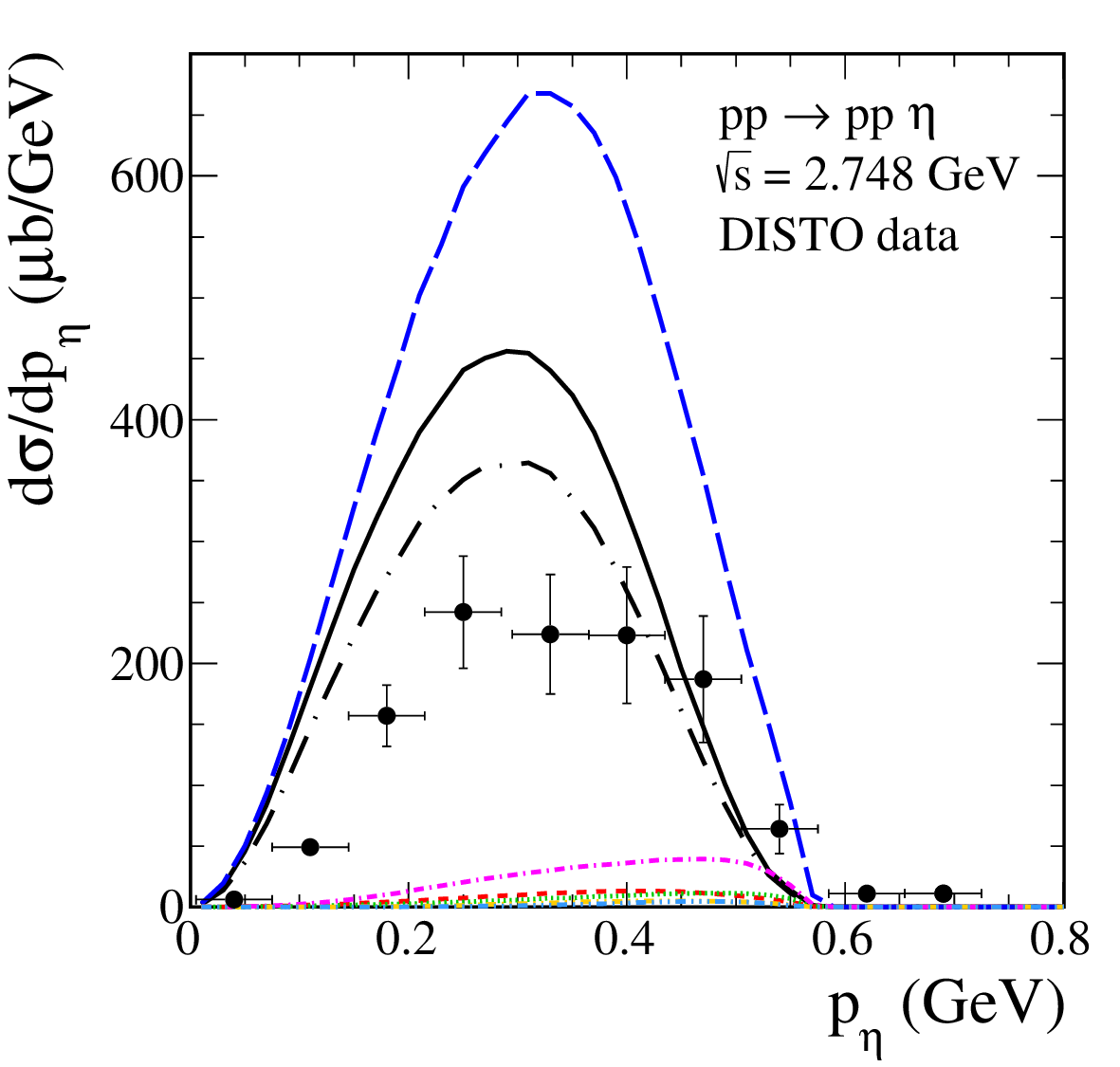}\\
\includegraphics[width=6.8cm]{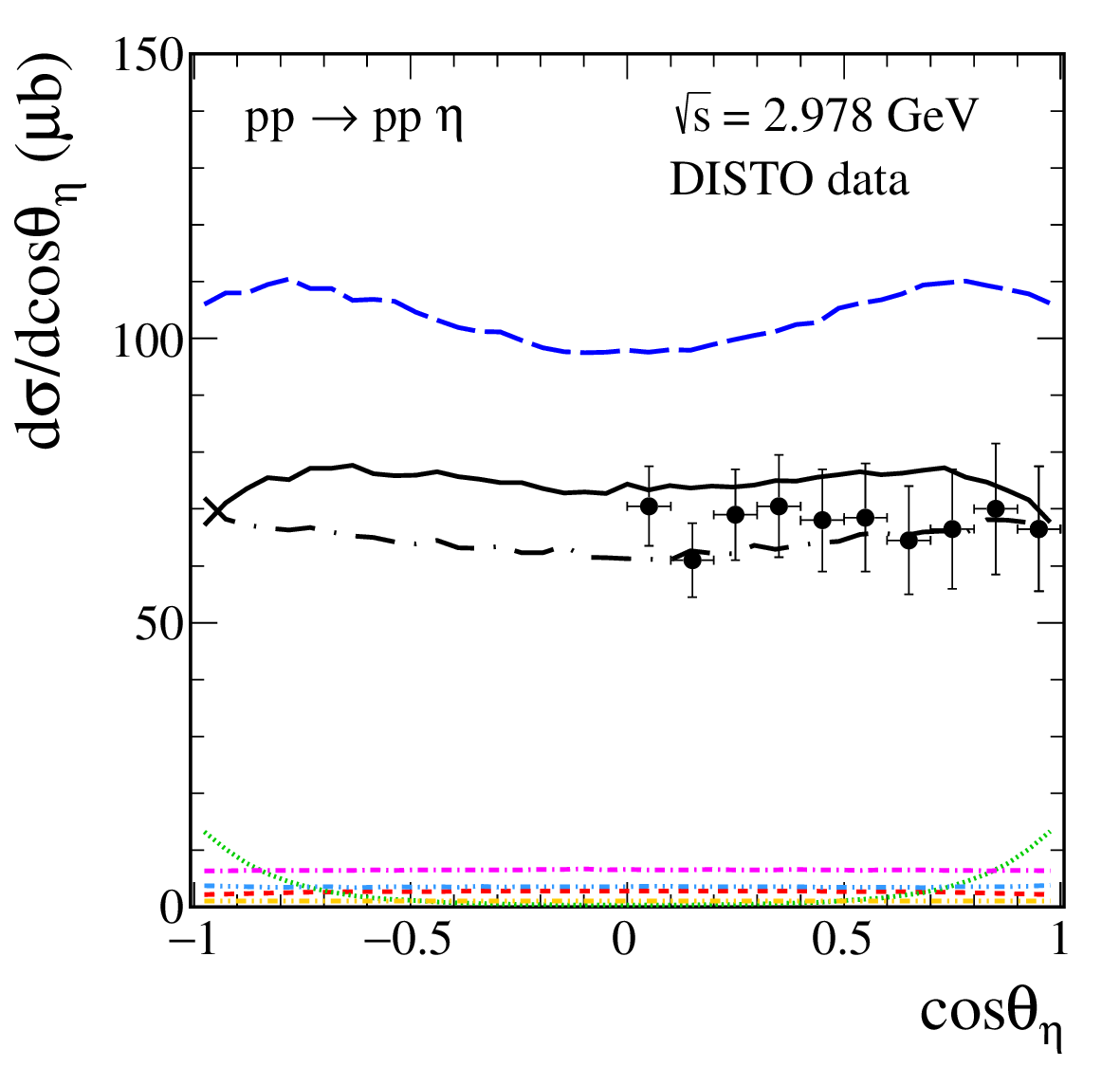}
\includegraphics[width=6.8cm]{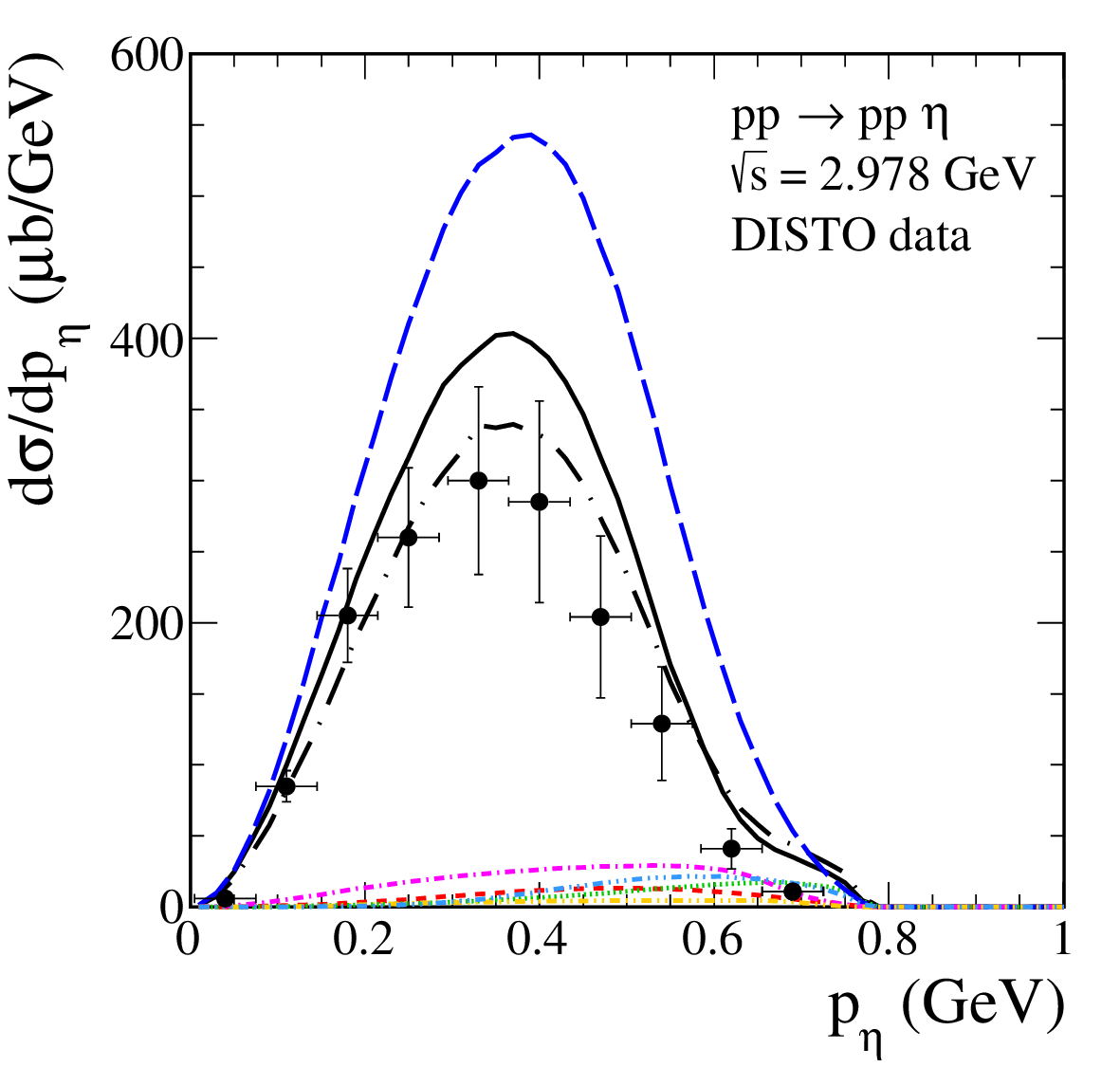}\\
\includegraphics[width=6.8cm]{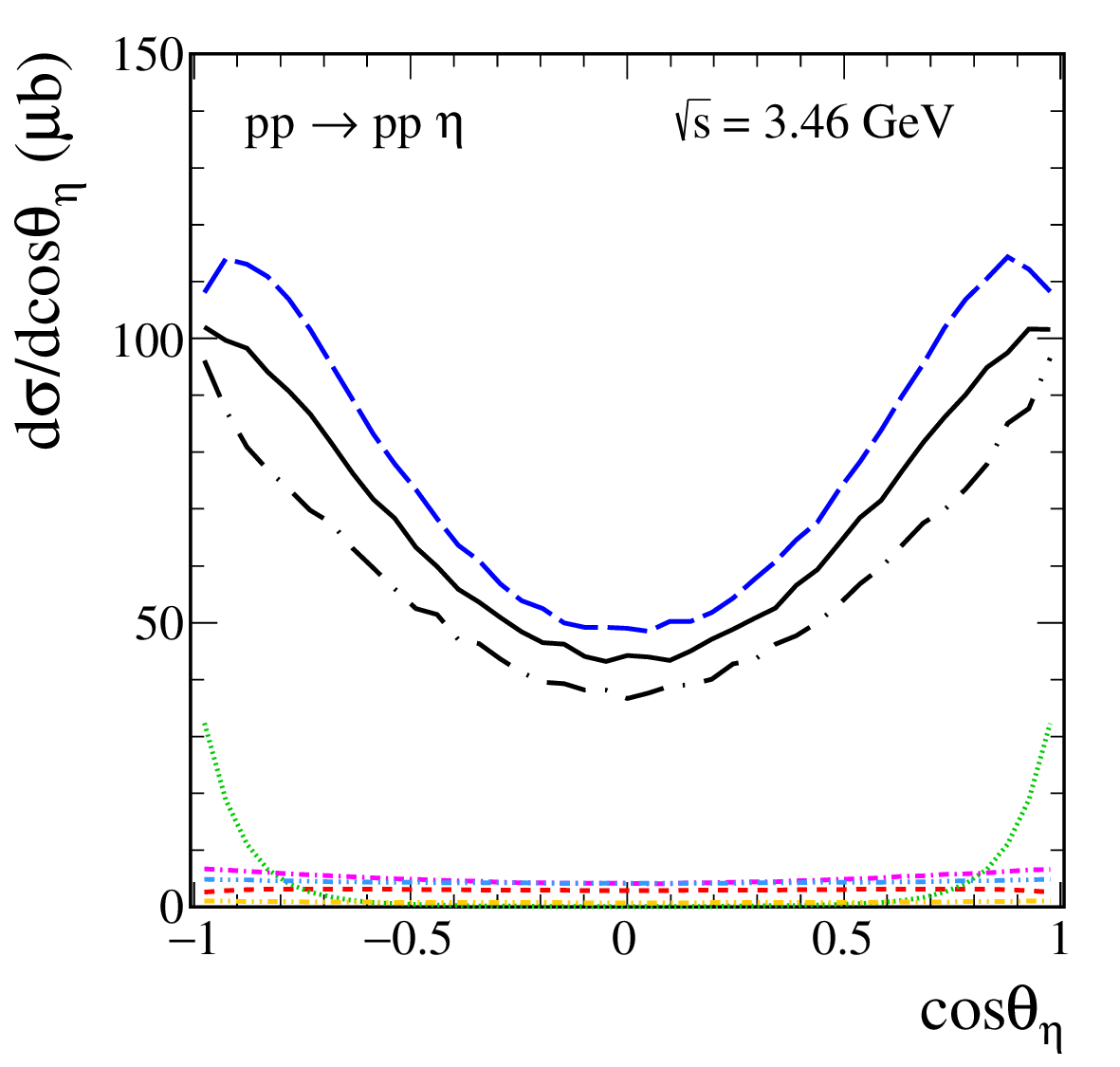}
\includegraphics[width=6.8cm]{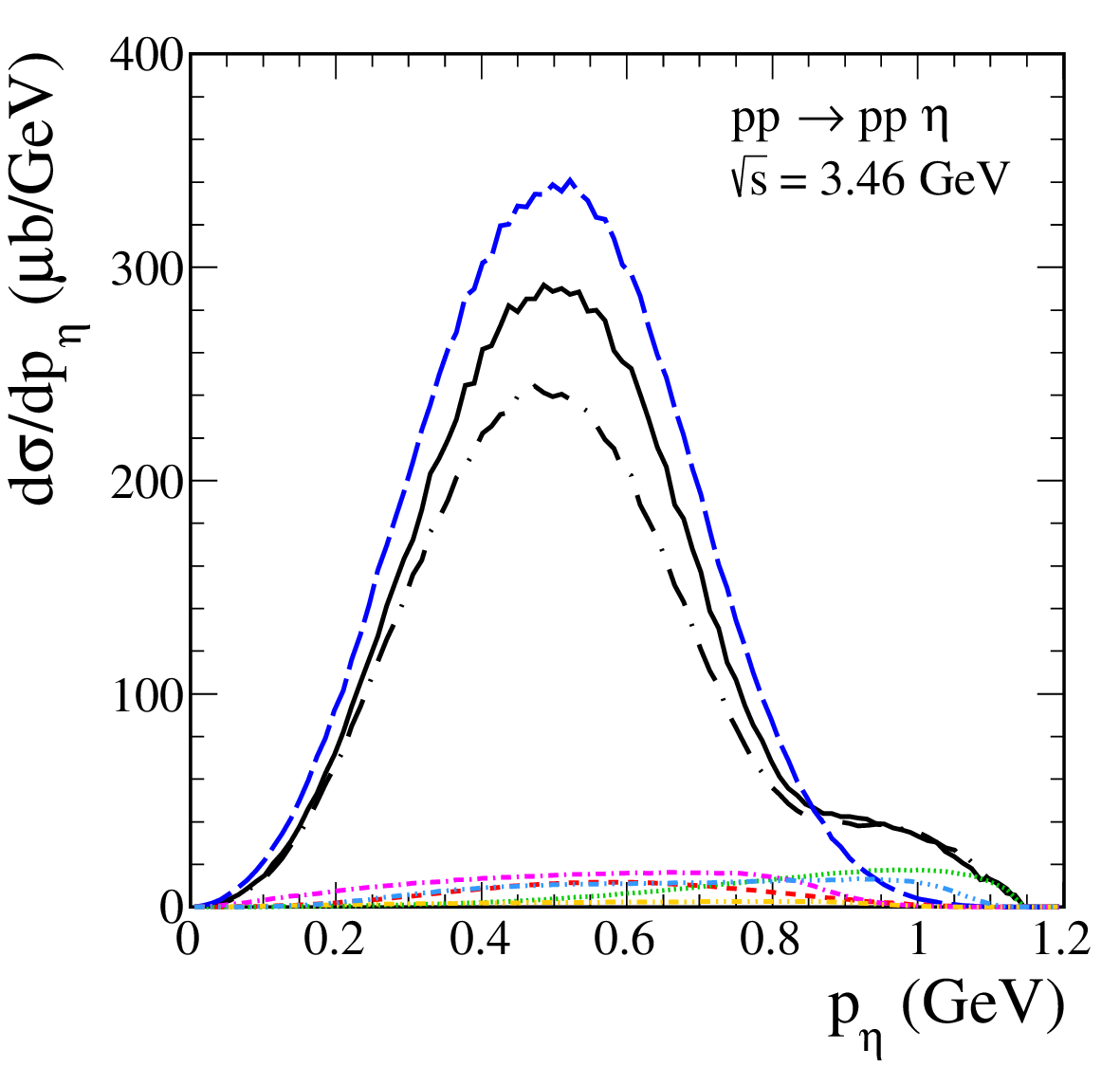}
\caption{Differential cross sections
$d\sigma/d\cos\theta_{\eta}$ and $d\sigma/dp_{\eta}$
for the $p p \to p p \eta$ reaction 
at $\sqrt{s} = 2.748$, 2.978, 3.46~GeV.
The experimental data points are from \cite{Balestra:2004kg}.
The black solid line presents the coherent sum for the model 1, while the black long-dashed-dotted line 
corresponds to calculation for the model 2.
The individual contributions for the model 1 are also shown.}
\label{fig:brems_1}
\end{figure}

\begin{figure}[!ht]
\includegraphics[width=6.8cm]{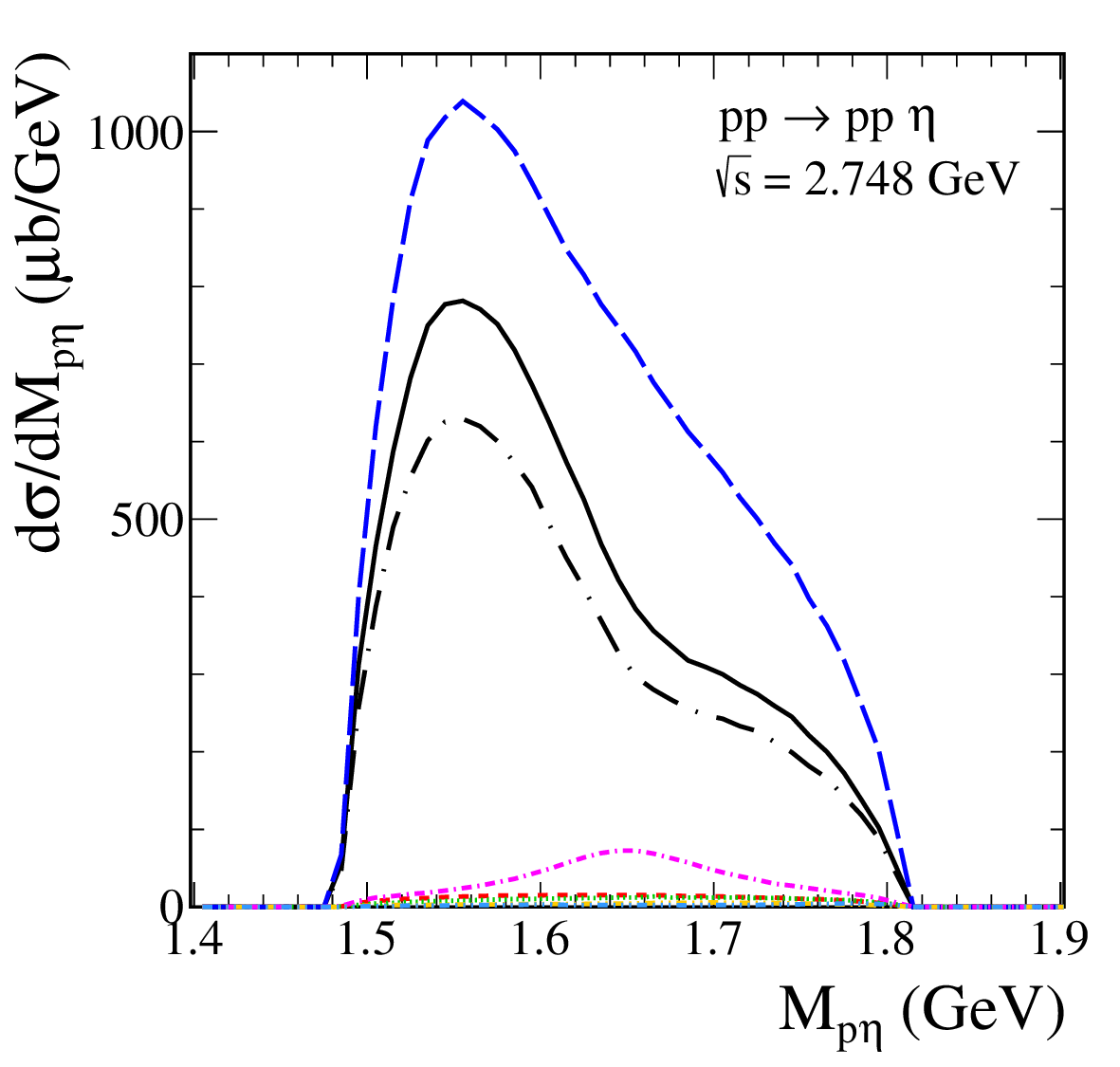}
\includegraphics[width=6.8cm]{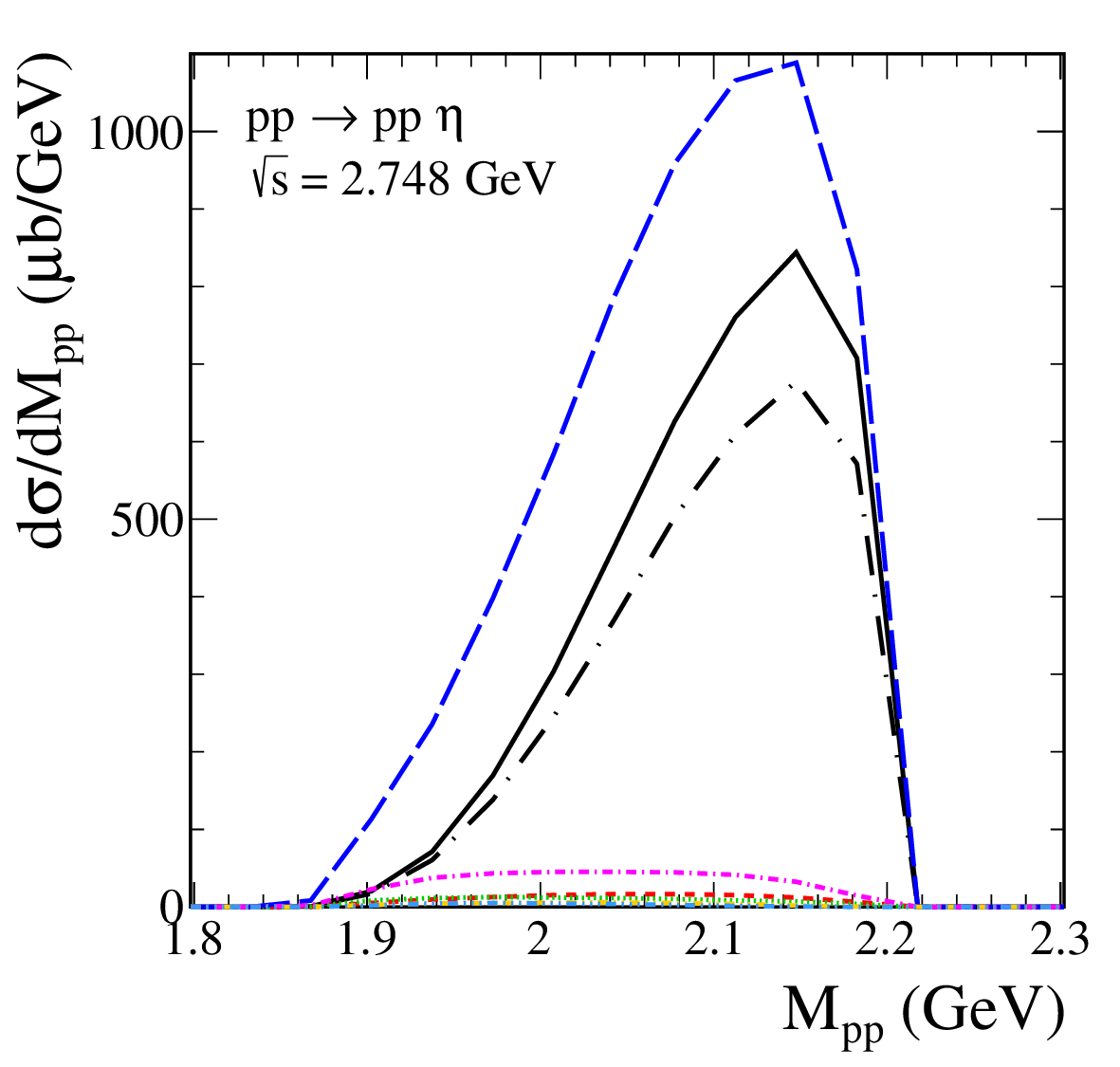}
\includegraphics[width=6.8cm]{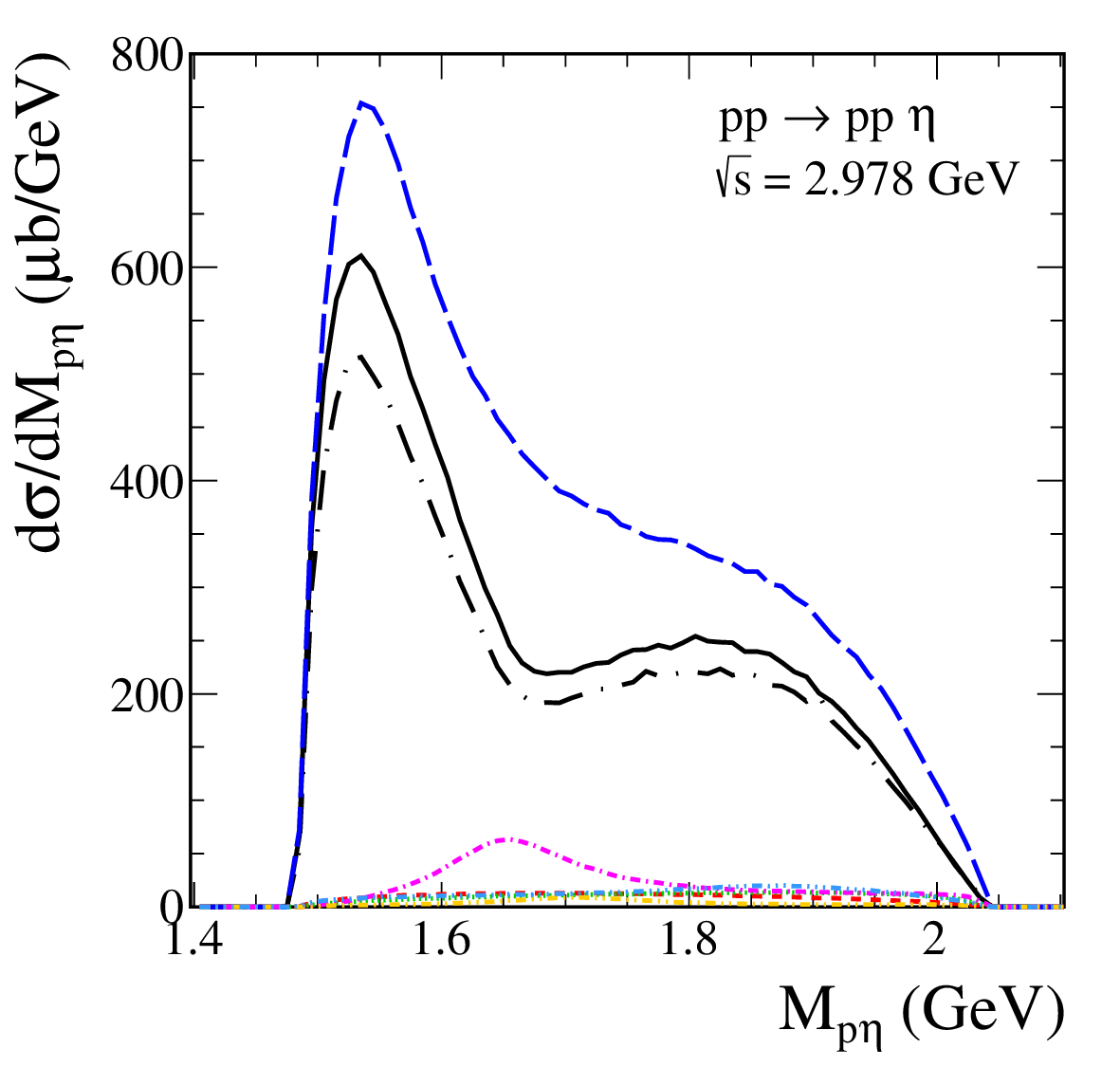}
\includegraphics[width=6.8cm]{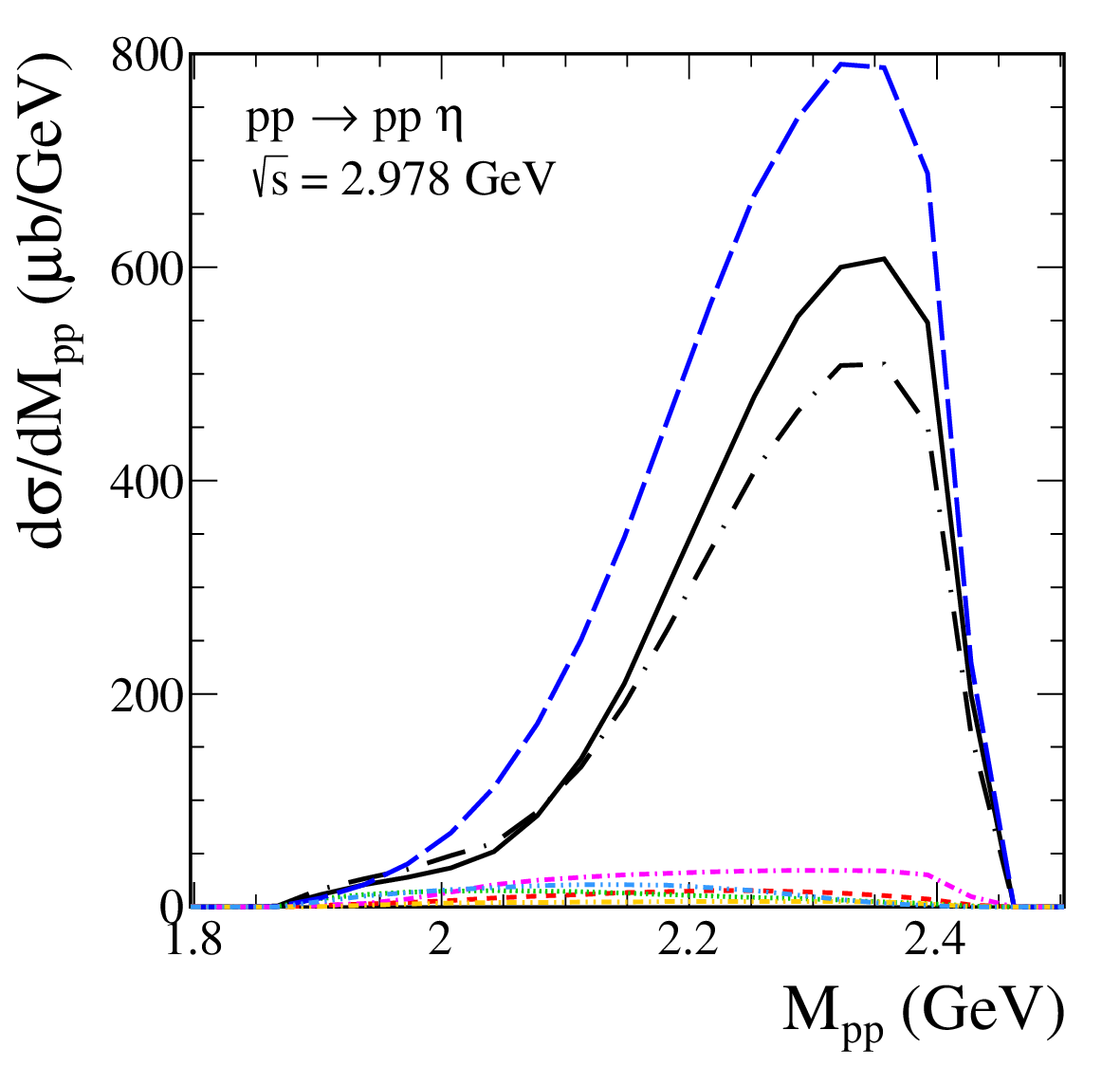}
\includegraphics[width=6.8cm]{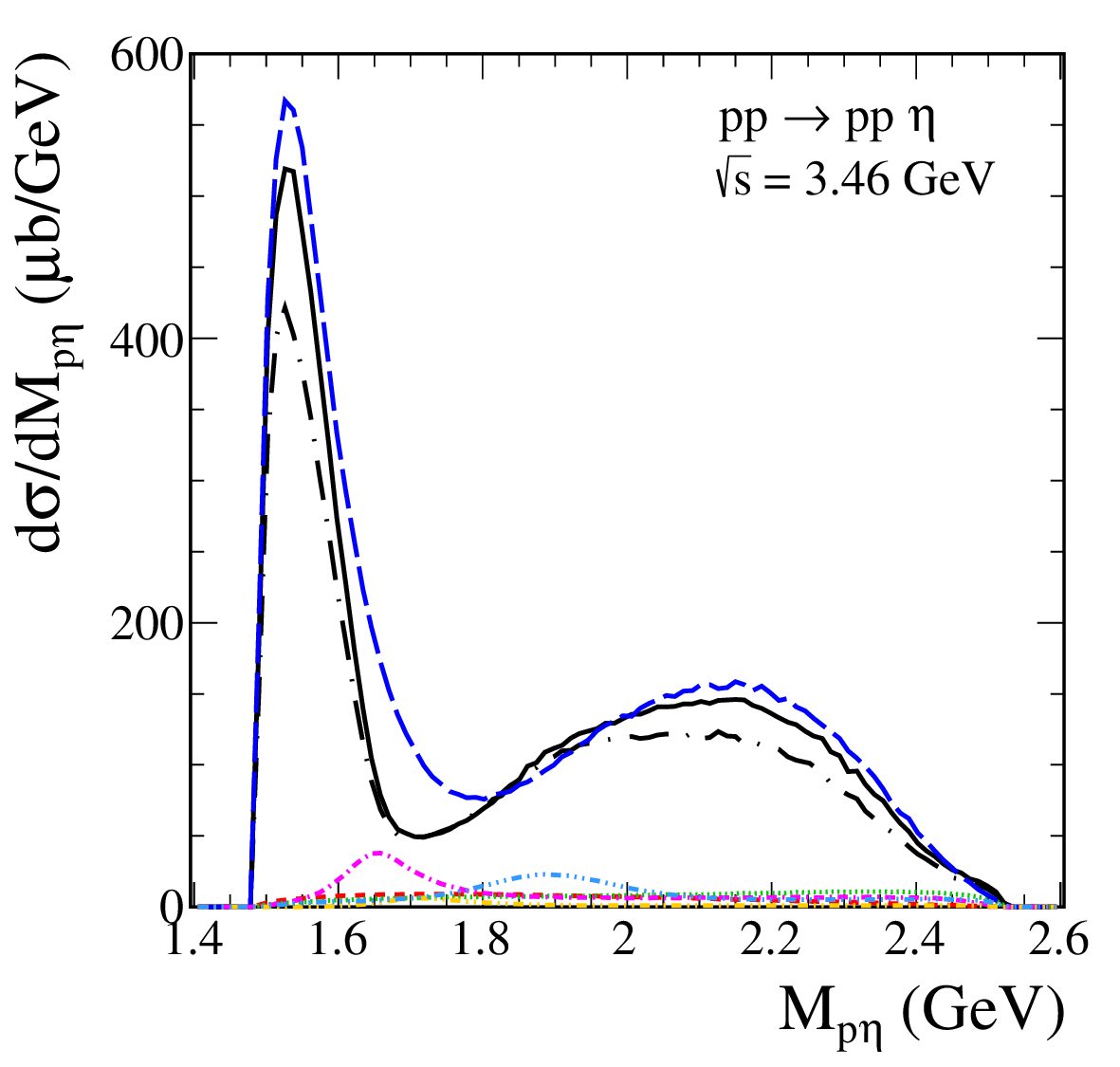}
\includegraphics[width=6.8cm]{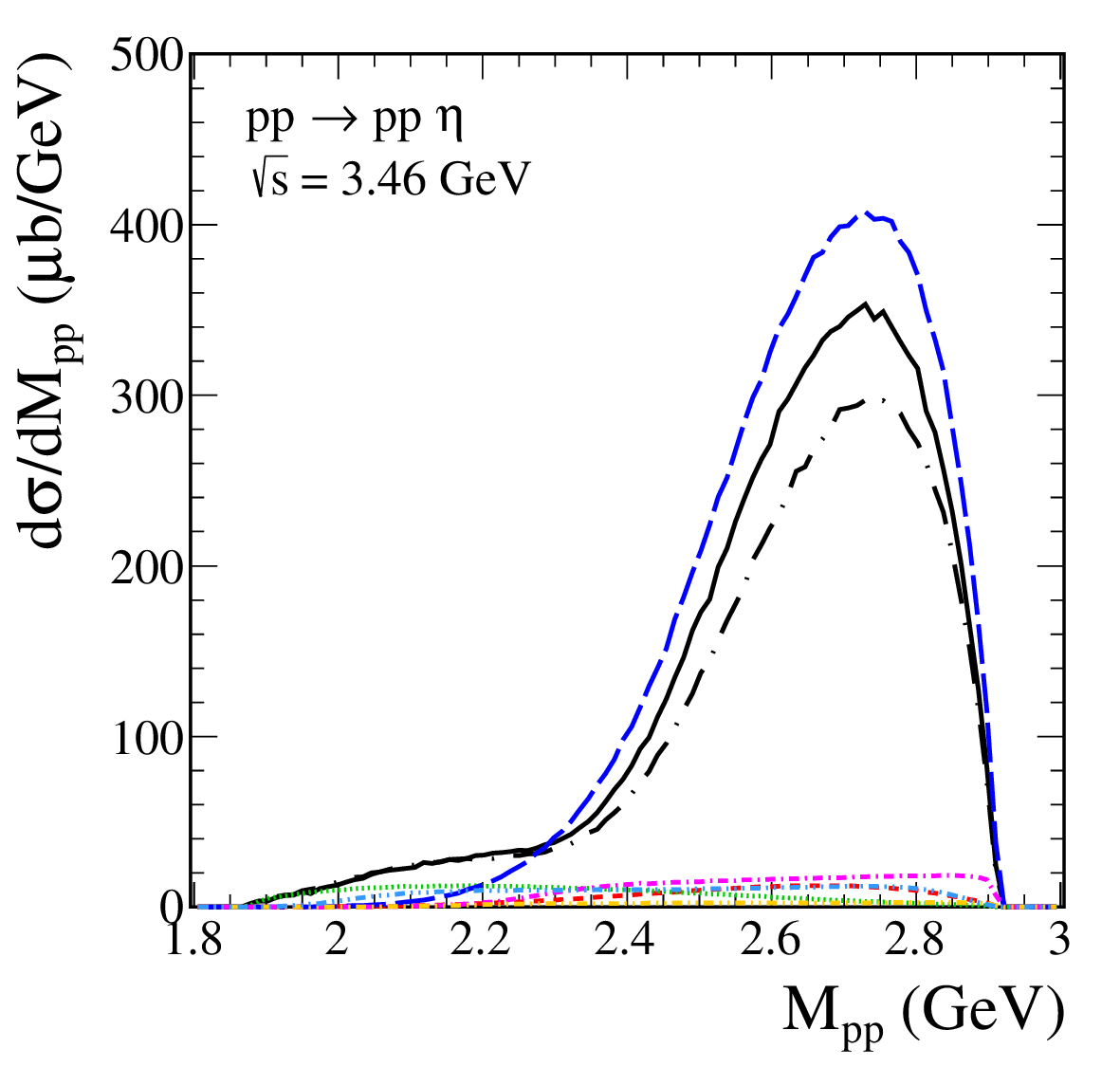}
\caption{Differential cross sections
$d\sigma/dM_{p\eta}$ and $d\sigma/dM_{pp}$
for the $p p \to p p \eta$ reaction 
at $\sqrt{s} = 2.748$, 2.978, 3.46~GeV.
The meaning of the lines
is the same as in Fig.~\ref{fig:brems_1}.}
\label{fig:brems_2}
\end{figure}

\begin{figure}[!ht]
\includegraphics[width=6.8cm]{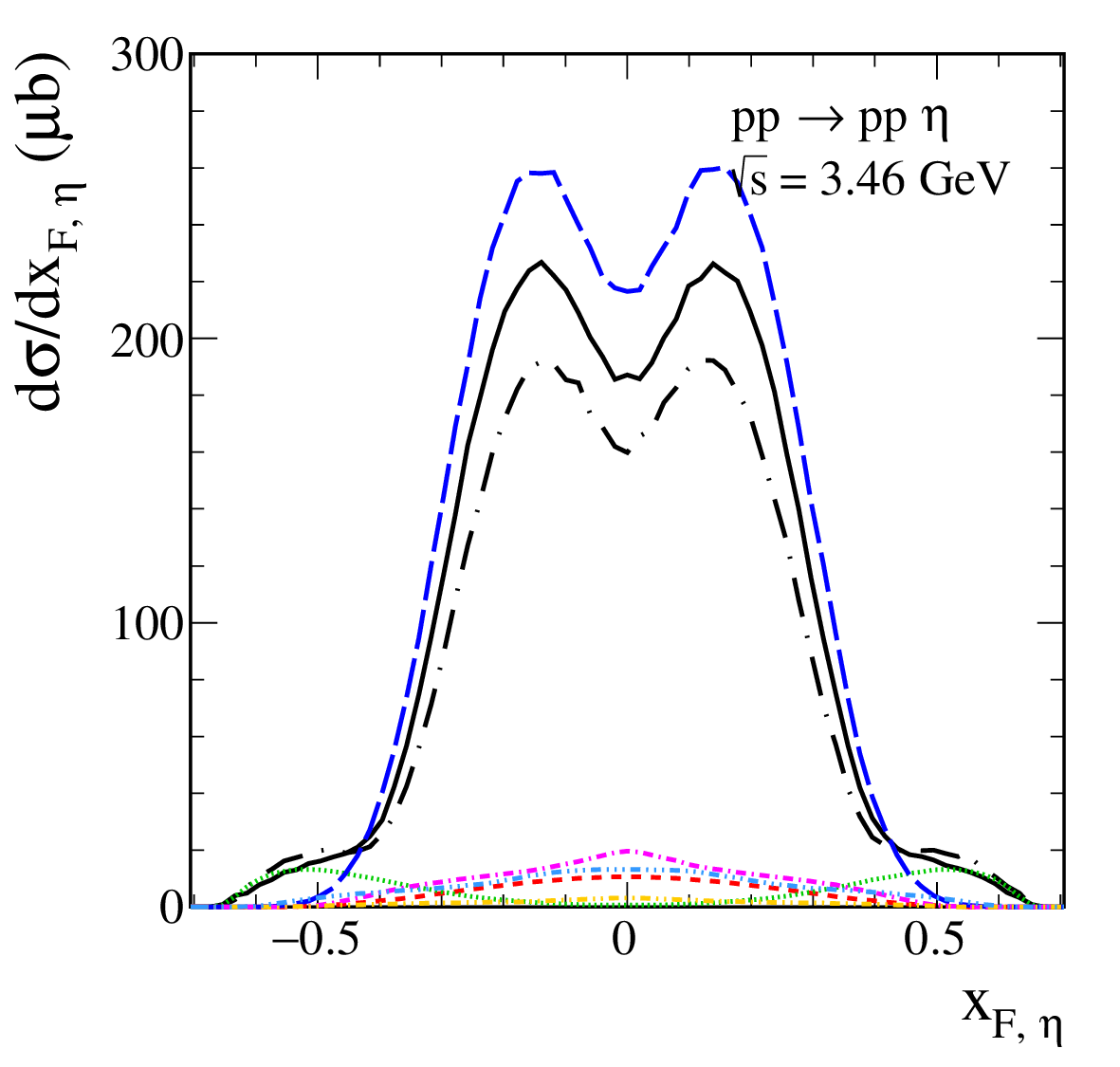}
\includegraphics[width=6.8cm]{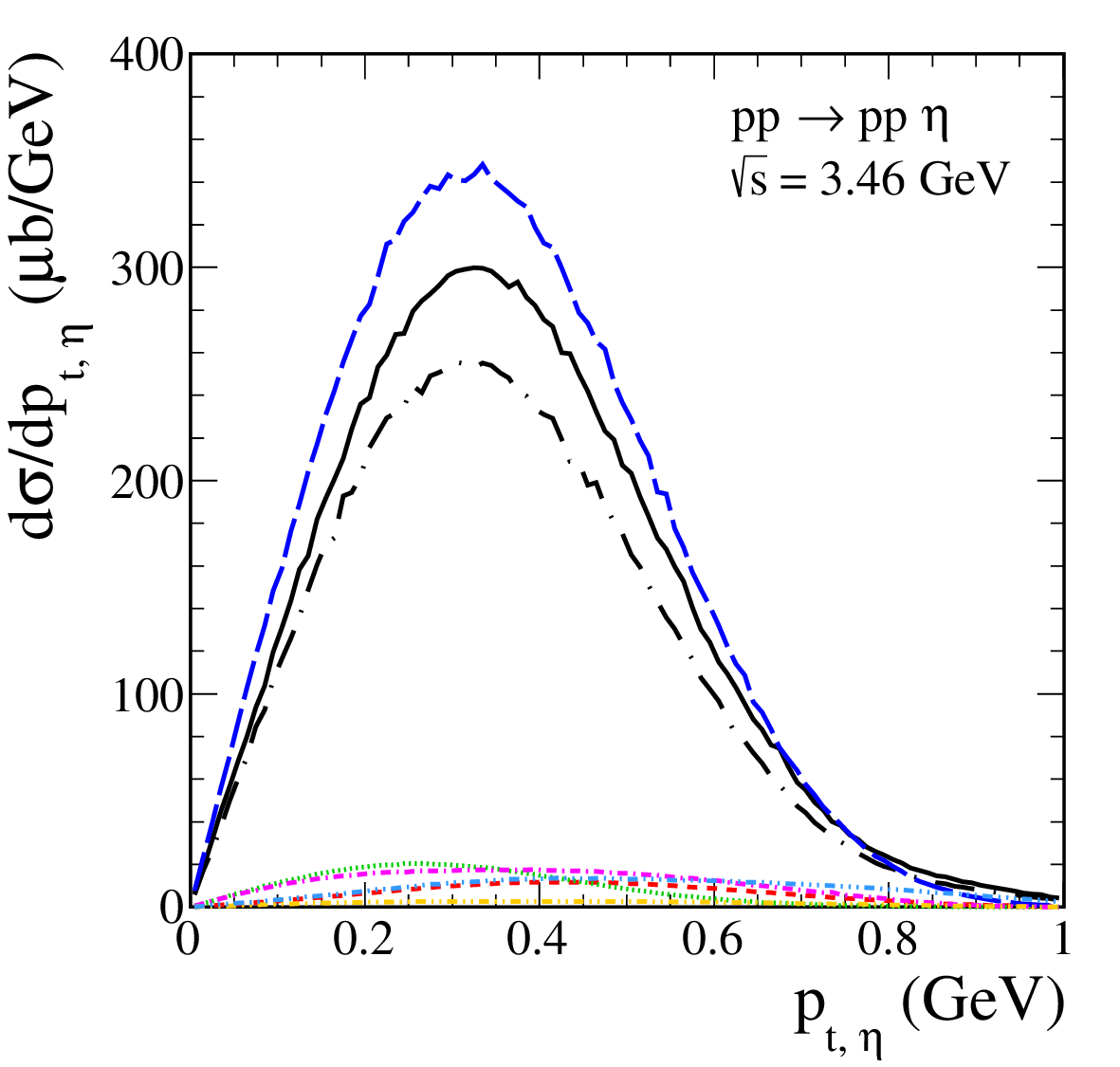}
\includegraphics[width=6.8cm]{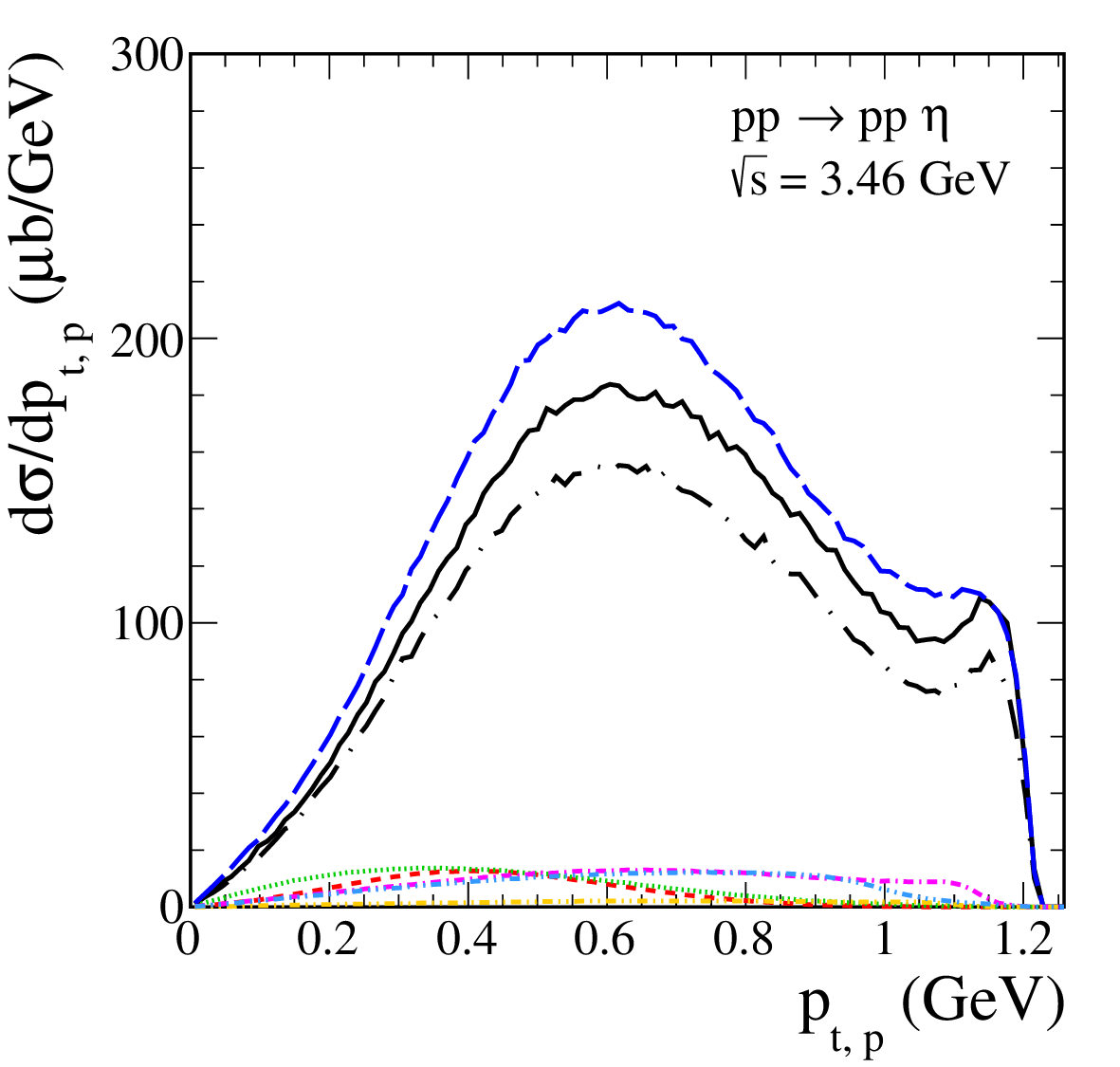}
\caption{Differential cross sections
for the $p p \to p p \eta$ reaction
at $\sqrt{s} = 3.46$~GeV.
The meaning of the lines
is the same as in Fig.~\ref{fig:brems_1}.}
\label{fig:brems_3}
\end{figure}

\begin{figure}[!ht]
\includegraphics[width=6.8cm]{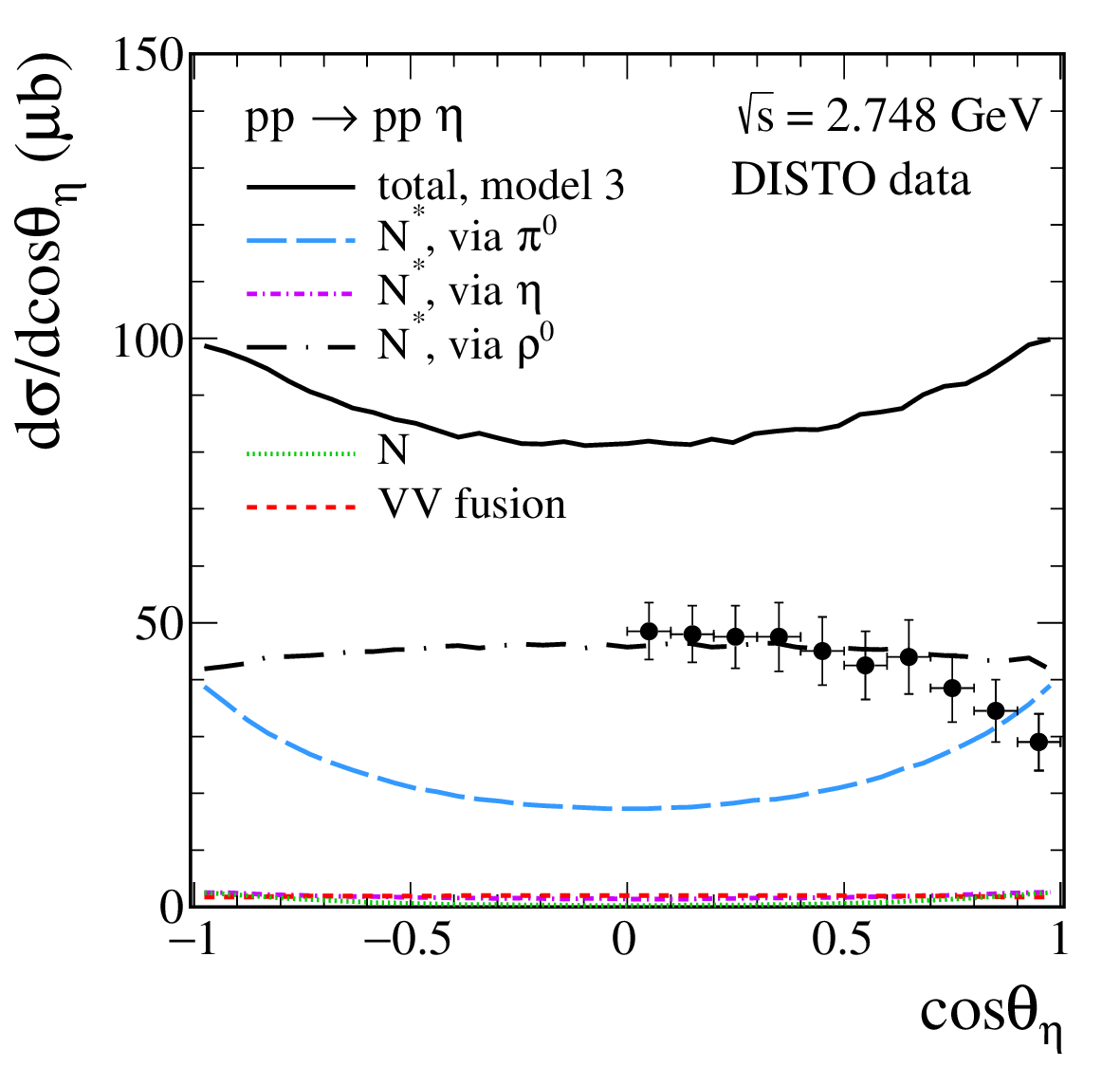}
\includegraphics[width=6.8cm]{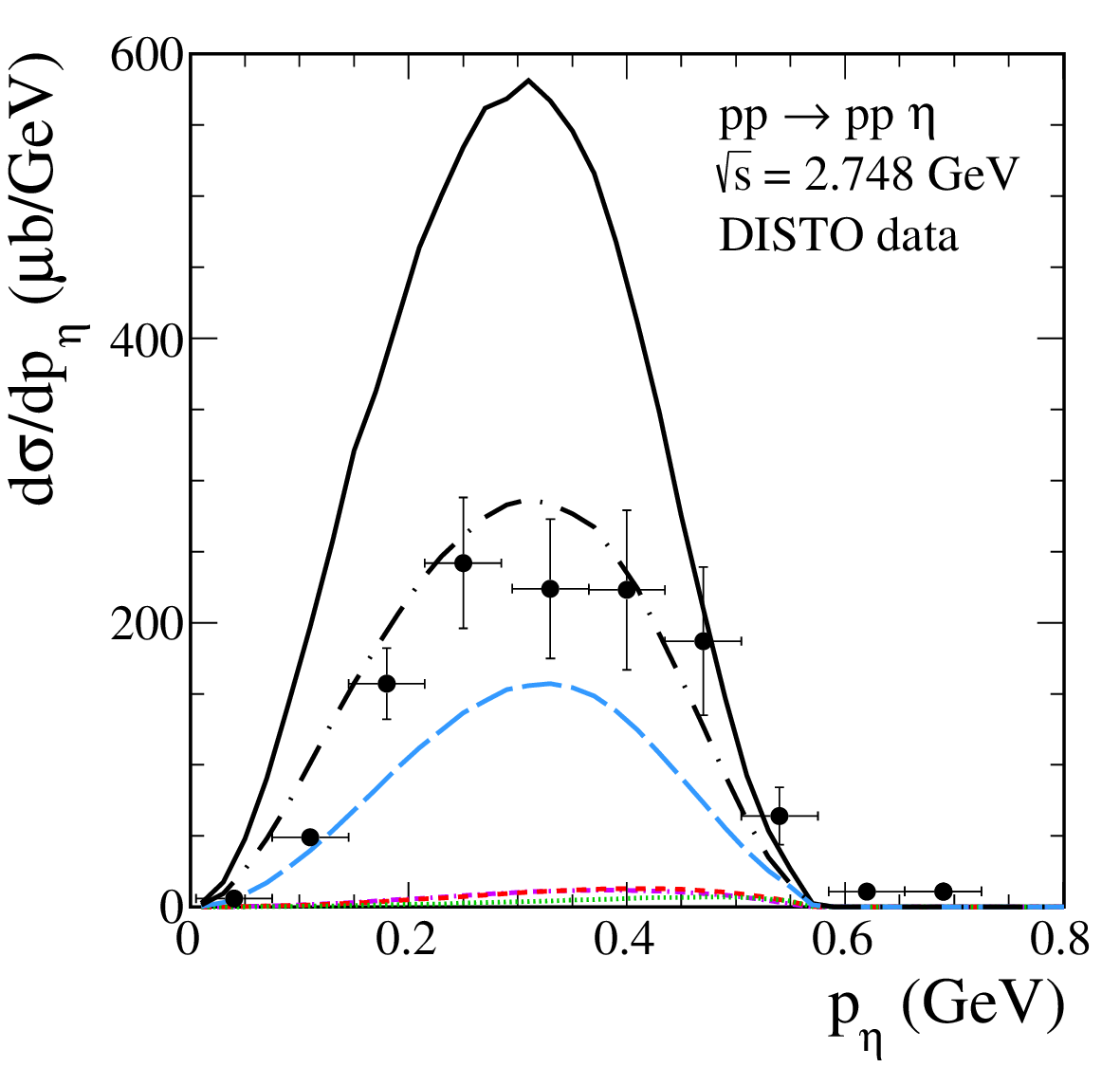}
\includegraphics[width=6.8cm]{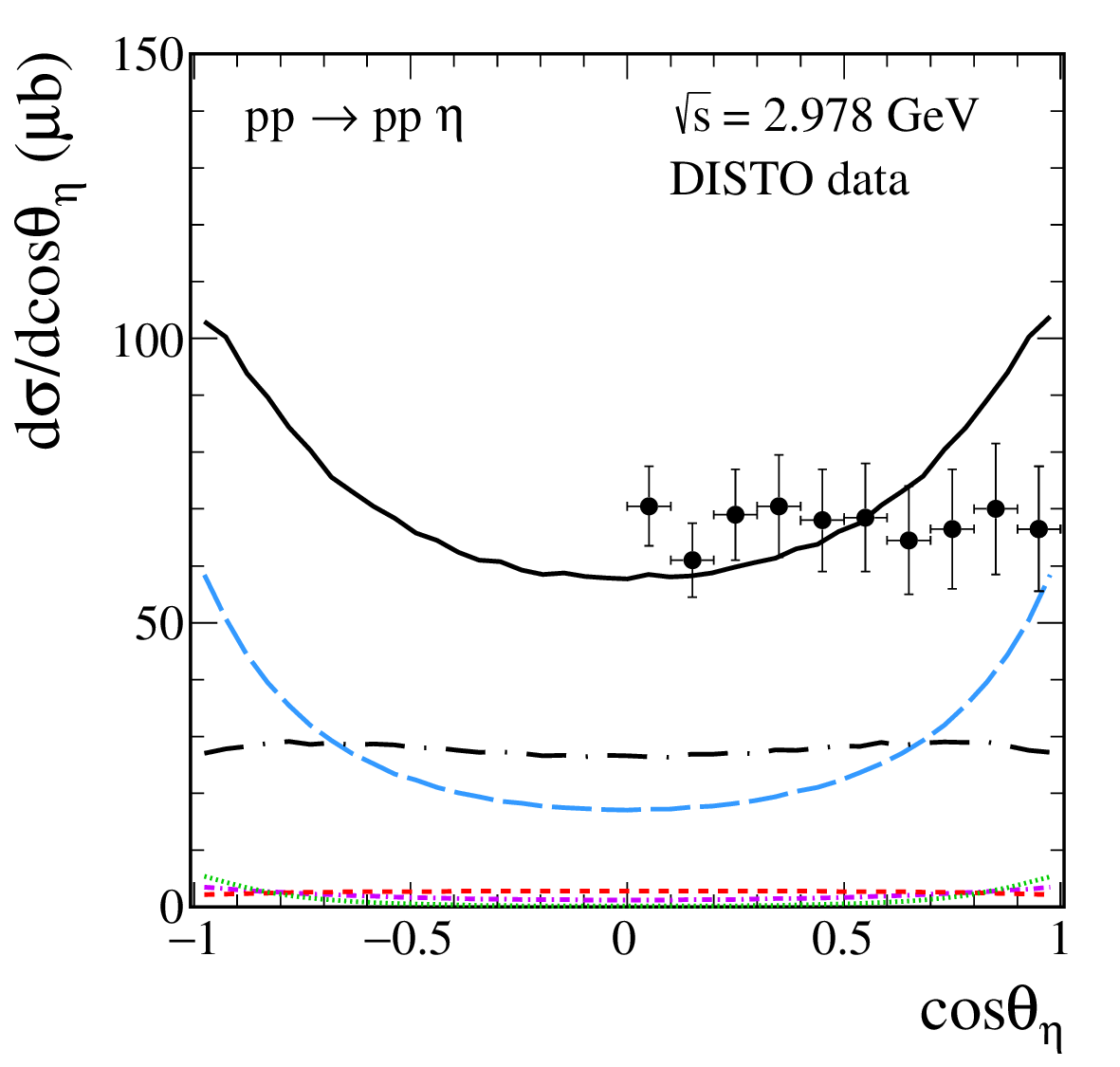}
\includegraphics[width=6.8cm]{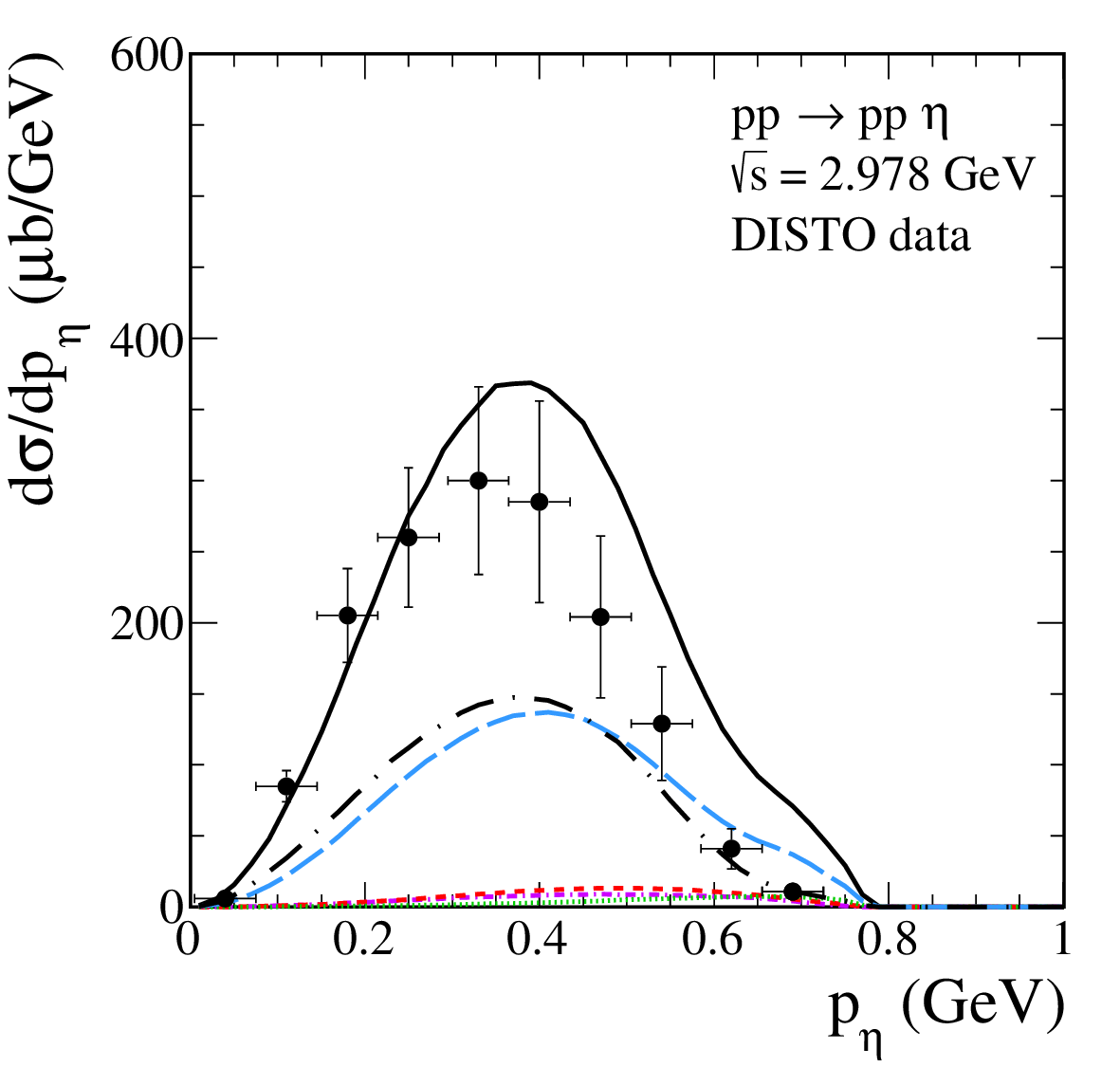}
\includegraphics[width=6.8cm]{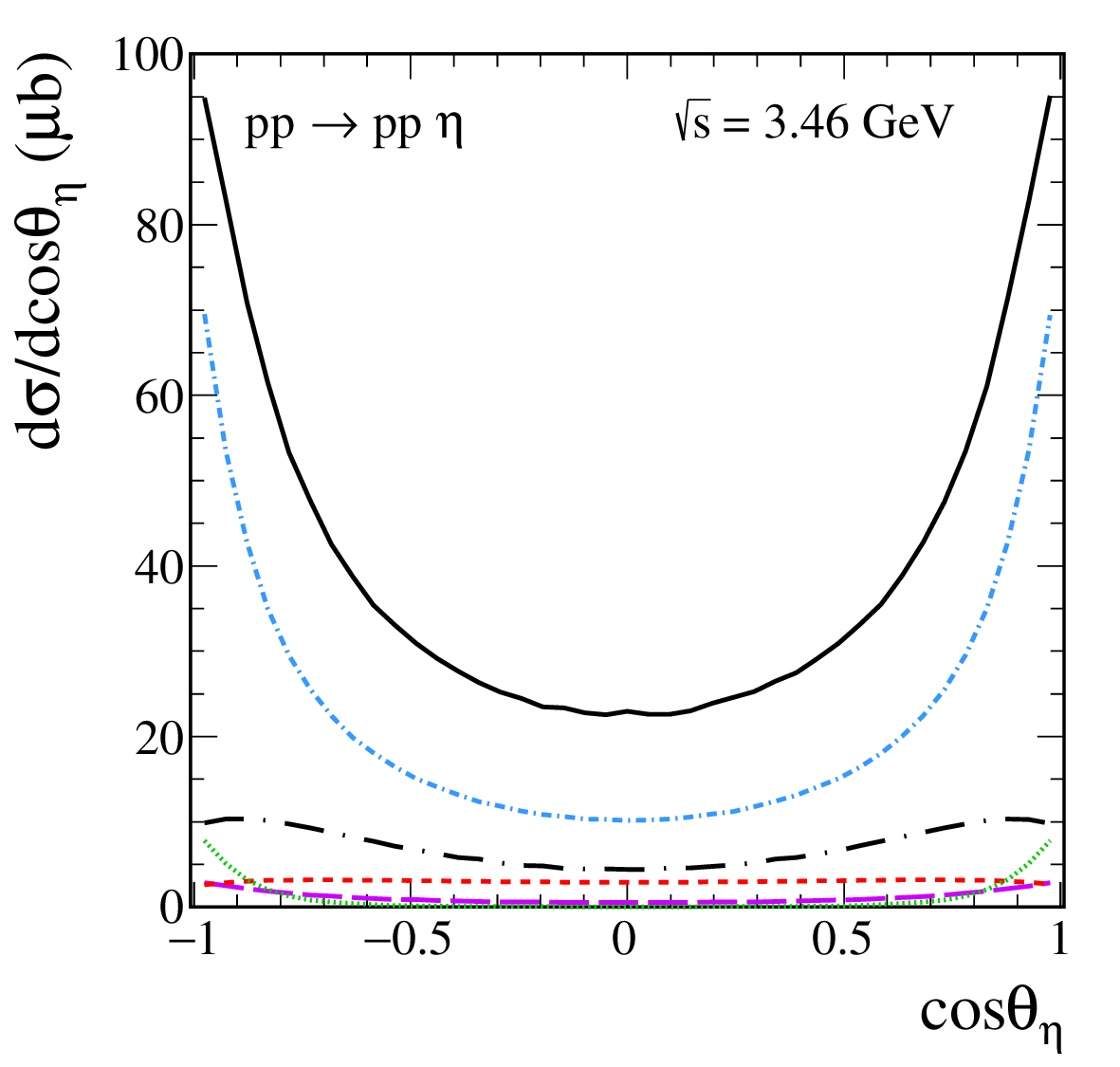}
\includegraphics[width=6.8cm]{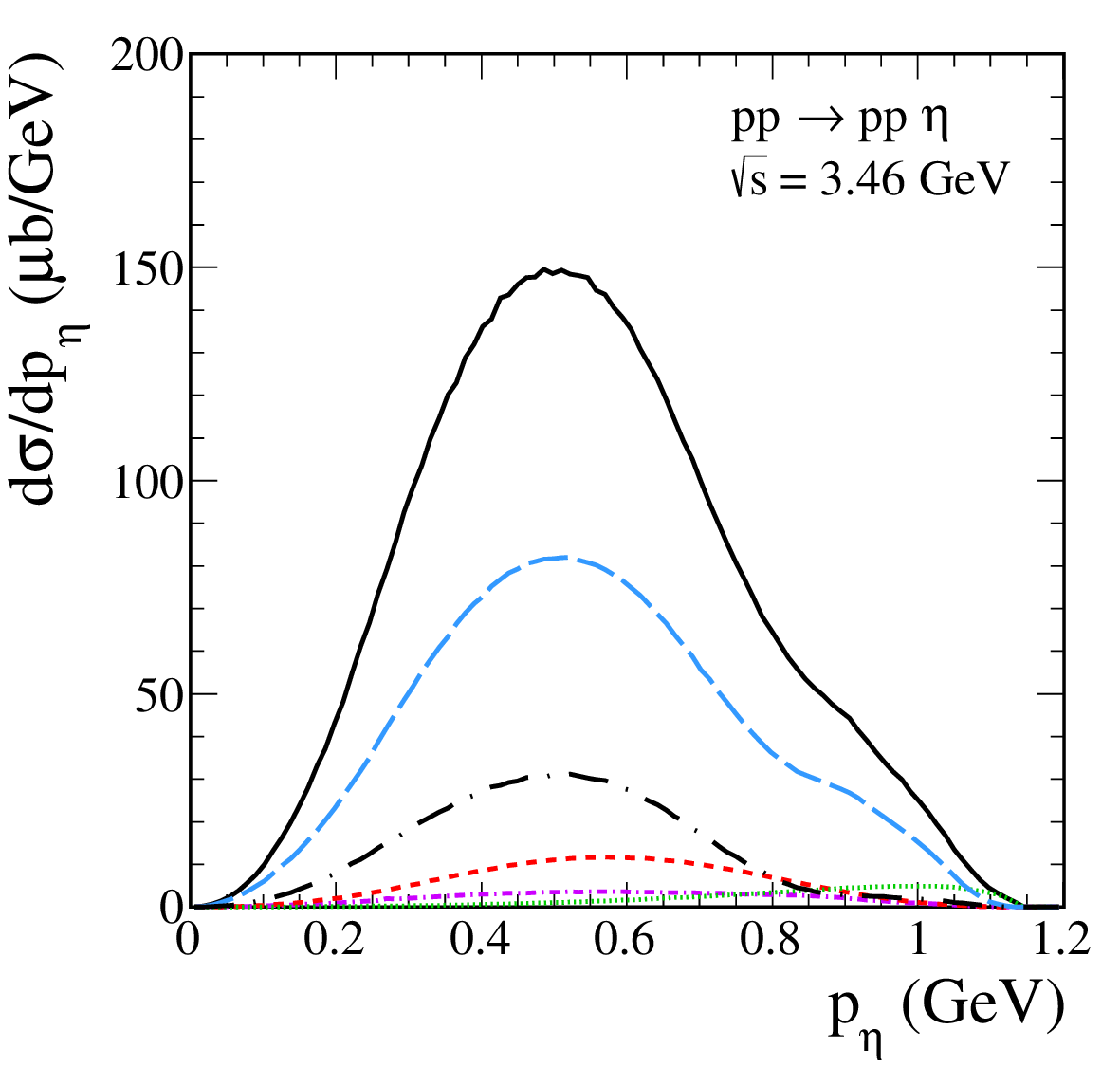}
\caption{Differential cross sections
$d\sigma/d\cos\theta_{\eta}$ and $d\sigma/dp_{\eta}$
for the $p p \to p p \eta$ reaction 
at $\sqrt{s} = 2.748$, 2.978, 3.46~GeV.
The experimental data points are from \cite{Balestra:2004kg}.
The results for model~3 are shown.
The black solid line represents the coherent sum 
of all contributions ($N$, $VV$ fusion, and $N^{*}$ terms). 
The blue long-dashed line 
represents the result for $N^{*}$ resonances 
excited via $\pi^{0}$ exchange,
the violet short-dashed-dotted corresponds to $N^{*}$ contribution
via $\eta$ exchange,
and the black long-dashed-dotted line 
corresponds to $N^{*}$ contribution via $\rho^{0}$ exchange.}
\label{fig:brems_4}
\end{figure}

\begin{figure}[!ht]
\includegraphics[width=6.8cm]{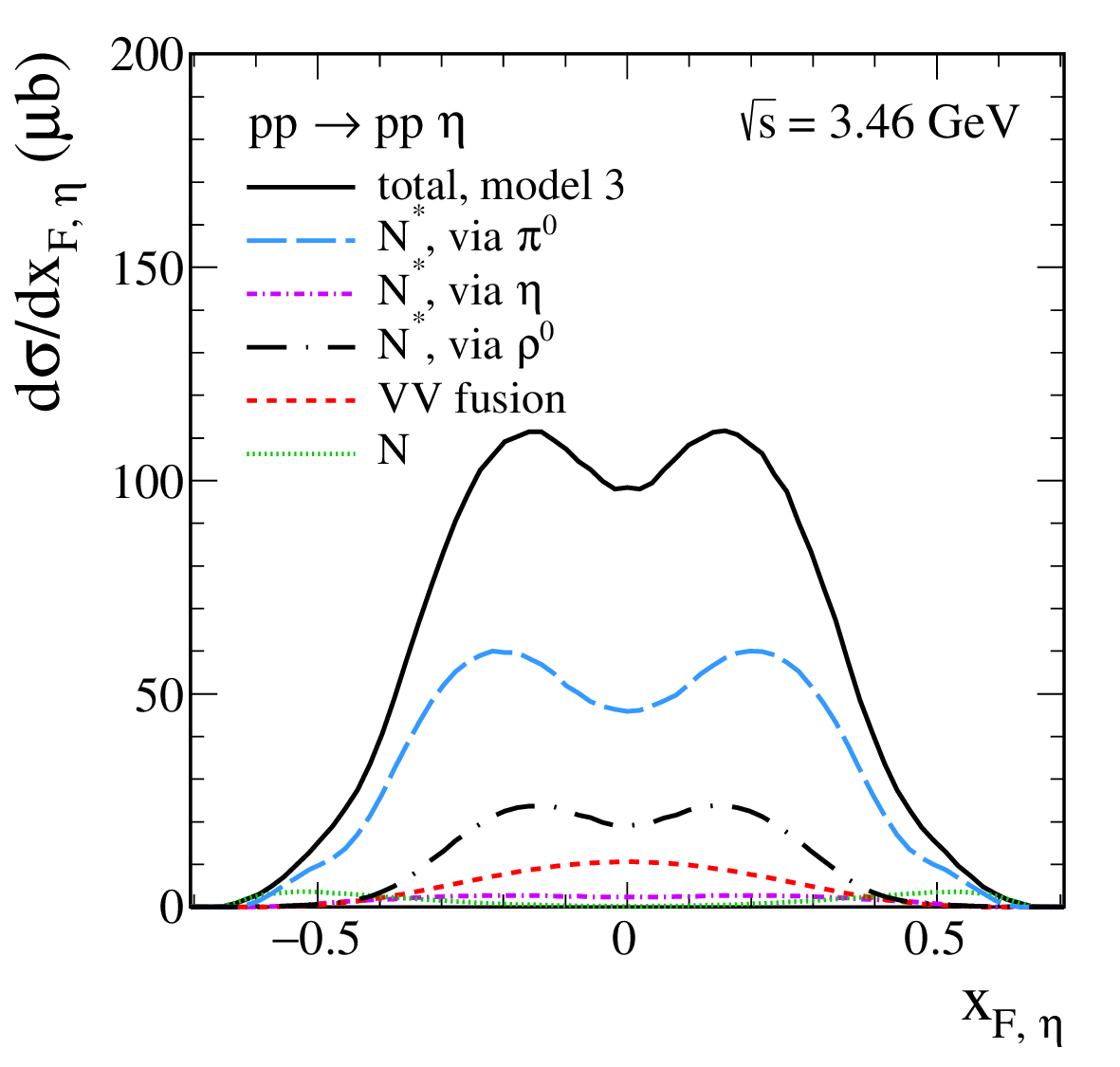}
\includegraphics[width=6.8cm]{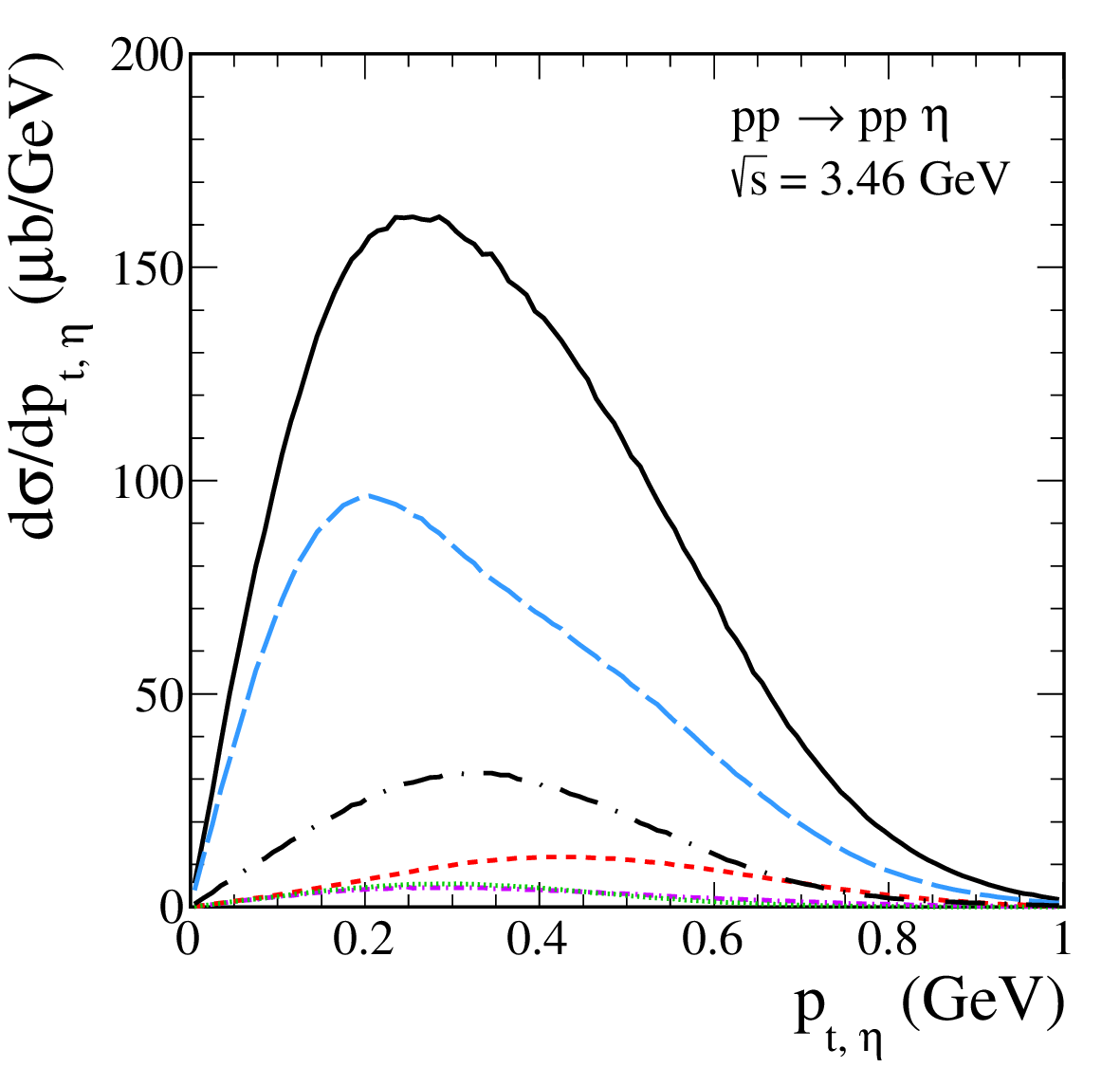}
\includegraphics[width=6.8cm]{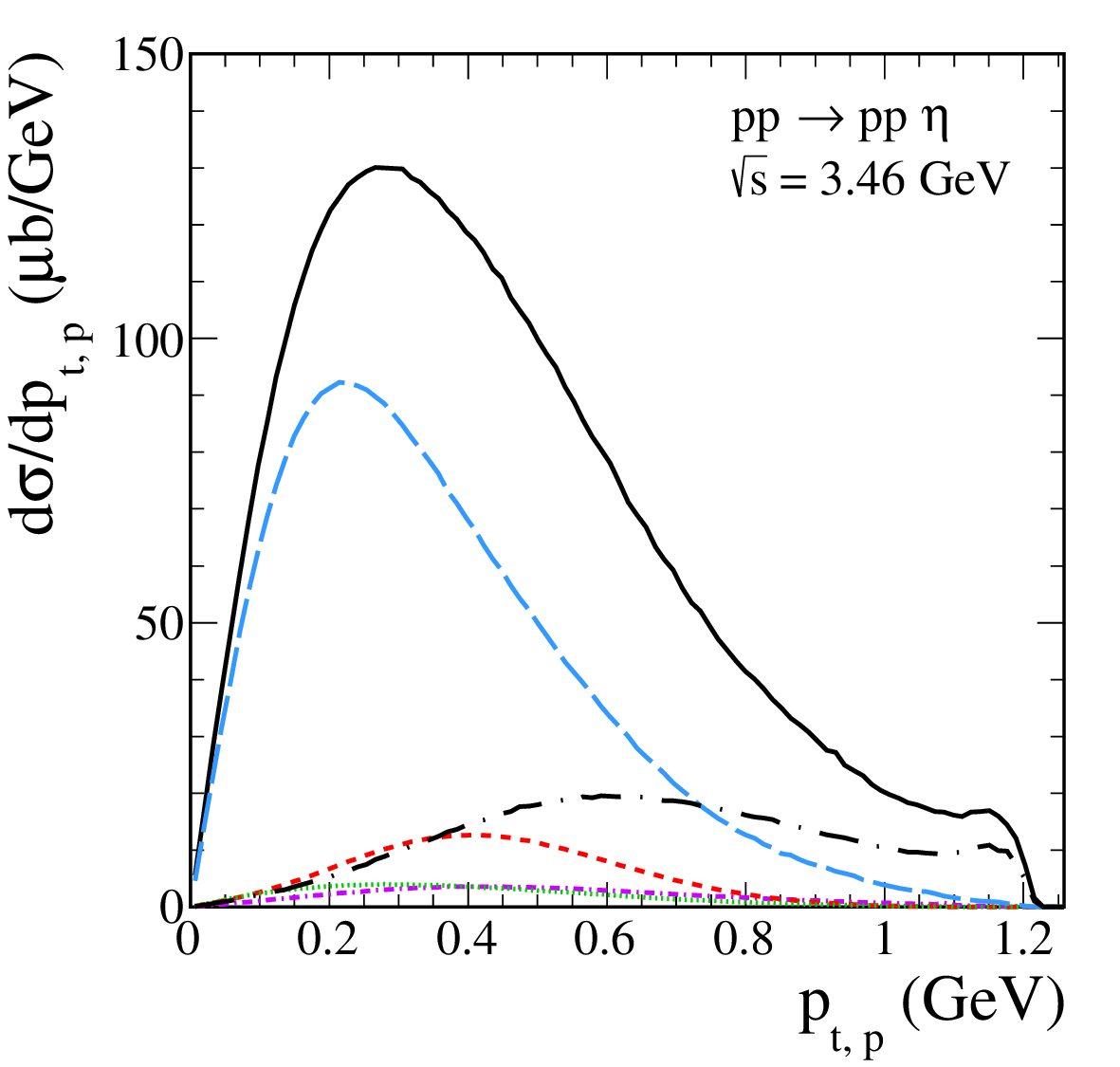}
\includegraphics[width=6.8cm]{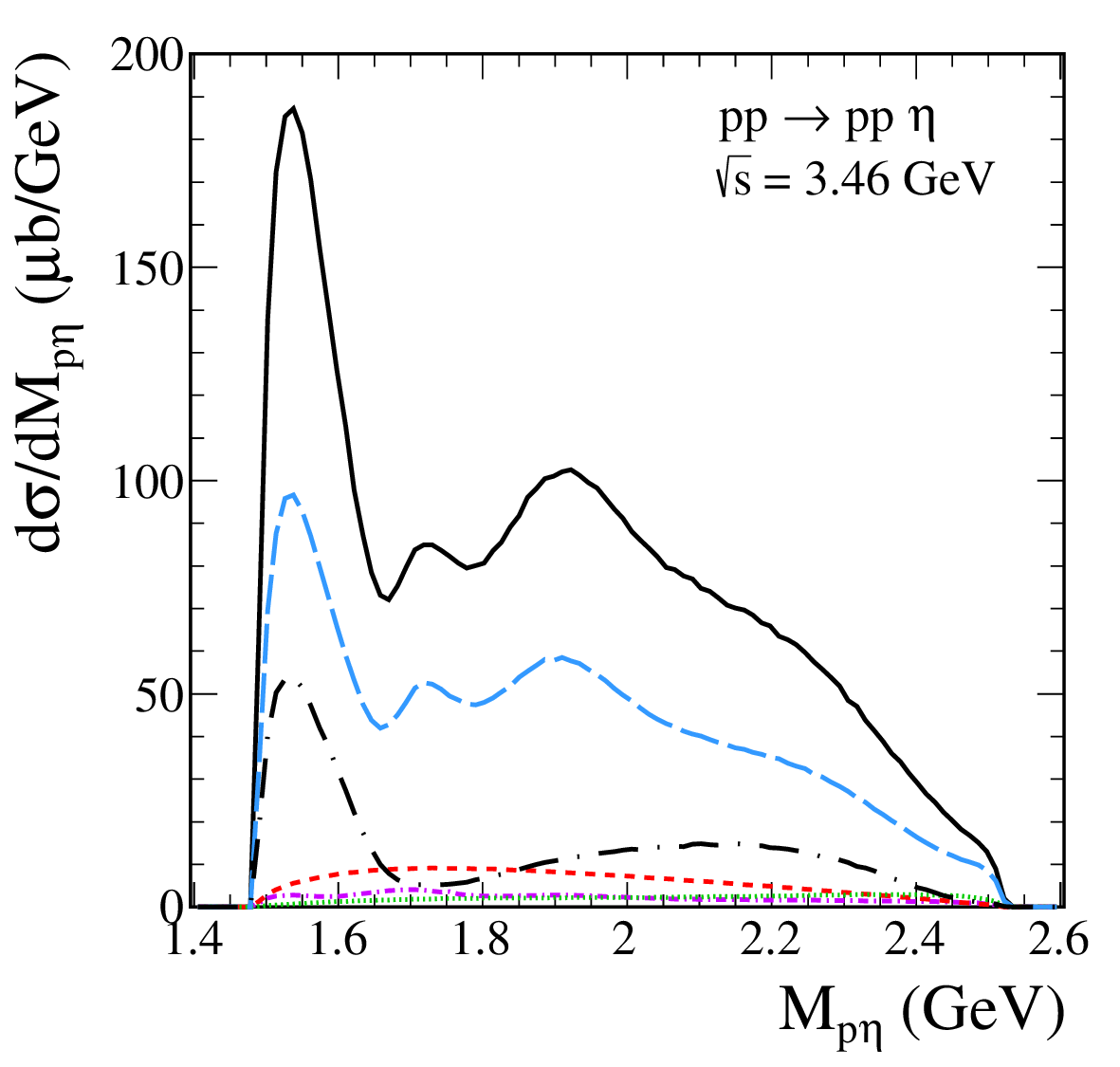}
\includegraphics[width=6.8cm]{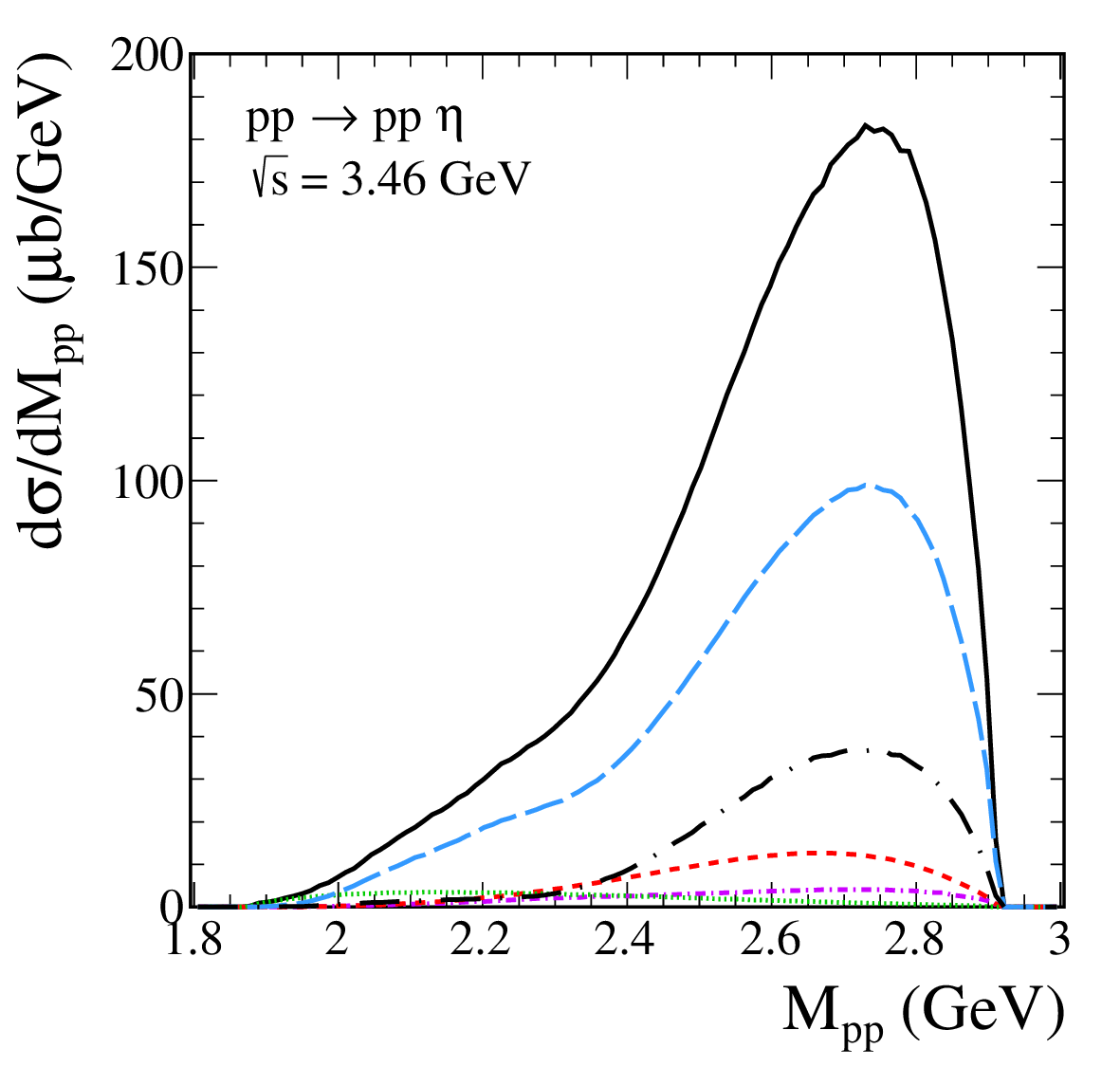}
\caption{Differential cross sections
for the $p p \to p p \eta$ reaction
at $\sqrt{s} = 3.46$~GeV.
The meaning of the lines
is the same as in Fig.~\ref{fig:brems_4}.}
\label{fig:brems_5}
\end{figure}

\begin{figure}[!ht]
\includegraphics[width=6.8cm]{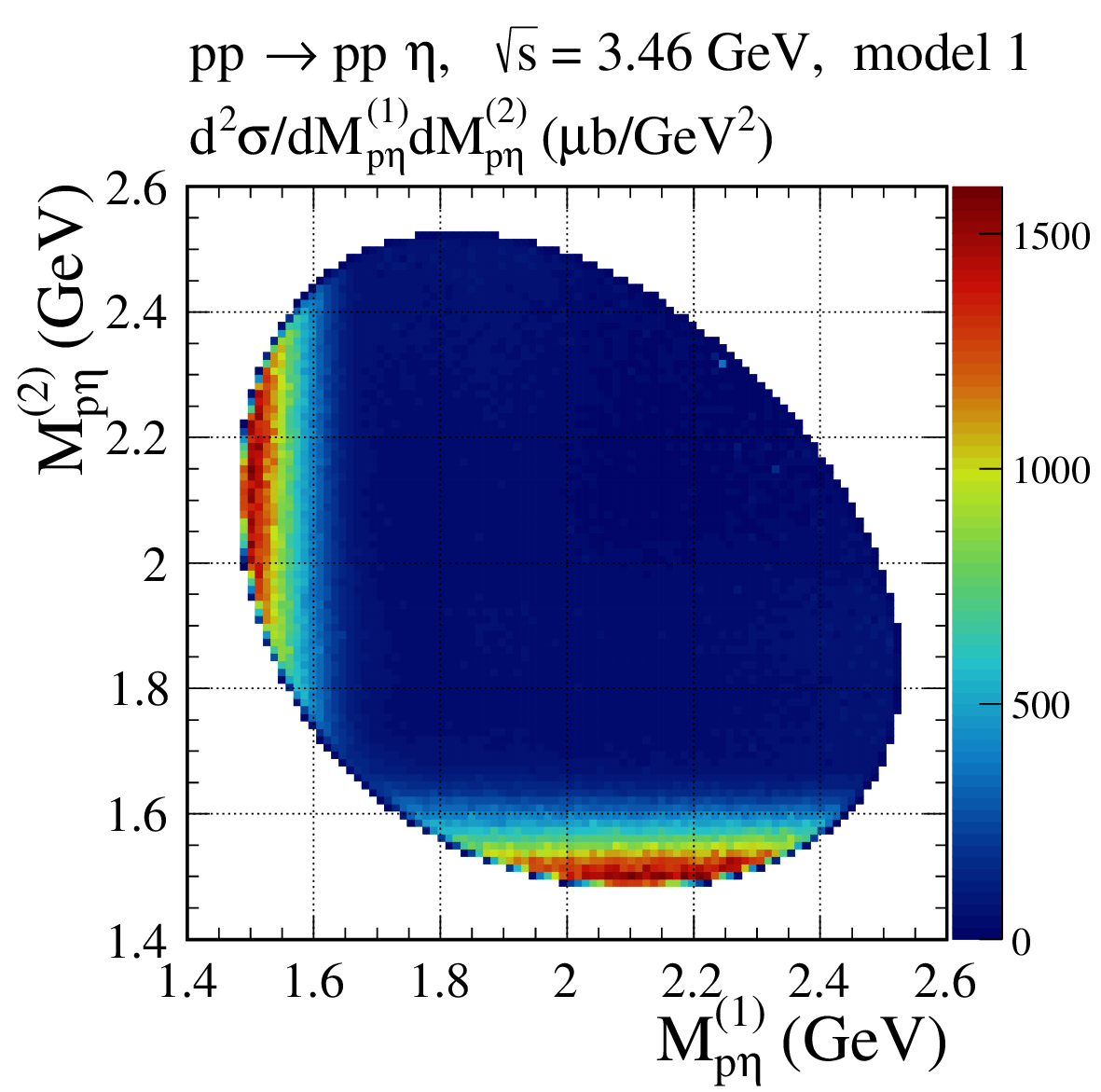}
\includegraphics[width=6.8cm]{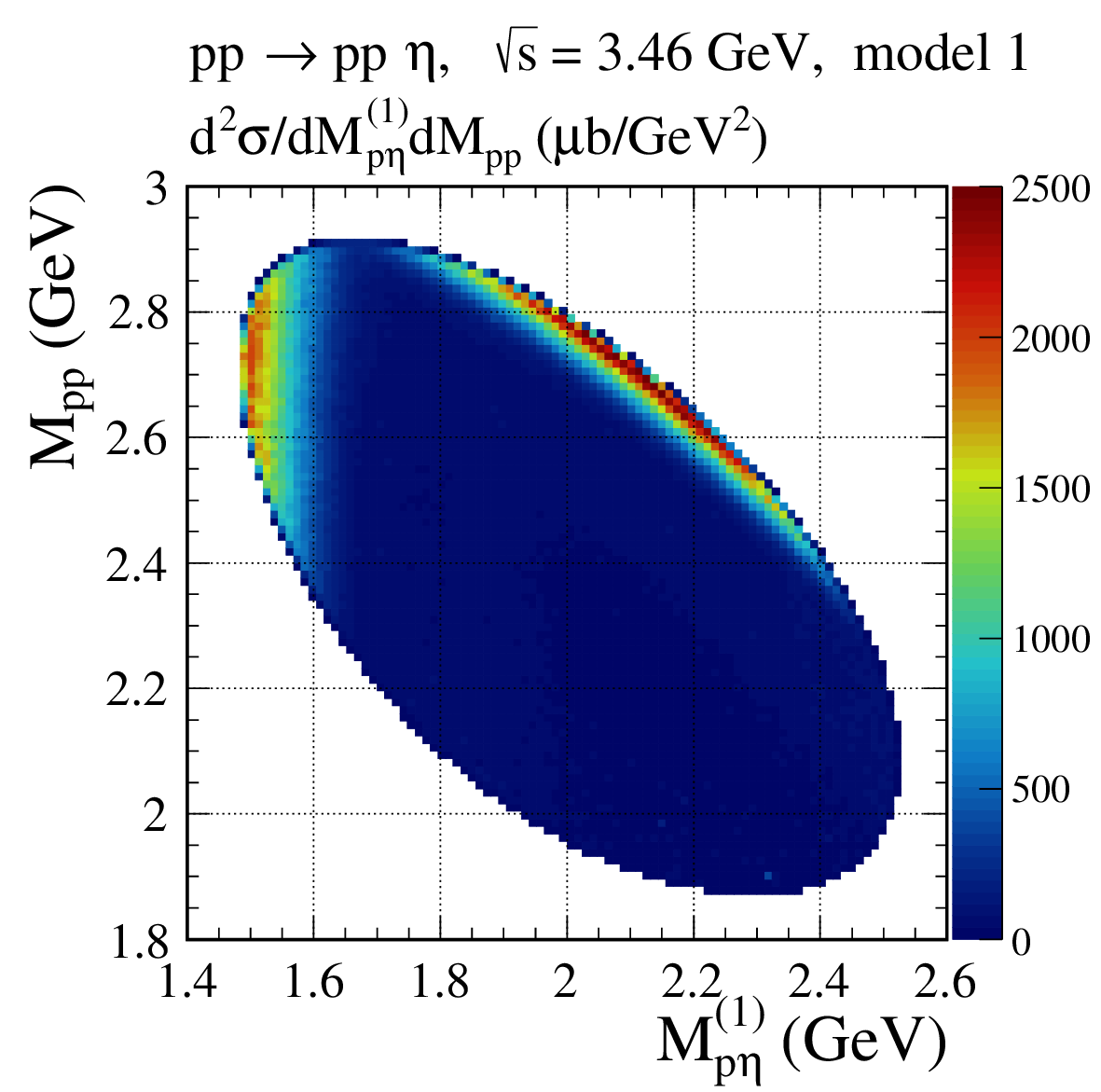}\\
\includegraphics[width=6.8cm]{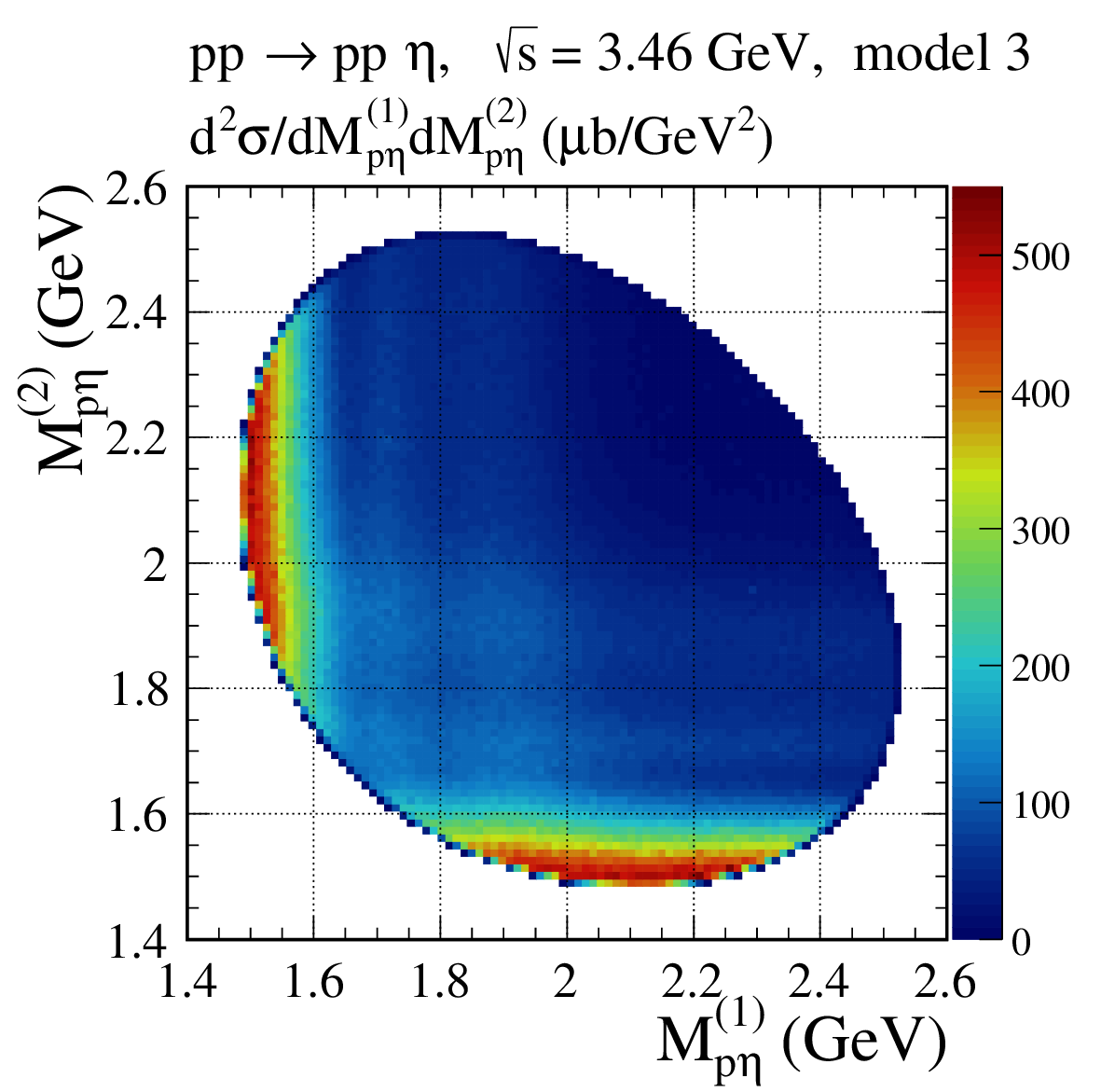}
\includegraphics[width=6.8cm]{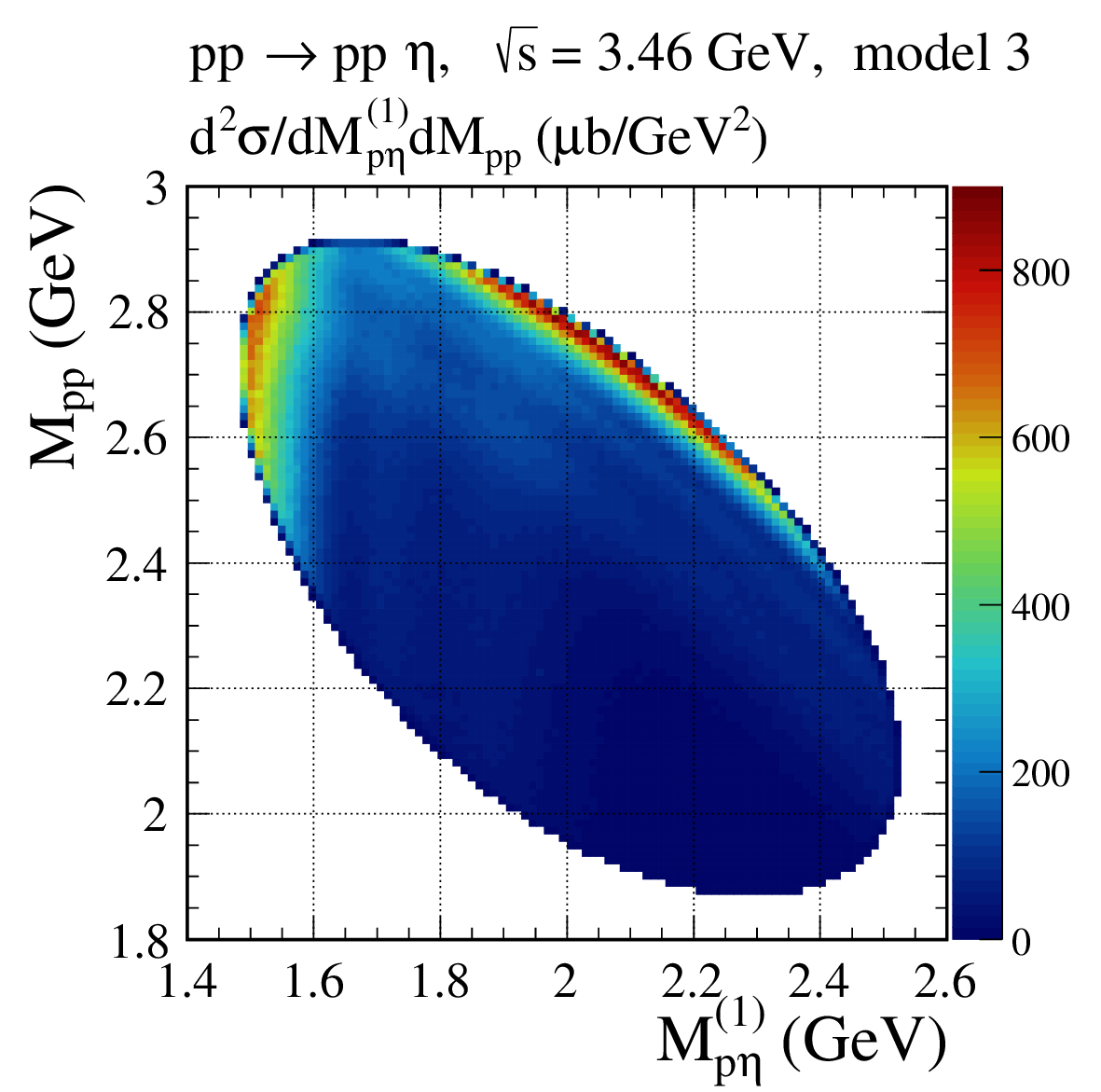}
\caption{The two-dimensional distributions in ($M_{p \eta}^{(1)}$, $M_{p \eta}^{(2)}$)
and in ($M_{p \eta}^{(1)}$, $M_{pp}$) for the $p p \to p p \eta$ reaction
for model~1 (top panels) and for model~3 (bottom panels) 
calculated for $\sqrt{s} = 3.46$~GeV.}
\label{fig:brems_6}
\end{figure}

\end{widetext}

\section{Summary and Conclusions}
\label{sec:conclusions}

The present study investigates the $\eta$ meson exclusive production in the $pp \to pp \eta$ reaction at energies relevant 
for the HADES, PANDA, and SIS100 experiments at GSI-FAIR.
The production mechanism was studied
by using an effective Lagrangian approach.
Concretely, the $\eta$-bremsstrahlung mechanism
with the intermediate proton exchange 
via the $\pi^{0}$, $\eta$, $\rho^{0}$, and $\omega$ exchanges, 
the mechanism with an excitation of nucleon resonances
[$N(1535)$, $N(1650)$, $N(1710)$, $N(1880)$]
via the pseudoscalar- or/and vector-meson exchanges
(depends on the model),
as well as the $\omega \omega$- and $\rho^0 \rho^0$-fusion mechanism
were considered.
The model results have been compared with 
the available DISTO \cite{Balestra:2004kg}
and HADES 
\cite{Teilab:2010gv,Teilab_thesis,HADES:2012aa} data.
The numerical analysis indicates that 
the $N(1535)$ contribution is the most important process 
at lower energies, but other contributions 
cannot be omitted due to interference effects. 
Resonances like $N(1650)$, $N(1710)$, $N(1880)$ 
and vector-meson fusion processes contribute 
significantly to the considered reaction,
but their relative importance has to be verified by experiment.

To determine the parameters of the model, 
the $\pi^{-} p \to \eta n$ and $\gamma p \to \eta p$ reactions 
were discussed.
An attempt was made to describe both integrated and differential
cross-section data for these processes, 
in which the nucleon resonances play a crucial role.
The use of reggeized-$\rho$ and -$\omega$ exchanges 
for the photoproduction of the $\eta$ meson is justified at higher energies and in the forward scattering region. 
There are new CLAS experimental data \cite{Hu:2020ecf}. 
By comparing the model to experimental data
for $\pi^{-} p \to \eta n$ and $\gamma p \to \eta p$,
the form factors and the cutoff parameters occurring in the vertices were fixed.
The coupling constants $g_{VV \eta}$ 
($V$ stands for $\rho^{0}$ or $\omega$) were extracted from the radiative decay rates of $V \to \eta \gamma$ using an effective Lagrangian approach and a vector-meson-dominance ansatz.
The $\pi N N^{*}$ and $\eta N N^{*}$ coupling constants
are known from the relevant partial decay widths; 
see Table~\ref{tab:table_par2} in Appendix~\ref{sec:appendixA}. 
The $\rho N N^{*}$ coupling constants can be constrained
by the available experimental data from
the radiative decays $N^{*} \to N \rho^{0} \to N \gamma$ as shown in Appendix~\ref{sec:appendixC}.
Additionally,
the $N(1535) \to N \rho^{0} \to N \pi^{+} \pi^{-}$ decay
was discussed.
These findings showed that the $\rho N N^{*}$ coupling constants 
are very sensitive to the off-shell effects of the intermediate $\rho^{0}$ meson.

The results presented in this work provide
valuable theoretical guidance for future experimental efforts aimed at measuring the $pp \to pp \eta$ reaction 
at intermediate energies. 
The way in which the nucleon resonances are excited 
is the reason for the discrepancy between the theoretical results
(models 1 -- 3) and the DISTO experimental data measured 
at proton-proton collision energies $\sqrt{s} = 2.748 - 2.978$~GeV.
This is mainly a matter of the $N(1535)$ resonance created 
by the exchange of virtual pseudoscalar and/or vector mesons.
The angular distribution of the $\eta$ meson 
($d \sigma/ d\cos\theta_{\eta}$) should be forward peaked 
if $\pi^{0}$ plus $\eta$ exchanges 
(or $\sigma$ exchange not included in the present calculations) 
are dominant processes in the bremsstrahlung mechanism.
When compared to the DISTO differential cross-section data,
one finds that a model with the dominance of the $\rho$ exchange 
is the preferred option.
The $VV$-fusion mechanism plays an important role 
at higher energies. 
To learn more about the relevant reaction mechanism, 
a detailed analysis of observables other than the total cross section, 
such as the invariant mass distributions of the $\eta p$ and $pp$ systems,
the polar angle and momentum distributions of the $\eta$ meson,
and the transverse momentum distribution of the proton,
seems necessary.


Total and differential cross sections for the $pp \to pp \eta$ reaction for the HADES experiment at $\sqrt{s} = 3.46$~GeV
are provided.
The calculations were also done at $\sqrt{s} = 5.0$~GeV (PANDA) 
and $\sqrt{s} = 8.0$~GeV (SIS100).
The results can serve as a prediction 
for the already performed HADES \cite{Trelinski_talk,Trelinski_Ciepal} 
and may offer a new perspective for planned PANDA and SIS100~\cite{SIS100} measurements at GSI.
Upcoming experiments in the intermediate energy range
will be crucial for validating theoretical predictions presented here,
and will be especially useful if they provide accurate data accounting for
differential distributions in several variables.
This is important for clarifying the production mechanism
in the considered kinematic range
and checking different models for the FSI and ISI effects (not included in the current analysis).
Further measurements would be useful to confirm the theoretical findings, particularly those concerning the role of excited $N^{*}$ resonances and $VV$-fusion processes, as well as the interplay between them.
Related works on the $pp \to pp \eta'$, 
$pp \to pp \omega$, and $pp \to pp \phi$ reactions 
are currently in progress and will appear elsewhere.

\acknowledgments
\vspace{-0.2cm}
I would like to thank
I.~Ciepa{\l}, P.~Salabura, A.~Szczurek,
and M.~Zieli{\'n}ski for useful comments.

\appendix

\section{The reaction $\pi^{-} p \to \eta n$, effective coupling Lagrangians, propagators, and form factors}
\label{sec:appendixA}

In this section, the $\pi^{-} p \to \eta n$ reaction
shown in Fig.~\ref{fig:diagrams_pimp_etan} is discussed.
The differential cross section is given by
\begin{eqnarray}
&& \frac{d\sigma}{d\Omega} = \frac{1}{64 \pi^{2} s} \frac{| \bk_{\eta} |}{| \bk_{\pi} |} \frac{1}{2} \sum_{{\rm spins}}|{\cal M}_{\pi^{-} p \to \eta n}|^{2}\,,
\label{dsig_dz_B1}
\end{eqnarray}
where $\bk_{\pi}$ and $\bk_{\eta}$ are the c.m. 
three-momenta of the initial $\pi^{-}$ 
and the final $\eta$ mesons, respectively.
The amplitude can be expressed as a sum of 
the $s$-, $t$-, and $u$-channel pole diagrams:
${\cal M}_{\pi^{-} p \to \eta n}
= {\cal M}_{s} + {\cal M}_{t} + {\cal M}_{u}$.
In the present work, 
the $n$ and neutral $N^{*}$ exchanges 
in the $s$ channel and the $p$ exchange in the $u$ channel
are taken into account.
For simplicity, the charged $N^{*}$ exchanges in the $u$ channel
and the $a_{0}$-meson exchange in the $t$-channel are neglected.
The $N^{*}$-resonance contributions 
in the $u$-channel
are suppressed by the kinematics
[see the discussion after Eqs.~(\ref{ff_baryon}), (\ref{ff_baryon_3half})].

\begin{widetext}

\begin{figure}[!ht]
\includegraphics[width=0.28\textwidth]{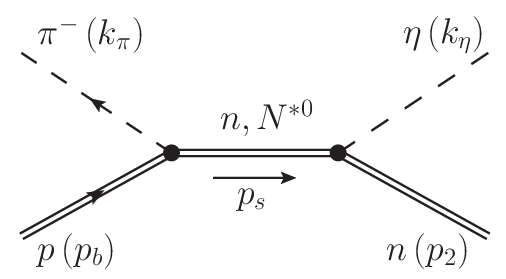}
\qquad
\includegraphics[width=0.28\textwidth]{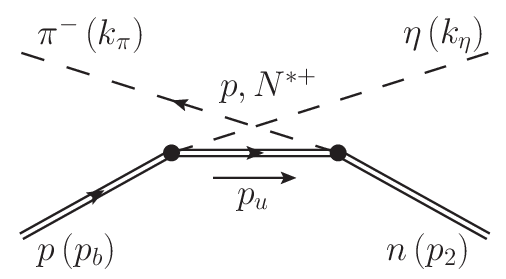}
  \caption{\label{fig:diagrams_pimp_etan}
  \small
Diagrams for the $\pi^{-} p \to \eta n$ reaction
with baryon exchanges in the $s$- and $u$-channel.}
\end{figure}

The amplitude ${\cal M}_{\pi^{-} p \to \eta n}$ reads
\begin{eqnarray}
{\cal M}_{\pi^{-} p \to \eta n} &=&
{\cal M}_{s}^{(n)} + 
{\cal M}_{u}^{(p)} + 
\sum_{N^{*}_{1/2}}
{\cal M}_{s}^{(N^{*}_{1/2})} +
\sum_{N^{*}_{3/2}} 
{\cal M}_{s}^{(N^{*}_{3/2})} \nonumber\\
&=&
\bar{u}_{n}(p_{2}, \lambda_{2})\,
\big\lbrace 
\Gamma^{(\eta n n)}(k_{\eta})\,
S^{(n)}(s)\, 
\Gamma^{(\pi^{-} p n)}(k_{\pi})
+
\Gamma^{(\pi^{-} n p)}(k_{\pi})\,
S_{F}^{(p)}(u)\, 
\Gamma^{(\eta p p)}(k_{\eta}) \nonumber\\
&&+
\sum_{N^{*}_{1/2}} 
\Gamma^{(\eta n N^{*}_{1/2})}(k_{\eta})\,
P^{(N^{*}_{1/2})}(s)\, 
\Gamma^{(\pi^{-} p N^{*}_{1/2})}(k_{\pi}) \nonumber\\
&&+
\sum_{N^{*}_{3/2}} 
\Gamma^{\mu\,(\eta n N^{*}_{3/2})}(k_{\eta})\,
P_{\mu \nu}^{(N^{*}_{3/2})}(s)\, 
\Gamma^{\nu\,(\pi^{-} p N^{*}_{3/2})}(k_{\pi})
\big\rbrace u_{p}(p_{b}, \lambda_{b})\,,
\label{pip_etan}
\end{eqnarray}
\end{widetext}
where $s = W_{\pi^{-} p}^{2} = p_{s}^{2}$ and $p_{s} = p_{b} + k_{\pi}$.
For the $u$ channel, 
the $\pi NN$ and $\eta NN$ vertices need to be interchanged
and there is an exchange of charge particles 
in the middle of the diagram, with 
$u = p_{u}^{2}$, $p_{u} = p_{b} - k_{\eta}$.
The $M N N$, $M N N^{*}_{1/2}$, and 
$M N N^{*}_{3/2}$
interaction vertices
could be read as below
with the coupling constants 
from Table~\ref{tab:table_par2}
taking into account isospin factor,
$g_{\pi^{-} p n} = \sqrt{2} g_{\pi N N}$ and
$g_{\pi^{-} p N^{*0}} = g_{\pi^{-} n N^{*+}} = \sqrt{2} g_{\pi N N^{*}}$.

The pseudoscalar-meson--nucleon coupling Lagrangians 
can be written as
\begin{eqnarray}
{\cal L}_{\pi NN}
&=& -\frac{g_{\pi NN}}{2m_{N}} 
\bar{N} \gamma_{5} \gamma_{\mu}
\partial^{\mu} (\btau\bPhi_{\pi}) N\,,
\label{piNN_Lagrangian} \\
{\cal L}_{\eta NN}
&=& -g_{\eta NN} \nonumber \\
&&\times
\bar{N} \left( i \gamma_{5} \lambda + 
(1 - \lambda)\frac{1}{2m_{N}}\gamma_{5} \gamma_{\mu}
\partial^{\mu} \right) \Phi_{\eta} N\,, 
\label{etaNN_Lagrangian} \nonumber \\
\end{eqnarray}
where $N$ and $\Phi$ denote the nucleon and meson fields, respectively.
The parameter $\lambda$ in (\ref{etaNN_Lagrangian})
controls the admixture of the two types of couplings:
pseudoscalar (PS) $(\lambda = 1)$ 
and pseudovector (PV) $(\lambda = 0)$.
Note that the $\pi NN$ coupling constant 
$g_{\pi NN}^{2}/4 \pi = 13.4 - 14.4$ 
\cite{Benmerrouche:1994uc,Machleidt:2000ge,Wang:2019bsf}
is much better determined than the $\eta NN$ coupling constant.
The $\pi NN$ coupling is preferred to be pseudovector \cite{Li:1998jd};
however, there is no compelling reason to select
the PV coupling rather than the PS form for the $\eta NN$ vertex 
\cite{Tiator:1994et,Kirchbach:1996kw}.
From fits to $\eta$ photoproduction of \cite{Tiator:1994et}
one can deduce that the PS coupling with 
$g_{\eta NN}^{2}/4 \pi = 0.4$ is preferred.
In \cite{Kirchbach:1996kw} the PV admixture was found.
More accurate data are needed for further study of the problem.
Different values of $\eta NN$ coupling have been determined/used 
in the literature, see e.g. 
\cite{Nakayama:2002mu,Nakayama:2008tg,Zhong:2011ti,Lu:2013jva}.
From \cite{Benmerrouche:1994uc,Machleidt:2000ge} one has
$g_{\eta NN}^{2}/4 \pi = 0.6 - 6.4$
and even smaller $g_{\eta NN}^{2}/4 \pi \lesssim 0.1$ 
\cite{Nakayama:2008tg,Tiator:2018heh}.
The coupling for the $\eta NN$ vertex is uncertain,
but significantly smaller than the $\pi NN$ coupling.
In this work, 
$g_{\pi NN}^{2}/4 \pi = 14.0$ in (\ref{piNN_Lagrangian}) is taken.
For the $\eta NN$ coupling (\ref{etaNN_Lagrangian}) 
the parameters are taken from \cite{Kirchbach:1996kw}:
$\lambda = 0.504$ and
$g_{\eta NN} \to g_{\eta} = g_{\eta NN} / \lambda =
f_{\eta NN} / (1 - \lambda) = 4.03$.

The $\Gamma^{(M N N^{*}_{1/2})}$ vertices involving 
spin-1/2 nucleon resonances 
obtained from the effective Lagrangians
are  
\cite{Benmerrouche:1994uc,Nakayama:1999jb,Nakayama:2002mu,Nakayama:2008tg}
\begin{eqnarray}
{\cal L}_{\pi N N^{*}_{1/2^{\mp}}}^{\rm PS}
&=& \pm ig_{\pi N N^{*}}\, 
\bar{N}^{*}
\begin{pmatrix}
1\\
\gamma_{5}
\end{pmatrix}
(\btau \bPhi_{\pi}) N + {\rm h.c.}\,, \qquad
\label{piNR_PS_Lagrangian}\\
{\cal L}_{\eta N N^{*}_{1/2^{\mp}}}^{\rm PS}
&=& \pm ig_{\eta N N^{*}}\, 
\bar{N}^{*}
\begin{pmatrix}
1\\
\gamma_{5}
\end{pmatrix}
\Phi_{\eta} N + {\rm h.c.}\,,
\label{etaNR_PS_Lagrangian}\\
{\cal L}_{\pi N N^{*}_{1/2^{\mp}}}^{\rm PV}
&=& \pm \frac{g_{\pi N N^{*}}}{m_{N^{*}} \mp m_{N}}\, 
\bar{N}^{*} 
\begin{pmatrix}
\gamma_{\mu}\\
\gamma_{5}\gamma_{\mu}
\end{pmatrix}
\partial^{\mu} (\btau \bPhi_{\pi}) N \nonumber\\
&&+ {\rm h.c.}\,,
\label{piNR_PV_Lagrangian}
\end{eqnarray}
\begin{eqnarray}
{\cal L}_{\eta N N^{*}_{1/2^{\mp}}}^{\rm PV}
&=& \pm \frac{g_{\eta N N^{*}}}{m_{N^{*}} \mp m_{N}}\, 
\bar{N}^{*} 
\begin{pmatrix}
\gamma_{\mu}\\
\gamma_{5}\gamma_{\mu}
\end{pmatrix}
\partial^{\mu} \Phi_{\eta} N \nonumber\\
&&+ {\rm h.c.}\,,
\label{etaNR_PV_Lagrangian} 
\end{eqnarray}
where the upper (lower) sign 
and factor in parentheses correspond to
negative-parity (positive-parity) resonances.
For the $\pi N N^{*}$ and $\eta N N^{*}$ vertices
one has the choice of \mbox{(pseudo-)scalar} (PS) or 
\mbox{(pseudo-)vector} (PV) couplings;
i.e., scalar and vector in the case of an odd parity resonance,
and pseudoscalar and pseudovector in the case of an even parity resonance.
The PS and PV couplings are equivalent when both baryons
are on their mass shell.

Note that in \cite{Feuster:1998cj} the PS coupling is used 
in the case of odd-parity $S_{11}$ resonances 
(with the incoming partial wave notation $L_{2I, 2S}$),
and the PV form in the case of even-parity $P_{11}$ resonances
(just as in the nucleon case).
As was mentioned there, this choice of PS-PV coupling
for the resonances is supported by the fact, 
that in the $\pi N \to \pi N$
partial wave data, the resonant structures are more pronounced in
the $S$ wave ($J^{P} = 1/2^{-}$) 
than in the $P$ wave ($J^{P} = 1/2^{+}$).
In \cite{Shyam:2007iz} the calculations for the $pp \to pp \eta$ reaction were done with two types of couplings, PS and PV, 
and little difference was found there;
see Fig.~10 of \cite{Shyam:2007iz}.
In principle one
can select a linear combination of both and fit the PS/PV
ratio to the data~\cite{Nakayama:2008tg}. 

The $\Gamma_{\mu}^{(M N N^{*}_{3/2})}$ vertices involving 
spin-3/2 nucleon resonances can be written as \cite{Benmerrouche:1994uc,Nakayama:2002mu,Kaptari:2007ss}
\begin{eqnarray}
{\cal L}_{\pi N N^{*}_{3/2^{\mp}}}
&=& \frac{g_{\pi N N^{*}}}{m_{\pi}}\, 
\bar{N}^{*\mu} \,
\Theta_{\mu \nu}(z)
\begin{pmatrix}
\gamma_{5}\\
1
\end{pmatrix}
\partial^{\nu} (\btau \bPhi_{\pi}) N \nonumber\\
&&+ {\rm h.c.}\,,
\label{piNR_Lagrangian_3half} \\
{\cal L}_{\eta N N^{*}_{3/2^{\mp}}}
&=& \frac{g_{\eta N N^{*}}}{m_{\eta}}\, 
\bar{N}^{*\mu} \,
\Theta_{\mu \nu}(z)
\begin{pmatrix}
\gamma_{5}\\
1
\end{pmatrix}
\partial^{\nu} \Phi_{\eta} N \nonumber\\
&&+ {\rm h.c.}\,,
\label{etaNR_Lagrangian_3half}
\end{eqnarray}
where 
$\Theta_{\mu \nu}(z) = g_{\mu \nu} 
- (A(1+4z)/2+z)\gamma_{\mu}\gamma_{\nu}$.
The choice of the so-called ``off-shell parameter'' 
$z$ is arbitrary and it is treated 
as a free parameter to be determined by fitting to the data;
see Table~XII of \cite{Feuster:1997pq} and
Table~V of \cite{Feuster:1998cj}.
Following Refs.~\cite{Nakayama:2002mu,Kaptari:2007ss},
for simplicity, the values $A = -1$ and $z = -1/2$ are adopted.

The partial decay widths of the nucleon resonances 
are calculated 
from the above Lagrangian couplings as follows:
\begin{eqnarray}
\Gamma(N^{*}_{1/2^{\mp}} \to N M) &=& 
f_{\rm ISO} 
\frac{g_{M N N^{*}}^{2}}{4 \pi} p_{N}
\frac{E_{N} \pm m_{N}}{m_{N^{*}}}\,,
\label{decay_width_Npi} \nonumber\\
\\
\Gamma(N^{*}_{3/2^{\mp}} \to N M) &=& 
f_{\rm ISO} 
\frac{g_{M N N^{*}}^{2}}{12 \pi} 
\frac{p_{N}^{3}}{m_{M}^{2}}
\frac{E_{N} \mp m_{N}}{m_{N^{*}}}\,,
\label{decay_width_Npi_3half}
\nonumber\\
\end{eqnarray}
where the upper (lower) sign corresponds t
o negative-parity (positive-parity) resonance.
Above,
$p_{N} = |\bp_{N}| = \lambda(m_{N^{*}}^{2}, m_{N}^{2}, m_{M}^{2})/ (2 m_{N^{*}})$ and 
$E_{N} = \sqrt{p_{N}^{2} + m_{N}^{2}}$
denote the absolute value of the three-momentum 
and energy of the nucleon in the rest frame of $N^{*}$, respectively.
Here, $\lambda$ denotes the K\"all{\'e}n function
with $\lambda(x,y,z) \equiv \sqrt{(x-y-z)^{2}-4yz}$.
Furthermore, the isospin factor $f_{\rm ISO}$
is equal to 3 for decays
into mesons with isospin one
($\pi$), and to 1 otherwise ($\eta$).
The absolute value of coupling constants $g_{M N N^{*}}$ ($M = \pi, \eta$), 
could be determined by the experimental decay widths
of $\Gamma(N^{*} \to N M)$
in the compilation of the PDG \cite{ParticleDataGroup:2024cfk};
see Table~\ref{tab:table_par2}.

The propagators $S_{F}$ and
$P^{(N^{*}_{1/2})}$
of the spin-1/2 for nucleon and resonances, respectively,
and 
$P_{\mu \nu}^{(N^{*}_{3/2})}$ of the spin-3/2 
resonances are expressed as
%
\begin{eqnarray}
iS_{F}(p^{2})
&=& i\frac{ \slash{p} \pm m_{N}}{p^{2} - m_{N}^{2}}\,,
\label{propagator_N}\\
iP^{(N^{*}_{1/2})}(p^{2})
&=& i\frac{ \slash{p} \pm m_{N^{*}}}{p^{2} - m_{N^{*}}^{2} + i m_{N^{*}} \Gamma_{N^{*}}(p^{2})}\,,
\label{propagator_Nresonance}\\
iP_{\mu \nu}^{(N^{*}_{3/2})}(p^{2})
&=& i\frac{ \slash{p} \pm m_{N^{*}}}{p^{2} - m_{N^{*}}^{2} + i m_{N^{*}} \Gamma_{N^{*}}(p^{2})} {\cal P}_{\mu \nu}(p)\,,
\label{propagator_Nresonance_3half} \nonumber\\
\\
{\cal P}_{\mu \nu}(p) &=&
-g_{\mu \nu} 
+ \frac{1}{3} \gamma_{\mu} \gamma_{\nu}
\pm \frac{1}{3 m_{N^{*}}} 
\left( \gamma_{\mu} p_{\nu} - \gamma_{\nu} p_{\mu} \right) \nonumber \\
&&
+ \frac{2}{3 m_{N^{*}}^{2}} p_{\mu} p_{\nu} \,,
\end{eqnarray}
where $\pm$ are for the particles and antiparticles, respectively.
$m_{N^{*}}$ and $\Gamma_{N^{*}}$ denote
the mass and total width of the $N^{*}$ resonance, respectively.
The complex term is introduced to take care of the finite width
$\Gamma_{N^{*}}$ of the unstable nucleon resonance.
For a resonance with mass near the $\eta$ production threshold 
the approximation of a constant width is not very good
and the energy-dependent width is needed; see, e.g.,
\cite{Drechsel:1998hk,Shyam:1999nm,Nakayama:2005ts,Tiator:2018heh}.
Using a formalism analogous to that in Refs.~\cite{Nakayama:2005ts,Tiator:2018heh},
one can restrict to the dominant decay channels of
$N^{*} \to \pi N$, $\eta N$, and $\pi \pi N$:
\begin{eqnarray}
\Gamma_{N^{*}}(s) &=& 
\Gamma_{\pi N}(s) 
+ \Gamma_{\eta N}(s) 
+ \Gamma_{\pi \pi N}(s) \,,\\
\Gamma_{\pi N}(s) &=& {\cal B}(N^{*} \to \pi N) \,
\Gamma_{N^{*}}
\left(\frac{k_{\pi}(s)}{k_{\pi, N^{*}}} \right)^{2 \ell + 1} \nonumber \\
&&\times
\left(\frac{X^{2} + k_{\pi, N^{*}}^{2}}{X^{2} + k_{\pi}^{2}(s)} \right)^{\ell} \,,
\label{Gamma_partial_piN}\\
\Gamma_{\pi \pi N}(s) &=& {\cal B}(N^{*} \to \pi \pi N)\,
\Gamma_{N^{*}}
\left(\frac{k_{2\pi}(s)}{k_{2\pi, N^{*}}} \right)^{2 \ell + 4}\nonumber \\
&&\times
\left(\frac{X^{2} + k_{2\pi, N^{*}}^{2}}{X^{2} + k_{2 \pi}^{2}(s)} \right)^{\ell + 2}\,.
\label{Gamma_partial_pipiN}
\end{eqnarray}
The width $\Gamma_{\eta N}$ has a similar dependence 
as $\Gamma_{\pi N}$.
Above $\ell$ is the orbital angular momentum 
of a resonance:
$\ell = 0$ for $N(1535)$, $N(1650)$, $N(1895)$;
$\ell = 1$ for $N(1710)$, $N(1720)$, $N(1880)$, $N(2100)$;
$\ell = 2$ for $N(1520)$, $N(1700)$, $N(1900)$.
$X$ is a damping parameter, which has been fixed in the present work to $X = 0.2$~GeV for all resonances.
${\cal B}(N^{*} \to \pi N)$ and $\Gamma_{N^{*}}$ 
are the $\pi N$ branching ratio and total width, respectively.
The c.m. momenta of pion and $\eta$ are denoted by
$k_{\pi}(s)$ and $k_{\eta}(s)$, respectively.
Here,
$k_{\pi}(s) = \sqrt{[s-(m_{N}+m_{\pi})^{2}][s-(m_{N}-m_{\pi})^{2}]/4s}$ 
and $k_{\pi,N^{*}} \equiv k_{\pi}(m_{N^{*}}^{2})$.
In (\ref{Gamma_partial_pipiN}),
$k_{2\pi}$ is the momentum of
the compound ($2 \pi$) system with mass $2 m_{\pi}$,
$k_{2\pi,N^{*}}$ is equal to 
$k_{2\pi}$ at $s= m_{N^{*}}^{2}$,
and
${\cal B}(N^{*} \to \pi \pi N)
= 1 - {\cal B}(N^{*} \to \pi N) - {\cal B}(N^{*} \to \eta N)$.
For the $N(1895)$, $N(1900)$, and $N(2100)$ resonances,
the relevant $\eta' N$ width is also taken into account.
The expression is similar to (\ref{Gamma_partial_piN})
but with ${\cal B}(N^{*} \to \pi N)
\to {\cal B}(N^{*} \to \eta' N)$, $m_{\pi} \to m_{\eta'}$.

Each vertex in (\ref{pip_etan})
is multiplied by a phenomenological cutoff function
\begin{eqnarray}
F_{B}(p^{2})
= \frac{\Lambda_{B}^{4}}{(p^{2} - m_{B}^{2})^{2} + \Lambda_{B}^{4}}\,,
\quad B = N, N^{*}_{1/2}\,,
\label{ff_baryon}
\end{eqnarray}
where $p^{2}$ denotes the four-momentum squared 
of 
off-shell baryon and $\Lambda_{B}$ is the cutoff parameter.
For spin-3/2 nucleon resonances
the multi-dipole form factor is used
and it is given by the following expression:
\begin{eqnarray}
&&F_{B}(p^{2})
=\frac{\Lambda_{B}^{4}}{(p^{2} - m_{B}^{2})^{2} + \Lambda_{B}^{4}}
\left( \frac{m_{B}^{2} \tilde{\Gamma}_{B}^{2}}
{(p^{2} - m_{B}^{2})^{2} + m_{B}^{2} \tilde{\Gamma}_{B}^{2}} \right)^{2}, \nonumber \\
&&B = N^{*}_{3/2}\,,
\label{ff_baryon_3half}
\end{eqnarray}
with $\tilde{\Gamma}_{B} = \Gamma_{B}/\sqrt{2^{1/3}-1}$;
see \cite{Vrancx:2011qv} where 
the spin-dependent hadronic form factor was introduced.
In order to reduce the number of free parameters,
for the cutoff parameters for \mbox{spin-1/2} and \mbox{spin-3/2} resonances, $\Lambda_{N^{*}} = 1.2$~GeV is used,
while $\Lambda_{N} = 1.0$~GeV
for the proton and neutron exchanges.

In considering of the nucleon and nucleon resonances
in the $s$ and $u$ channels (see Fig.~\ref{fig:diagrams_pimp_etan})
some additional comments are in order.
While the four-momenta squared of transferred $s$-channel baryons is
\mbox{$p_{s}^{2} = W_{\pi^{-} p}^{2} > (m_{N} + m_{\eta})^{2}
\gtrsim 2.21$~GeV$^{2}$}, this is not the case for transferred
$u$-channel baryons where one has
\mbox{$p_{u}^{2} < (m_{N} - m_{\eta})^{2} \lesssim 0.15$~GeV$^{2}$}.
The form factors (\ref{ff_baryon}) and (\ref{ff_baryon_3half})
used to describe the $p_{s,u}^{2}$ dependence of baryons
are defined so that they reach unity if 
$p_{s,u}^{2} = m_{B}^{2}$, i.e., 
the four-momentum squared is equal to the baryon mass
squared $m_{B}^{2}$.
These factors have a large impact on the cross section result, 
since $\sigma_{\pi^{-}p \to \eta n}$ is proportional to $(F_{B}(p^{2}))^{4}$, and the exchanged baryons are relatively
away from their mass shell ($p^{2} \neq m_{B}^{2}$).
In practice, due to the form factors according to my parametrization, 
the $N^{*}$ contributions in the $u$ channel are largely suppressed.
To be more precise, for the $N(1535)$ contribution,
the $s$-channel term
gives $\sigma = 3.184$~mb
for $W_{\pi^{-} p} = 1.53$~GeV
and $\sigma = 0.791$~mb for $W_{\pi^{-} p} = 1.65$~GeV,
while the $u$-channel term
gives only $\sigma = 0.062$~$\mu$b and $\sigma = 0.106$~$\mu$b, respectively.
For higher mass resonances, there is also a similar order of magnitude difference between these two terms.
This is the reason why 
the $N^{*}$ exchanges in the $u$ channel can be neglected.
For the nucleon $(N = n, p)$ contributions,
the situation is different, partly due to the form-factor functions, but also due to the smaller mass in the propagators.
There, the $s$-channel term gives $\sigma = 1.7$~$\mu$b
for $W_{\pi^{-} p} = 1.53$~GeV
and $\sigma = 1.4$~$\mu$b for $W_{\pi^{-} p} = 1.65$~GeV,
while the $u$-channel term
gives greater contributions $\sigma = 17.7$~$\mu$b and $\sigma = 33.9$~$\mu$b, respectively.
Adding coherently these two contributions, one finds
$\sigma = 12.5$~$\mu$b for $W_{\pi^{-} p} = 1.53$~GeV
and
$\sigma = 31.6$~$\mu$b for $W_{\pi^{-} p} = 1.65$~GeV.
Since these terms are of similar order 
and interfere destructively
especially in the low energy region, they cannot be neglected.
The $u$-channel proton-exchange contribution plays an important role mainly in the backward scattering region, 
$\cos\theta \approx -1$
(see the upper right panel of Fig.~\ref{fig:B2}).

In the present calculation,
to estimate the potential role of $N(1650)$, $N(1710)$,
and other resonances,
the values of $g_{MNN^{*}}$ coupling constants are taken
from Table~\ref{tab:table_par2}.
Note that the branching ratios determine
only the square of the corresponding coupling constants,
thus their signs must be fixed from independent studies.
For example, for this purpose, 
the parameters of \cite{Shyam:2007iz}
follow the predictions of \cite{Feuster:1998cj} 
(see Table~IV of \cite{Feuster:1998cj}).

There are a few remarkable observations to the $N \eta$ channel.
Note that from the PDG \cite{ParticleDataGroup:2024cfk} the 
$N(1535) \to N \pi, N \eta$ branching ratios are $\approx 0.42$
(see Table~\ref{tab:table_par2}).
The following $N^{*} \to N \eta$ branching ratios
can also be found in \cite{CBELSATAPS:2019ylw}:
$0.41 \pm 0.04$ for $N(1535)$, 
$0.33 \pm 0.04$ for $N(1650)$,
$0.18 \pm 0.10$ for $N(1710)$.
In Table~II of \cite{Shyam:2007iz} 
(Table~1 of \cite{Ouyang:2009kv})
to estimate the $g_{\eta NN^{*}}$ coupling constants
for the $N(1650)$ and $N(1710)$
smaller values 0.03--0.1 (0.065)
and 0.06 (0.062), respectively, were used.
As a result, the predictions presented there 
for the above resonances may be underestimated.

Figures~\ref{fig:B1} and \ref{fig:B2} shows the
total and differential cross-sections, respectively,
for the reaction $\pi^{-} p \to \eta n$.
Theoretical results for two models 1 and 2 are presented.
Model~1 uses the PS-type couplings
(\ref{piNR_PS_Lagrangian}), (\ref{etaNR_PS_Lagrangian})
for all $N^{*}$ resonances.
Model~2 is similar to model~1
with the difference that for model~2 
the PV-type couplings
(\ref{piNR_PV_Lagrangian}), (\ref{etaNR_PV_Lagrangian}) are used
for all even-parity resonances included in the calculation.
In both cases, the values for $g_{MNN^{*}}$ 
were taken from Table~\ref{tab:table_par2}
with the upper values for the $N(1710)$.
The experimental data are from \cite{Bulos:1969ofe,Deinet:1969cd,Richards:1970cy,Nelson:1972pra,Feltesse:1975nz,Chaffee:1975zj,Debenham:1975bj,Brown:1979ii,Crouch:1980vw,Prakhov:2005qb}
(detailed analysis of the data obtained before 1980 
can be found in the review by Clajus and Nefkens \cite{Clajus_Nefkens} 
and they were also discussed in Sec.~III~C of \cite{Durand:2008es}).
It can be seen from Fig.~\ref{fig:B1}
that the $N(1535)$ resonance gives a major contribution
to the $\pi^{-} p \to \eta n$ reaction.
Clearly, the complete result indicates a large interference effect
between different contributions, 
for example between $N(1535)$ and $N(1650)$.
From the upper right panel of Fig.~\ref{fig:B2} one can see
that angular distribution of the $N(1535)$ contribution is flat.
The $N(1520)$ contribution is very small, but it interferes with other contributions and is therefore necessary to describe the shape of the angular distribution at low energies.
The $u$-channel $N$ contribution
plays an important role at the backward angle
$\cos\theta \approx -1$.
Model results for this reaction around $W = 1.6$~GeV indicate an important contribution from $N(1710)$ 
(see also \cite{Shrestha:2012va}). 
At higher energies, 
the reaction mechanism is still under discussion; see
\cite{Zhong:2007fx,Durand:2008es,Nakayama:2008tg,He:2008uf,Ronchen:2012eg,Shrestha:2012va,Lu:2013jva,Xiao:2016dlf}.

\begin{widetext}

\begin{table}[!ht]
\centering
\caption{Coupling constants for the $MNN^{*}$ 
($M = \pi, \eta, \eta'$) vertices
obtained from (\ref{piNR_PV_Lagrangian})--(\ref{decay_width_Npi_3half}).
Shown are the hadronic Breit-Wigner parameters for nucleon resonances,
and the branching ratios (${\cal B}$)
for adopted values and in the brackets from
the PDG \cite{ParticleDataGroup:2024cfk}.
The coupling constants $g_{MNN^{*}}$ are dimensionless.
The symbol ``$(-)$'' indicates
the negative sign of the $g_{\eta N N(1650)}$ coupling constant.}
\label{tab:table_par2}
\begin{tabular}{c|c|c|c|c|c}
\hline 
\hline 
$N^{*} \, J^{P}$ & Mass (MeV)&
Width (MeV) &
Decay channel &
${\cal B}$ (\%) [PDG]&
$g_{MNN^{*}}^{2}/4 \pi$ \\
\hline
$N(1520) \,3/2^{-}$ & 1515 & 110 & $\pi N$   
& 65 [$60\pm5$]  
& 0.204\\
          &&       & $\eta N$  
& 0.08 [$0.08\pm0.01$] 
& 3.945\\
          &&       & $\pi\pi N$  
& 34.92 [$30 \pm 5$] 
& \\
\hline
$N(1535) \,1/2^{-}$ & 1530 & 150 & $\pi N$   
& 50 [$42 \pm 10$] 
& 0.040\\
          &&       & $\eta N$  
& 42 [$42.5 \pm 12.5$] 
& 0.290\\
          &&       & $\pi\pi N$  
          & 8 [$17.5\pm 13.5$] 
          & \\
\hline
$N(1650) \,1/2^{-}$ & 1650 & 125 & $\pi N$   & 60 [$60 \pm 10$] 
& 0.037\\
          &&       & $\eta N$  & 25 [$25 \pm 10$] 
          & $(-)$ 0.076 \\
          &&       & $\pi\pi N$  & 15 [$39 \pm 19$] 
          & \\
\hline 
$N(1700) \,3/2^{-}$ & 1720 & 200 
& $\pi N$   & 12 [$12 \pm 5$]  
& 0.021\\
          &&       
& $\eta N$  & 2 [seen]\footnote{Note that 
${\cal B}(N(1700) \to \eta N) = (1 \pm 1) \, \%$ from \cite{CBELSATAPS:2019ylw}.} 
& 0.916\\
          &&       & $\pi\pi N$  & 86 [$> 89$] 
          & \\
\hline
$N(1710) \,1/2^{+}$ & 1710 & 140
    & $\pi N$   & 12.5 -- 20 [$12.5 \pm 7.5$] & 0.101 -- 0.161\\
&&  & $\eta N$  & 30 -- 50 [$30 \pm 20$] & 2.021 -- 3.368\\
&&  & $\pi\pi N$  & 57.5 -- 30 [$31 \pm 17$]
          & \\
\hline 
$N(1720) \,3/2^{+}$ & 1720 & 250 
& $\pi N$   & 11 [$11 \pm 3$]  
& 0.002\\
          &&       
& $\eta N$  & 3 [$3 \pm 2$]
& 0.079\\
          &&       & $\pi\pi N$  & 86 [$> 50$]  
          & \\

\hline 
$N(1880) \,1/2^{+}$ & 1880 & 300 
&    $\pi N$     & 17 [$17 \pm 14$] & 0.198 \\
&& & $\eta N$    & 28 [$28 \pm 27$] & 1.798 \\
&& & $\pi\pi N$  & 55 [$> 32$]      & \\
\hline 
$N(1895) \,1/2^{-}$ & 1900 & 120 
& $\pi N$   & 5 [$10 \pm 8$] & 0.0025\\
&&       & $\eta N$  & 30 [$30 \pm 15$] 
          & 0.058 \\
&&       & $\eta' N$  & 25 [$25 \pm 15$] 
          & 0.510 \\
          &&       & $\pi\pi N$  & 40 [$45.5 \pm 28.5$] 
          & \\
\hline 
$N(1900) \,3/2^{+}$ & 1920 & 200 
& $\pi N$   & 10.5 [$10.5 \pm 9.5$] & 0.001\\
&&       & $\eta N$  & 8 [$8 \pm 6$] 
          & 0.064 \\
&&       & $\eta' N$  & 6 [$6 \pm 2$] 
          & 9.862 \\
          &&       & $\pi\pi N$  & 75.5 [$> 56$] 
          & \\
\hline 
$N(2100) \,1/2^{+}$ & 2100 & 260 
& $\pi N$   & 20 [$20 \pm 12$] & 0.138\\
&&       & $\eta N$  & 25 [$25 \pm 20$] 
          & 0.750 \\
&&       & $\eta' N$  & 8 [$8 \pm 3$] 
          & 0.942 \\
          &&       & $\pi\pi N$  & 47 [$> 55$] 
          & \\
\hline
\hline 
\end{tabular}
\end{table}


\begin{figure}[!ht]
\includegraphics[width=6.8cm]{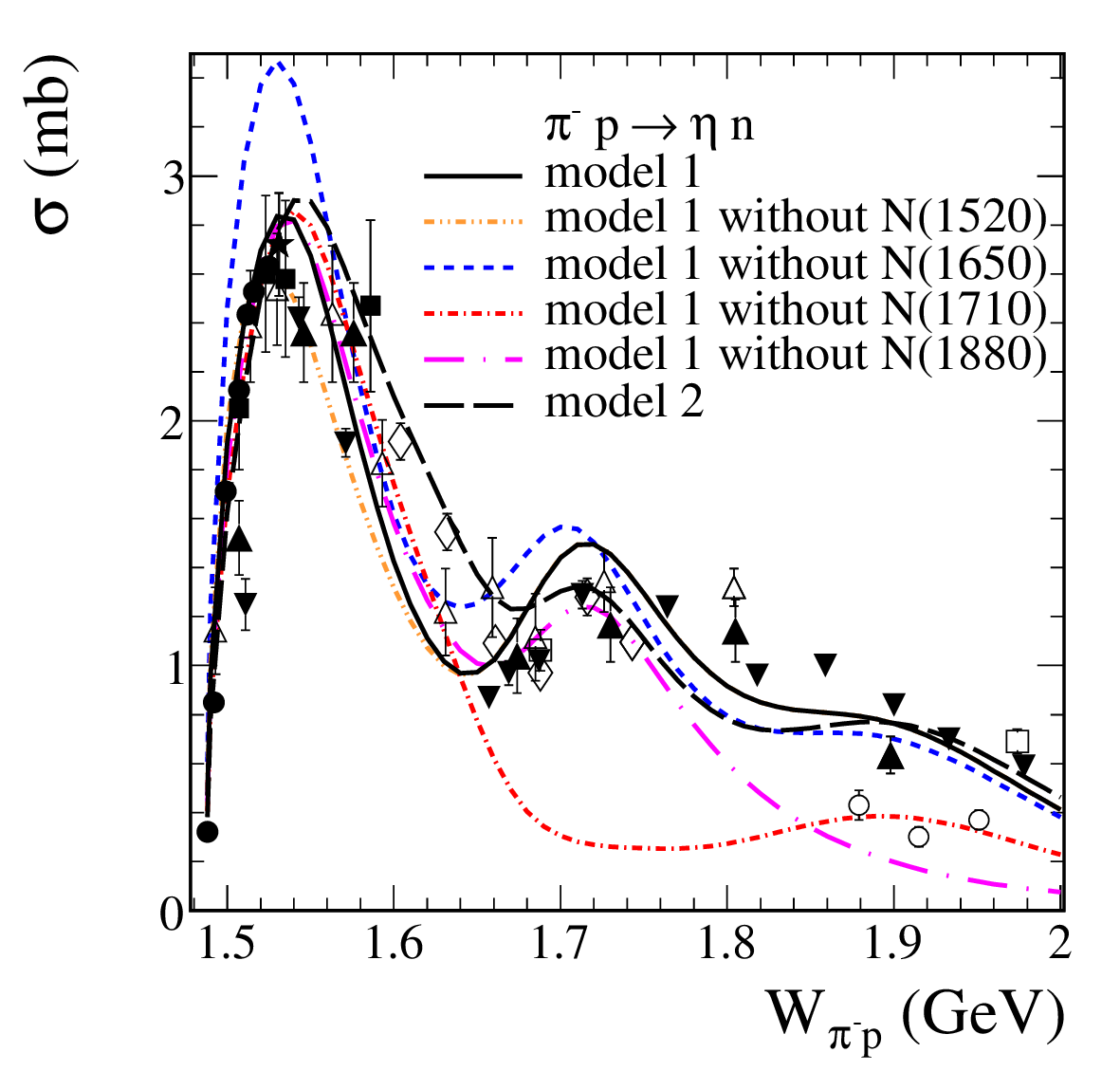}
\includegraphics[width=6.8cm]{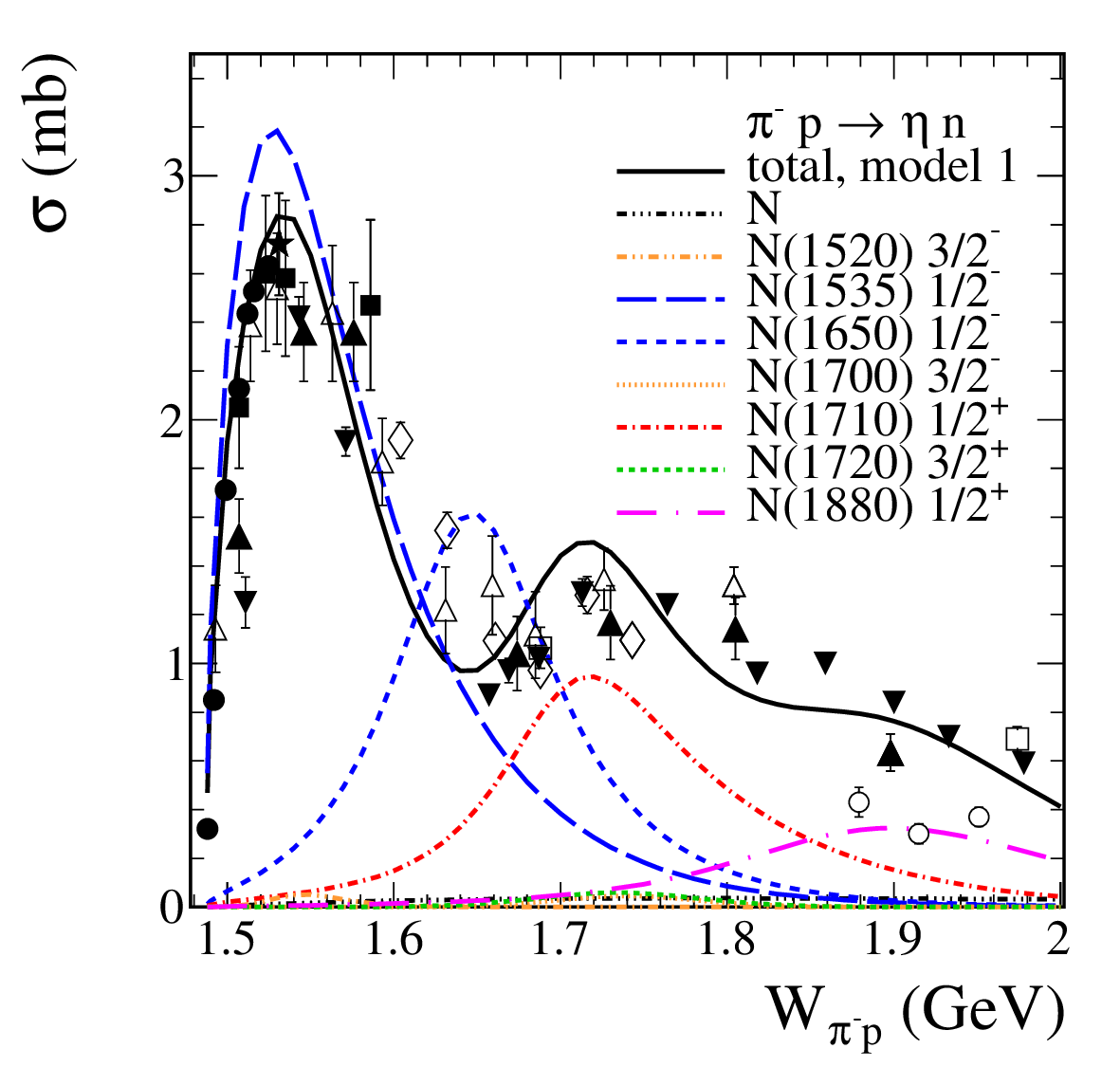}
\caption{\label{fig:B1}
\small
Total cross section for the reaction $\pi^{-} p \to \eta n$
as a function of the c.m. energy $W_{\pi^{-} p}$.
Several $N^{*}$ contributions, shown in the right panel, were included.
The coherent sum of all
contributions is shown by the solid and long-dashed lines, corresponding to model 1 and 2, respectively.
Data are from \cite{Bulos:1969ofe} ($\bigtriangleup$),
\cite{Deinet:1969cd} ($\blacksquare$),
\cite{Richards:1970cy} ($\blacktriangle$),
\cite{Nelson:1972pra} ($\square$),
\cite{Feltesse:1975nz} ($\bigstar$),
\cite{Chaffee:1975zj} ($\lozenge$),
\cite{Brown:1979ii} ($\blacktriangledown$),
\cite{Crouch:1980vw} ($\circ$),
\cite{Prakhov:2005qb} ($\bullet$).}
\end{figure}

\begin{figure}[!ht]  
\center
\includegraphics[width=6.8cm]{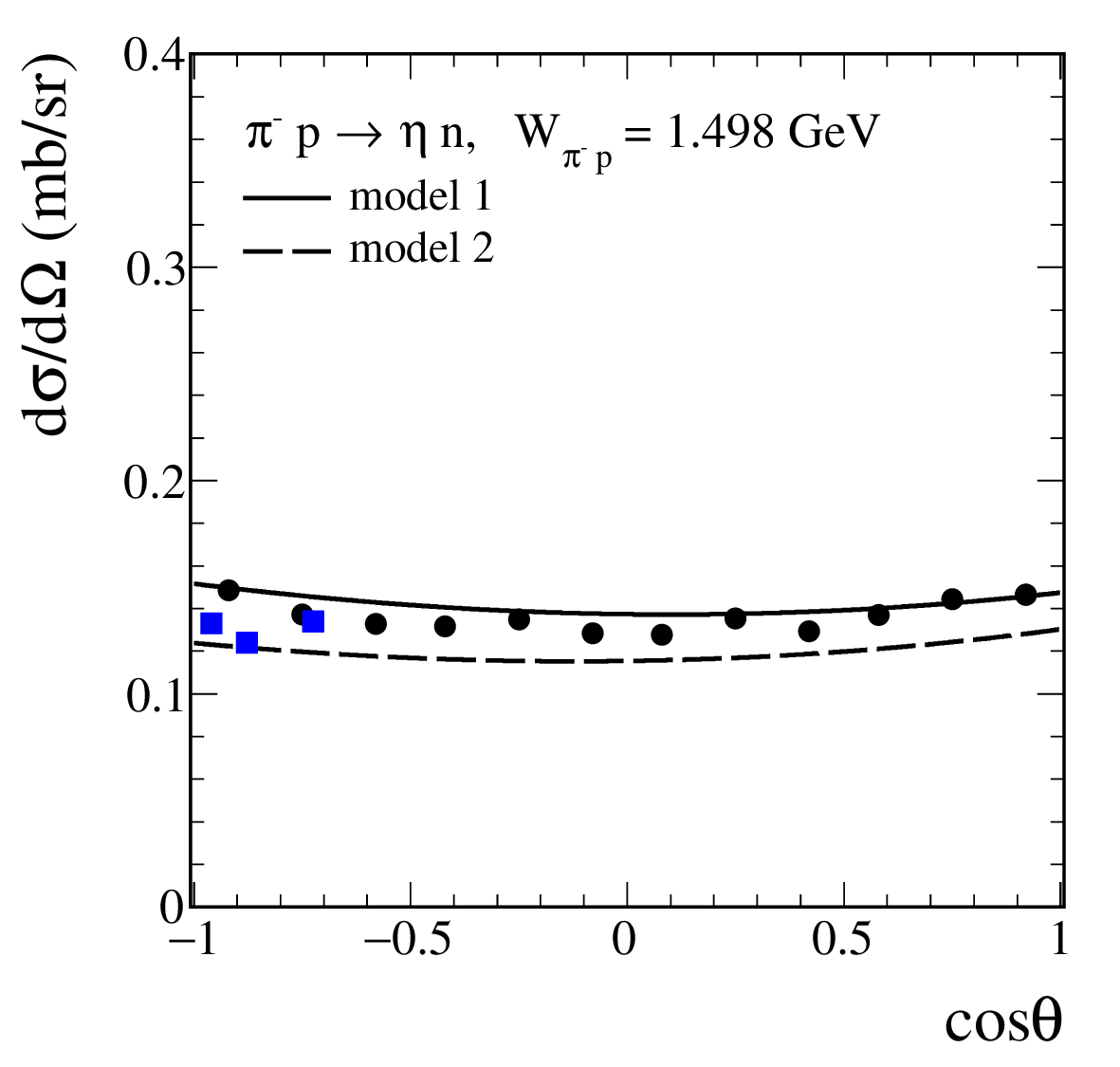}
\includegraphics[width=6.8cm]{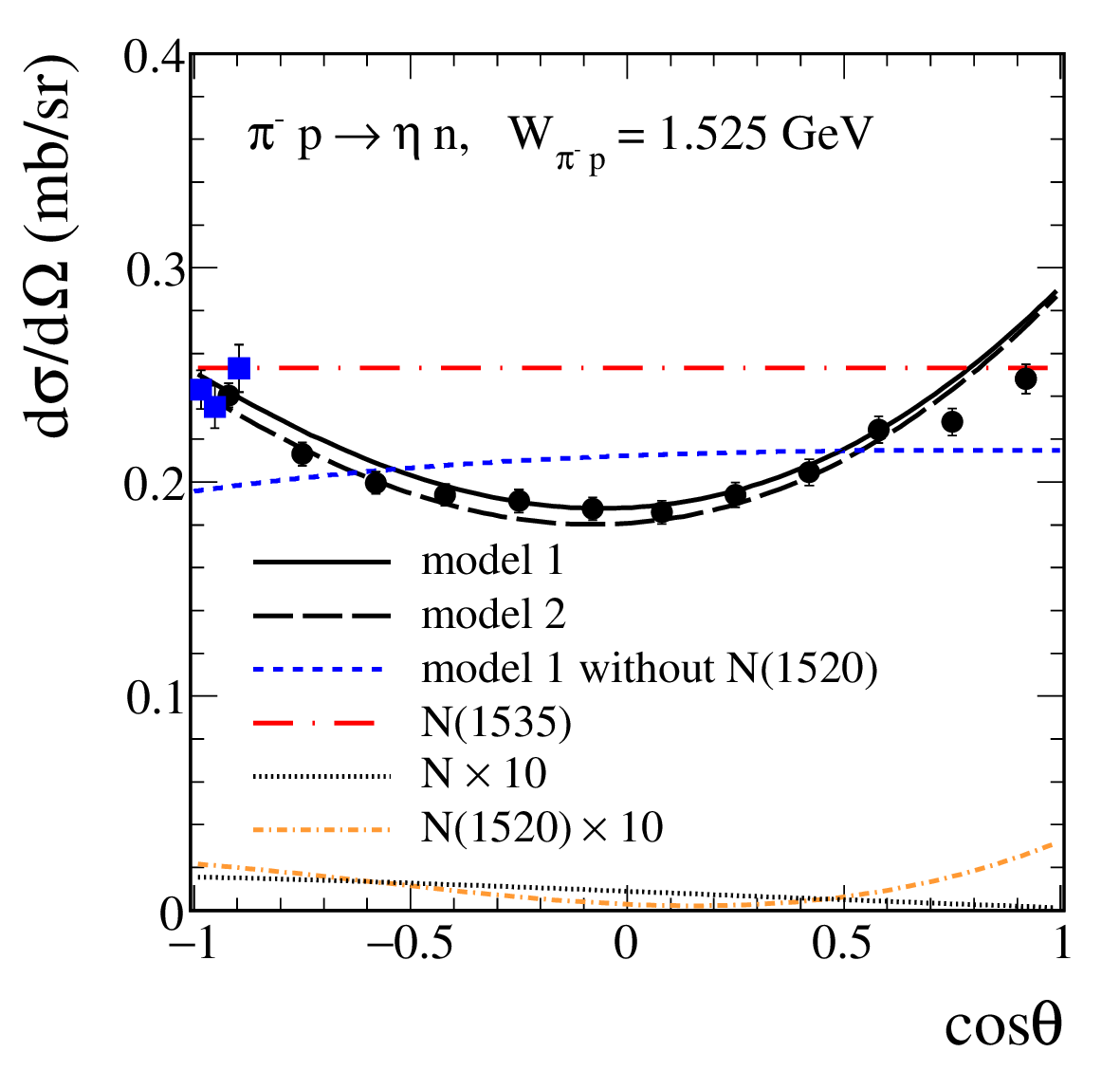}
\includegraphics[width=6.8cm]{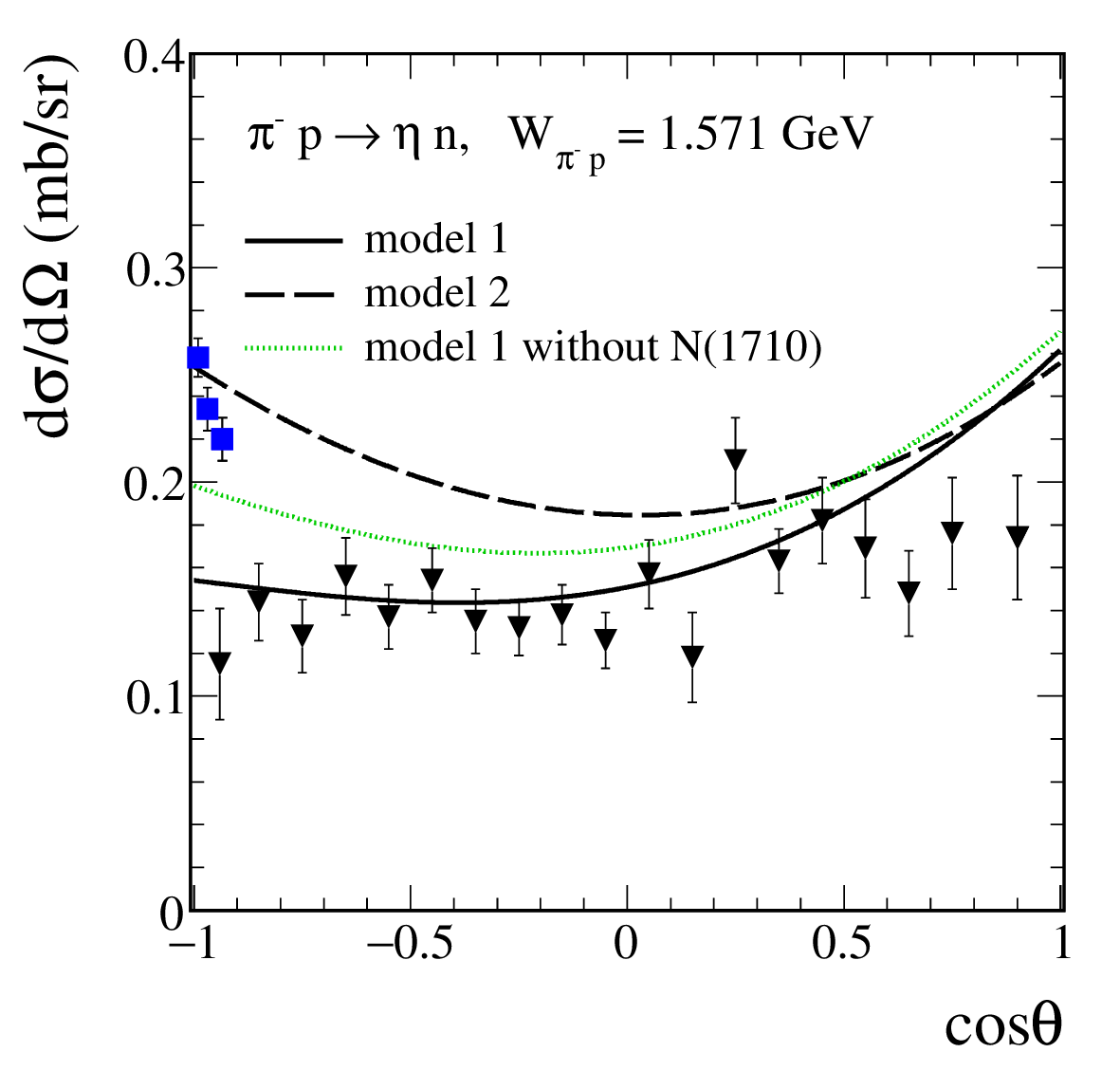}
\includegraphics[width=6.8cm]{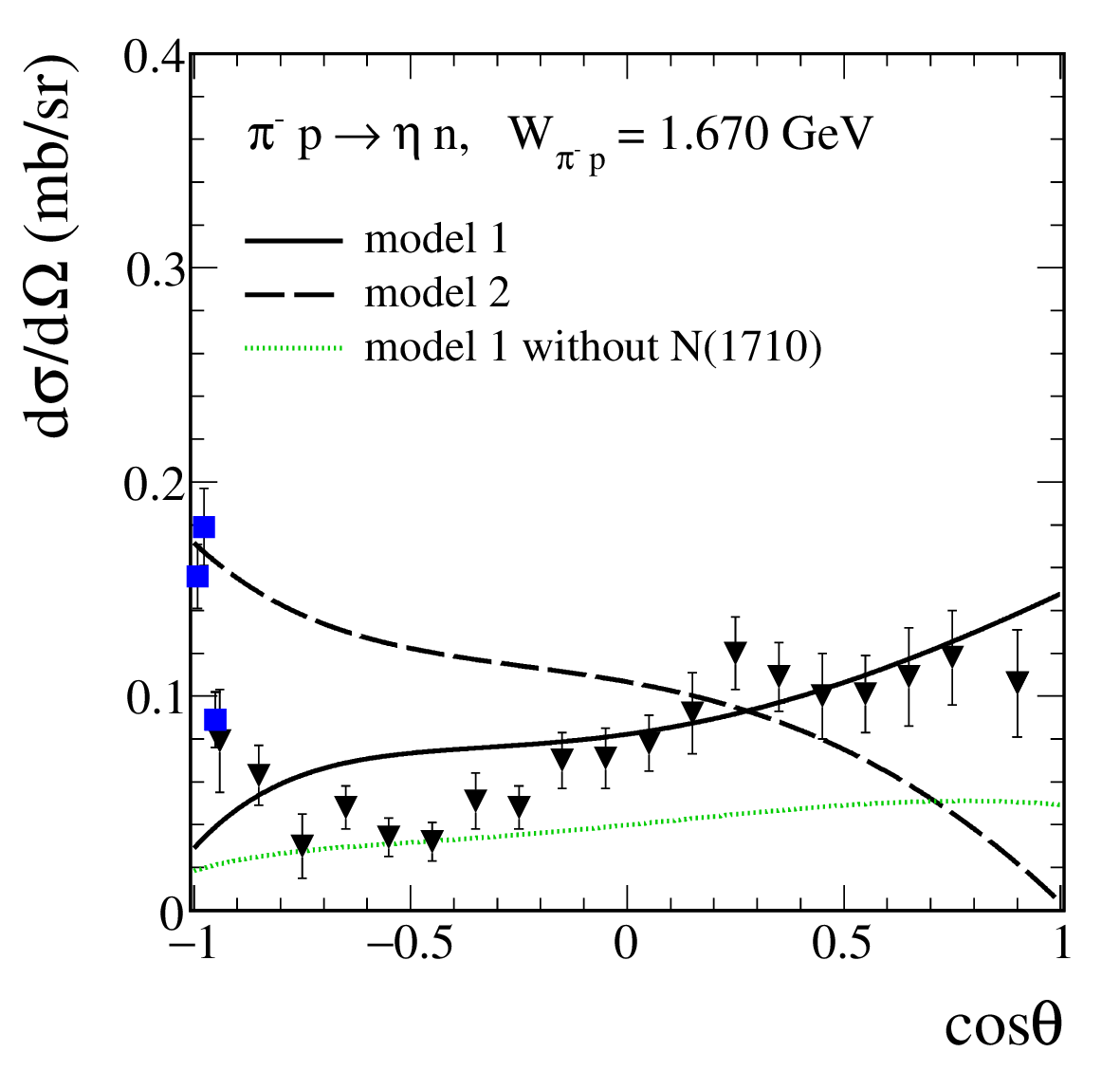}
  \caption{\label{fig:B2}
  \small
The differential cross sections 
$d\sigma/d\cos\theta$ for $\pi^{-} p \to \eta n$
calculated for various $W_{\pi p}$.
Data are from \cite{Prakhov:2005qb} ($\bullet$),
\cite{Brown:1979ii} ($\blacktriangledown$),
and in the backward scattering region
from \cite{Debenham:1975bj} ({\color{blue} $\blacksquare$}).
The solid and dashed lines represent the results for model 1 and model 2, respectively.
In the right top panel,
the $N$ and $N(1520)$ contributions multiplied by 
a factor 10 to be visible.
In the bottom panels the green dotted lines correspond
to the model without $N(1710)$.}
\end{figure}

\end{widetext}

\section{Photoproduction of the $\eta$ meson via reggeized-vector-meson exchanges}
\label{sec:appendixB}

\begin{figure}[!ht]
\includegraphics[width=0.28\textwidth]{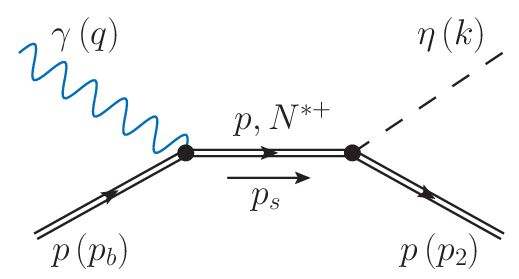}
\includegraphics[width=0.28\textwidth]{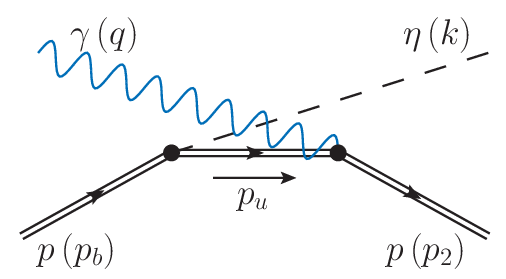}
\includegraphics[width=0.31\textwidth]{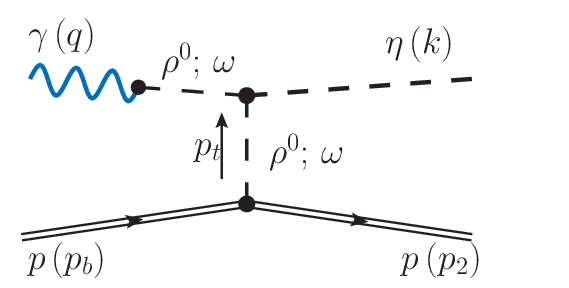}
  \caption{\label{fig:diagrams_eta_photoproduction}
  \small
Diagrams for photoproduction of the $\eta$ meson.}
\end{figure}

The unpolarized differential cross section for the $\gamma p \to \eta p$ reaction is given by
\begin{eqnarray}
&& \frac{d\sigma}{d\Omega} = \frac{1}{64 \pi^{2} s} \frac{| \bk |}{| \bq |} \frac{1}{4} \sum_{{\rm spins}}|{\cal M}_{\gamma p \to \eta p}|^{2}\,, 
\nonumber \\
&& d\Omega = \sin\theta \,d\theta \,d\phi \,.
\label{dsig_dz}
\end{eqnarray}
The considerations are made in the center-of-mass (c.m.) frame,
$s$ is the invariant mass squared of the $\gamma p$ system,
and $\bq$ and $\bk$ are 
the c.m. three-momenta of the initial photon and 
final $\eta$ meson, respectively.
Taking the direction of $\bq$ as a $z$ axis,
the polar and azimuthal angles of $\bk$
are defined as $\theta$ and $\phi$, respectively.
The amplitude ${\cal M}_{\gamma p \to \eta p}$
contains the processes shown in
Fig.~\ref{fig:diagrams_eta_photoproduction},
that is,
the $s$- and $u$-channel $p$ and $N^{*}$ exchanges and
the $t$-channel vector-meson $(V = \rho^{0}$, $\omega$) exchanges.
In principle the contact-type interaction current 
that ensures gauge invariance should also be considered; 
see the discussion, e.g., in \cite{Huang:2011as}.
From a pragmatic point of view, however, only 
the tensor-type coupling in $\gamma pp$ vertex
is kept, while
the vector-type coupling is omitted.

The standard kinematic variables are
\begin{eqnarray}
&&s = W_{\gamma p}^{2} = (p_{b} + q)^{2} = (p_{2} + k)^{2}\,, \nonumber \\
&&p_{s} = p_{b} + q = p_{2} + k  \,, \quad s = p_{s}^{2}\,, 
\nonumber \\
&&p_{u} = p_{b} - k = p_{2} - q  \,, \quad u = p_{u}^{2}\,, 
\nonumber \\
&&p_{t} = p_{b} - p_{2} = k - q  \,, \quad t = p_{t}^{2}\,.
\label{2to2_kinematic}
\end{eqnarray}

The amplitude for the $\gamma p \to \eta p$ reaction
reads
\begin{eqnarray}
{\cal M}_{\gamma p \to \eta p} =
{\cal M}_{s}^{(p)} + 
{\cal M}_{u}^{(p)} + 
\sum_{N^{*}_{1/2}}
{\cal M}_{s}^{(N^{*}_{1/2})} +
\sum_{V}
{\cal M}_{t}^{(V)} \,.\nonumber \\
\label{amp_photo}
\end{eqnarray}
It was checked numerically that
the contributions from the $u$-channel $N^{*}$ exchanges are small. 
The reason for this is explained in the previous section.
Thus, these terms can be safely neglected in the considerations.
Therefore, in the current analysis, only
the $s$-channel $N^{*}$ resonances
with $J^{P} = 1/2^{\pm}$ are considered.
To calculate the amplitudes with the $N^{*}$ and $V$ (vector-meson) exchanges,
the vector-meson-dominance (VMD) approach is used.
Amplitudes with the virtual proton and $N^{*}$ resonances 
are treated analogously as for the $\pi^{-} p \to \eta n$
reaction discussed in Appendix~\ref{sec:appendixA}.
Therefore, the focus will now be on the $\eta$-photoproduction 
process via the $V$ exchanges.

The amplitude via the vector-meson exchanges includes two terms
\begin{eqnarray}
{\cal M}_{t}^{(V)} = 
{\cal M}^{(\rho \, {\rm exchange})} + {\cal M}^{(\omega \, {\rm exchange})}\,.
\label{amp_sum_photo}
\end{eqnarray}
The generic amplitude with $V = \rho^{0}, \omega$,
for the diagram in Fig.~\ref{fig:diagrams_eta_photoproduction},
can be written as 
\begin{eqnarray}
{\cal M}^{(V \, {\rm exchange})}
&=&
(-i)  
\epsilon^{(\gamma)\,\mu''}(\lambda_{\gamma})\,
i\Gamma^{(\gamma \to V)}_{\mu'' \mu'}(q)\,
i\Delta^{(V)\,\mu' \mu}(q)\nonumber \\ 
&& \times 
i\Gamma_{\mu \nu}^{(VV \eta)}(q, p_{t})
i\tilde{\Delta}^{(V)\,\nu \nu'}(s,t)\nonumber \\ 
&& \times  
\bar{u}(p_{2}, \lambda_{2}) 
i\Gamma_{\nu'}^{(V pp)}(p_{2},p_{b}) 
u(p_{b}, \lambda_{b}) \,, \nonumber
\label{gamp_Mp}\\
\end{eqnarray}
where $p_{b}$, $p_{2}$ and $\lambda_{b}$, $\lambda_{2} = \pm \frac{1}{2}$ 
denote the four-momenta and helicities of the incoming and outgoing protons.
In the equation above, 
reggeized treatment of $\rho$ and $\omega$ exchanges is used; 
see Eq.~(\ref{reggeization_2}).
For the building blocks of the amplitude (\ref{gamp_Mp}) 
see Sec.~\ref{sec:formalism}.
The VMD relations 
for the $\gamma V$ transition vertices and $\gamma_{V}$ couplings
are given in Eqs.~(3.23)--(3.25) of \cite{Ewerz:2013kda}.
The above amplitude then can be expressed as follows:
\begin{eqnarray}
{\cal M}^{(V  \, {\rm exchange})}
&=&
\dfrac{e}{\gamma_{V}}
\epsilon^{(\gamma)\, \mu}(\lambda_{\gamma})\,
\Gamma_{\mu \nu}^{(VV \eta)}(q, p_{t})\,
\tilde{\Delta}_{T}^{(V)}(s,t)\nonumber \\ 
&& \times  
\bar{u}(p_{2}, \lambda_{2}) 
\Gamma^{(V pp)\,\nu}(p_{2},p_{b}) 
u(p_{b}, \lambda_{b})\,. \nonumber
\label{gamp_Mp_aux}\\
\end{eqnarray}
Here
\begin{eqnarray}
\frac{4 \pi}{\gamma_{\rho}^{2}} = 0.496 \pm 0.023\,, \quad
\frac{4 \pi}{\gamma_{\omega}^{2}} = 0.042 \pm 0.0015\,,
\label{A5}
\end{eqnarray}
where $\gamma_{\rho} > 0$ and $\gamma_{\omega} > 0$.

The $g_{\omega \omega \eta}$ and $g_{\rho \rho \eta}$
coupling constants
occurring in the $VV \eta$ vertices (\ref{VVM})
were adjusted to the experimental decay widths
$\Gamma(\omega \to \eta \gamma)$ and 
$\Gamma(\rho^{0} \to \eta \gamma)$,
respectively.
The amplitude for the reaction
$V (k_{V}, \epsilon^{(V)}) 
\to \eta (k_{\eta}) \gamma (k_{\gamma},\epsilon^{(\gamma)})$
is given by
\begin{eqnarray}
{\cal M}^{(V \to \eta \gamma)}
&=&
-\dfrac{e}{\gamma_{V}}\,\frac{g_{VV \eta}}{2 m_{V}} 
(\epsilon^{(\gamma)\, \mu}(\lambda_{\gamma}))^{*}
\epsilon^{(V)\, \nu}(\lambda_{V}) \nonumber \\
&& \times 
\varepsilon_{\mu \nu \alpha \beta}
k_{\gamma}^{\alpha} k_{V}^{\beta}
F^{(VV \eta)}(0, m_{V}^{2}, m_{\eta}^{2})
\,,
\label{V_etagam_decay}
\end{eqnarray}
where
%
$F^{(VV \eta)}(k_{\gamma}^{2} = 0, k_{V}^{2} = m_{V}^{2}, 
k_{\eta}^{2} = m_{\eta}^{2}) = 1$.
%
In the calculations, $\Gamma_{\rho} = 148$~MeV,
$m_{\rho} = 775$~MeV, 
$\Gamma_{\omega} = 8.68$~MeV,
$m_{\omega} = 783$~MeV,
and the central values of branching fractions 
${\cal B}(V \to \eta \gamma)$
are taken from the PDG \cite{ParticleDataGroup:2024cfk}.
Estimated values are
$|g_{\rho \rho \eta}| = 12.36$ and
$|g_{\omega \omega \eta}| = 12.16$.
It was assumed in the calculations that these coupling constants are positive.

The form factor in the $VV \eta$ vertex
$i\Gamma_{\mu \nu}^{(VV \eta)}(q, p_{t})$
in (\ref{gamp_Mp_aux}) is
%
\begin{eqnarray}
F^{(VV \eta)}(q^{2}, p_{t}^{2}, k^{2}) 
= F^{(VV \eta)}(0, p_{t}^{2}, m_{\eta}^{2}) 
= F_{V}(p_{t}^{2})
\label{F_VV_aux1}
\end{eqnarray}
and 
\begin{eqnarray}
F_{V}(t) = 
\frac{\Lambda_{V}^{2}-m_{V}^{2}}{\Lambda_{V}^{2}-t}\,.
\label{F_VV_aux2}
\end{eqnarray}
From comparison of the model to the $\eta$-meson
angular distributions from the CLAS experiment
one can extract the cutoff parameter 
$\Lambda_{V} = \Lambda_{V,\,{\rm mon}}$.
Details will be given when discussing 
differential distributions below.

\begin{figure}[!ht]
\includegraphics[width=6.8cm]{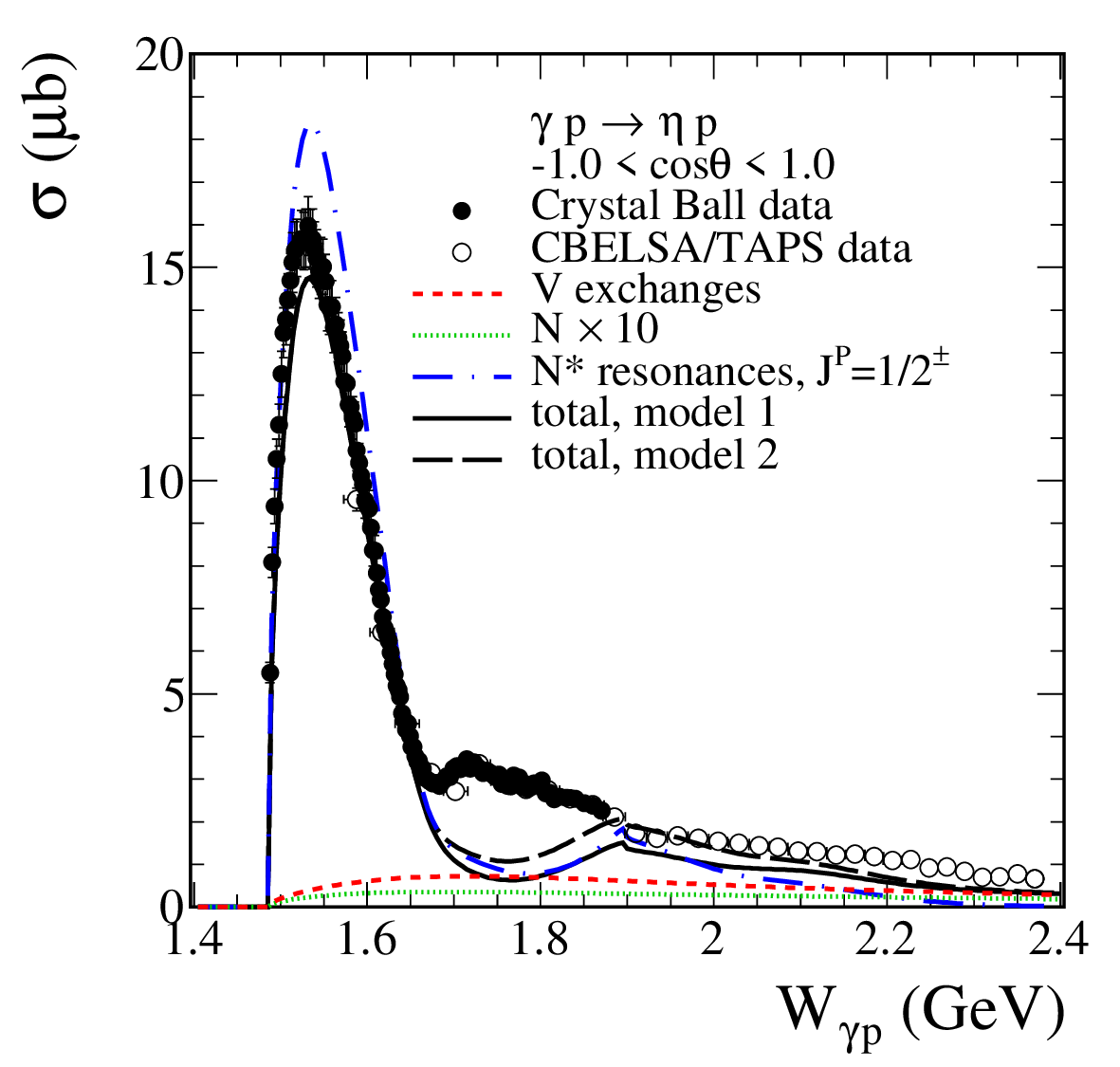}
\includegraphics[width=6.8cm]{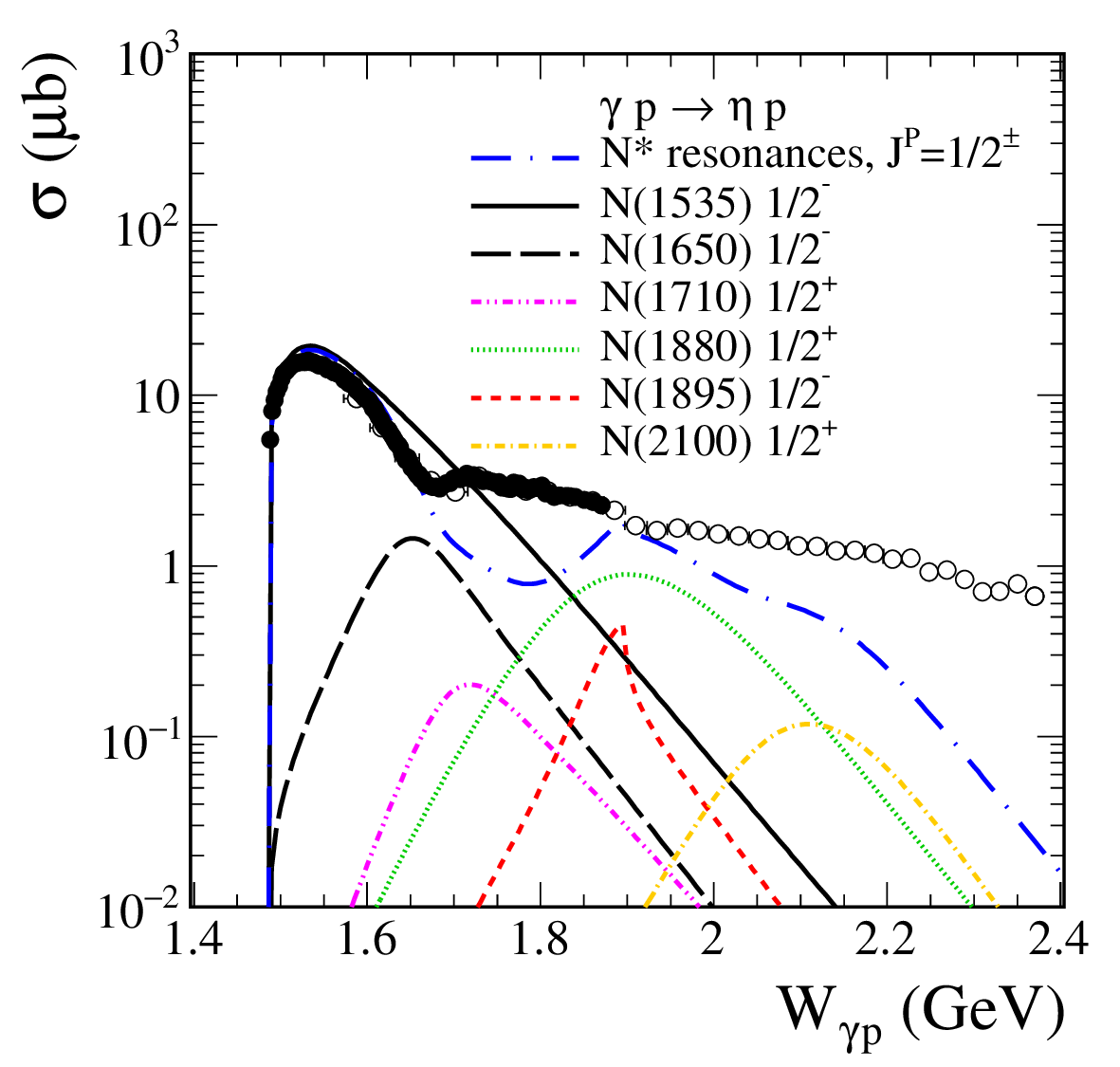}
\caption{\label{fig:A4}
\small
The elastic $\eta$ photoproduction cross section
as a function of the center-of-mass energy $W_{\gamma p}$ for $-1.0 < \cos\theta < 1.0$.
The data from Crystal Ball ($\bullet$) \cite{McNicoll:2010qk}
and CBELSA/TAPS ($\circ$) \cite{Crede:2009zzb} experiments are shown.
Several contributions were included.
The coherent sum of all
contributions is shown by the solid and long-dashed lines, corresponding to model~1 and model~2, respectively.
In the top panel, 
the $N$ contribution is multiplied by a factor of 10 
to make it visible.
In the bottom panel, the $N^{*}$ resonances
included in the calculation are shown.}
\end{figure}

Figure~\ref{fig:A4} shows the integrated cross sections for
the $\gamma p \to \eta p$ reaction
for two models, 1 and 2 
(explained in the previous section),
together with experimental data.
It can be seen from the calculation results 
that the dominant production mechanism is via $N(1535)$ resonance.
At low energies, also the $N(1650)$ play an important role, 
while at higher energies the $N(1880)$ resonance 
is especially pronounced.
From the top panel,
one can see that the $V$-exchange contribution
has different energy dependence of cross section
and it plays some role rather at higher energies.
There, the $\rho$-exchange term 
is much larger than the $\omega$-exchange term
due to larger coupling constants
both for the $\gamma \to V$ transition vertex (\ref{A5}) 
and for the tensor coupling in the $V$-proton vertex (\ref{Vpp_couplings}).
There is a large interference between the two components.

Figure~\ref{fig:A1} shows the calculated results 
for the $\gamma p \to \eta p$ reaction
together with selected experimental data on $d\sigma/d\cos\theta$
from \cite{McNicoll:2010qk,Crede:2009zzb,Williams:2009yj,Hu:2020ecf}.
The new CLAS data \cite{Hu:2020ecf} 
are shown in the bottom panels, (c) -- (e),
as the black solid circles ($\bullet$).
They are consistent with earlier CLAS results \cite{Williams:2009yj} 
shown in the panel (b)
but extend the energy range beyond 
the nucleon resonance region into the Regge regime,
i.e. large $W_{\gamma p}$ and $\cos\theta$.
As illustrated in panel (c) for $W_{\gamma p} \approx 2.38$~GeV
and in the very forward region
the CBELSA/TAPS data \cite{Crede:2009zzb} 
appear to be higher than the CLAS data \cite{Hu:2020ecf}.
The $V$-exchange contribution describes
the new CLAS data \cite{Hu:2020ecf}
within their uncertainties 
only in the forward scattering region ($0.5 < \cos\theta < 1.0$)
and for $W_{\gamma p} \gtrsim 2.4$~GeV.
The reggeized-vector-meson-exchange model 
is not expected to describe 
the experimental data at lower energies. 
Indeed, this region is dominated by nucleon resonances;
see Fig.~\ref{fig:A4} and the discussions in
\cite{Chiang:2002vq,Nakayama:2008tg}.
It can be seen from Fig.~\ref{fig:A1} that 
with the increase of energy, the contributions of
$u$-channel proton exchange and $t$-channel $V$ exchange
become more significant 
at the backward angle and forward angle respectively.

\begin{widetext}

\begin{figure}[!ht]  
\center
(a)\includegraphics[width=6.8cm]{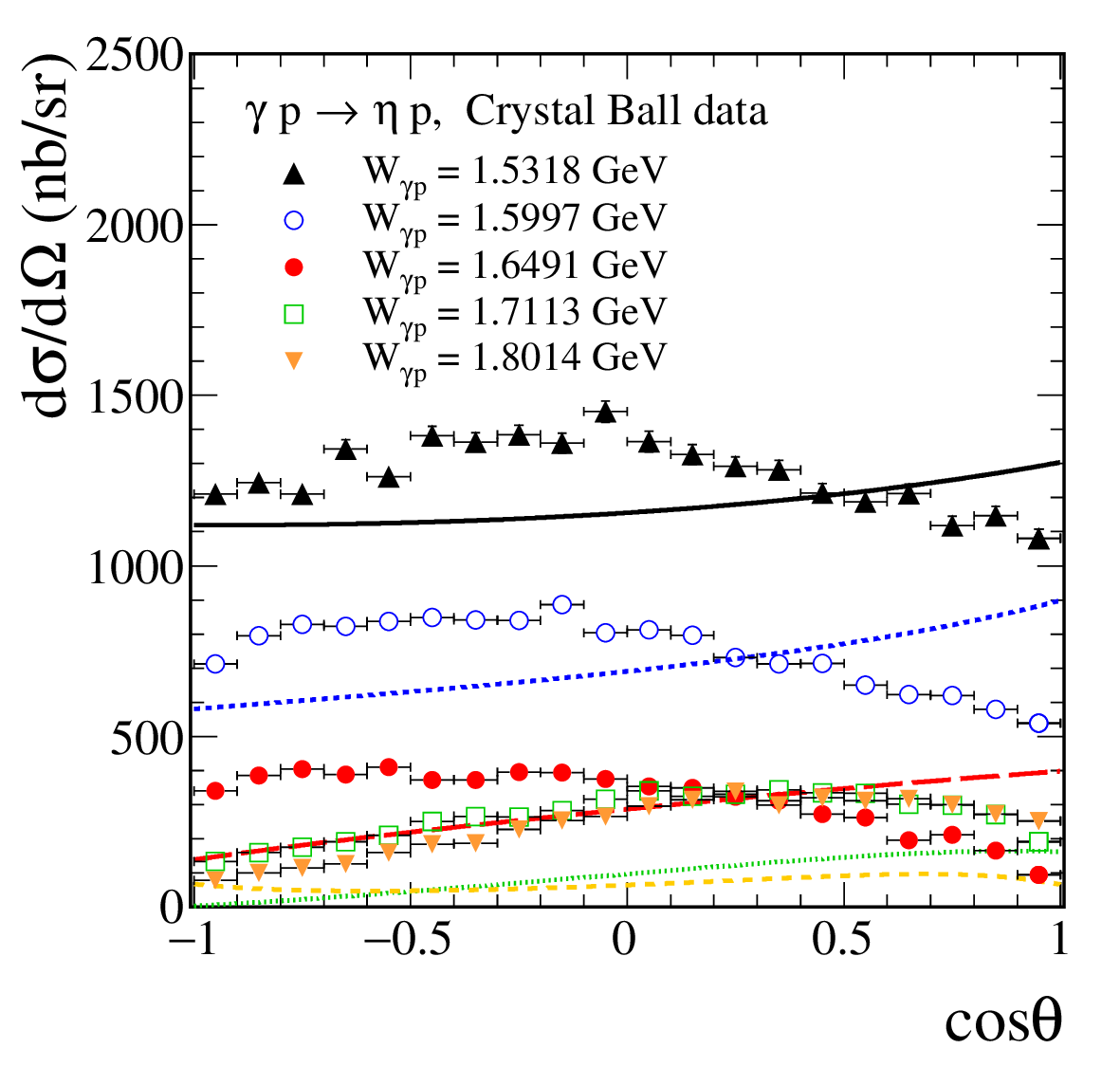}
(b)\includegraphics[width=6.8cm]{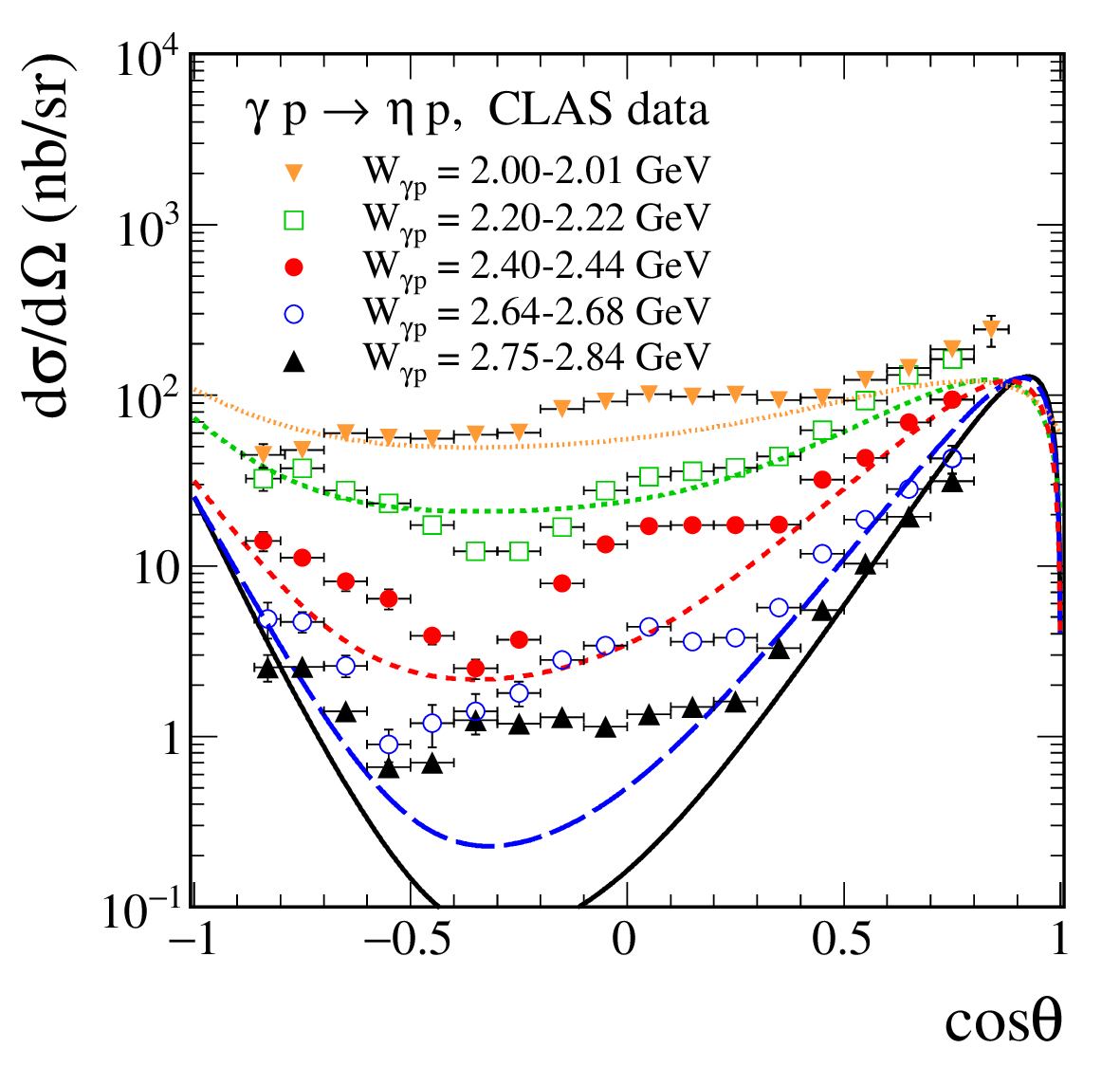}
(c)\includegraphics[width=6.8cm]{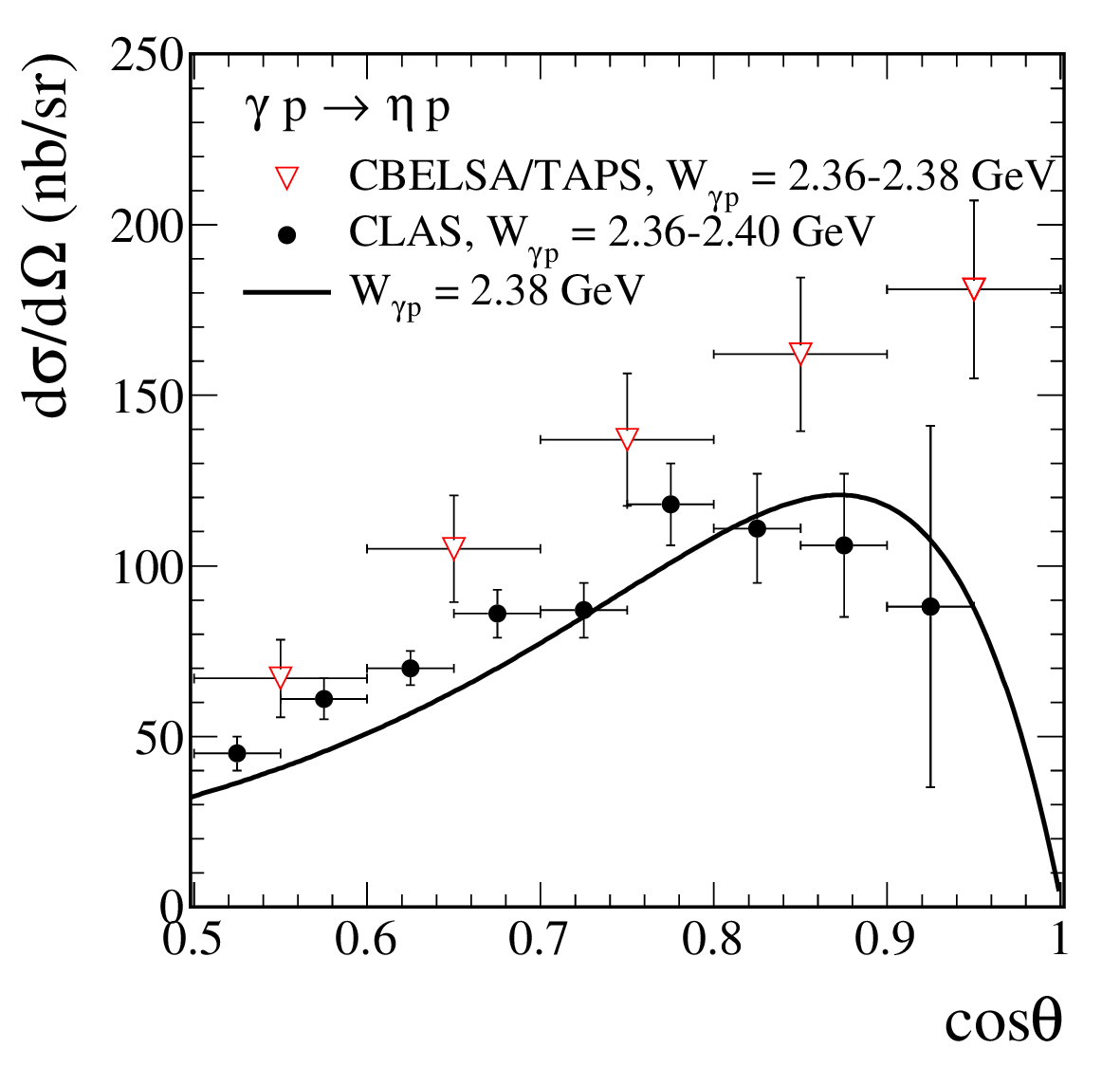}
(d)\includegraphics[width=6.8cm]{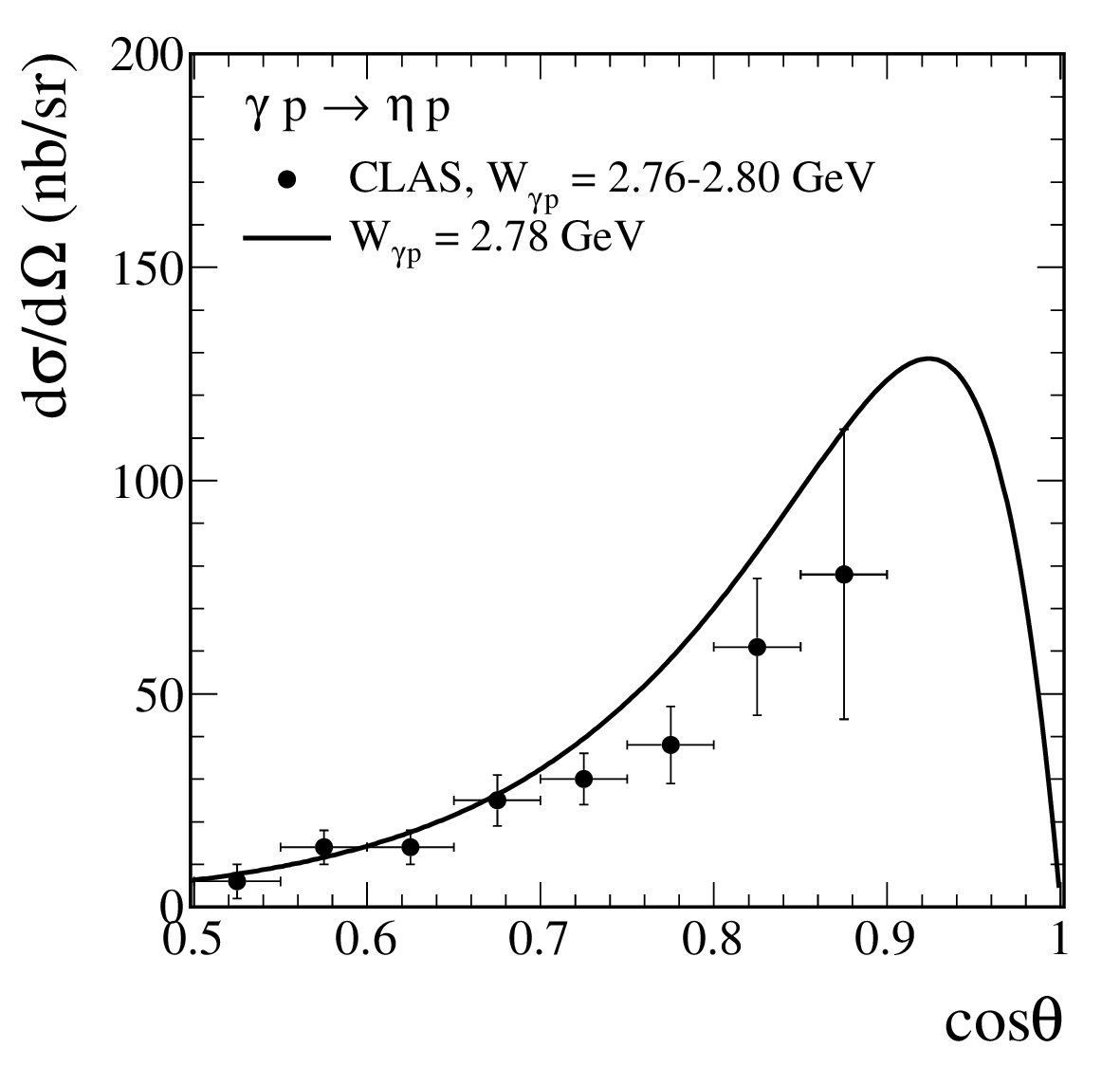}
(e)\includegraphics[width=6.8cm]{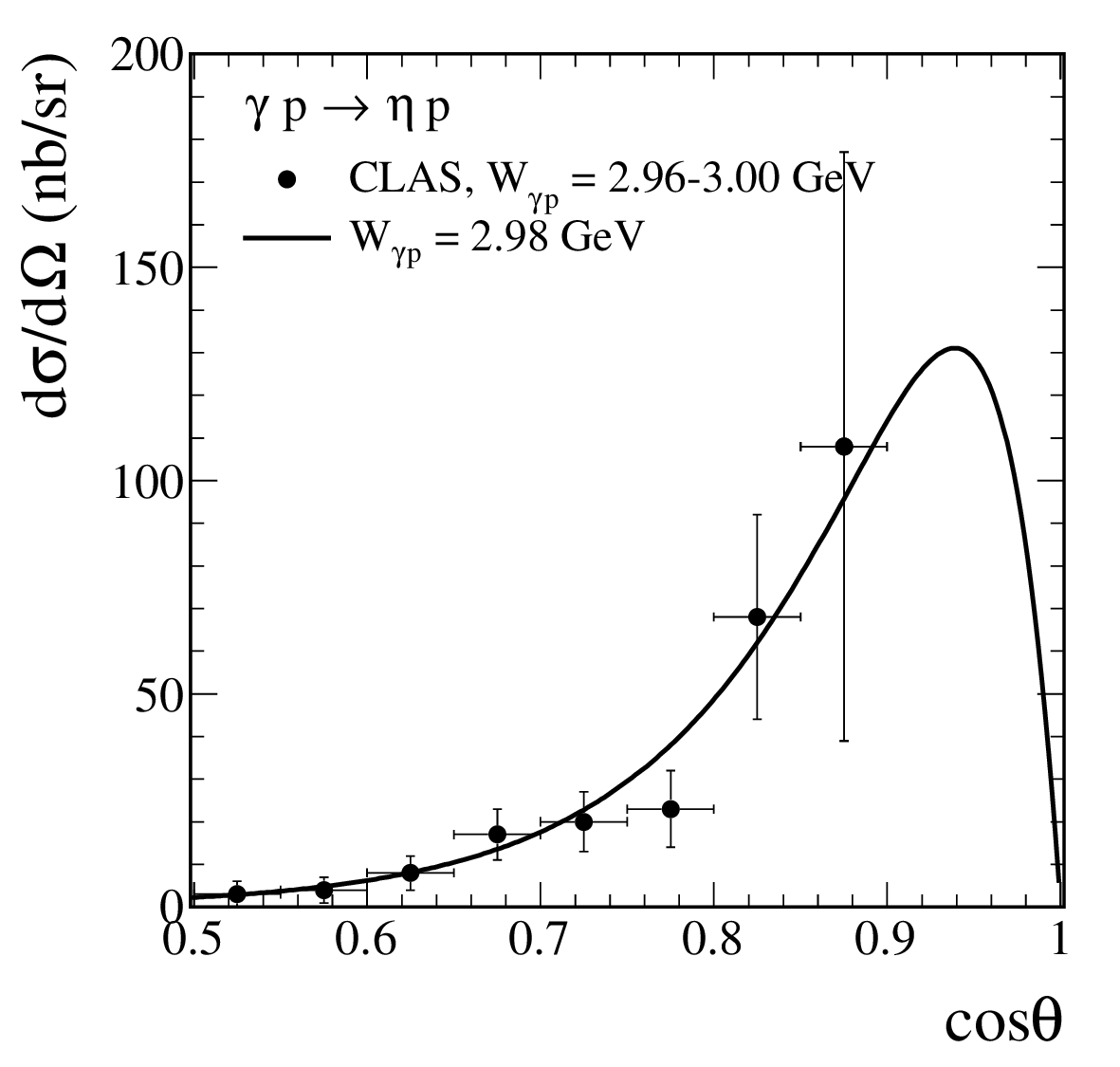}
  \caption{\label{fig:A1}
  \small
The differential cross sections 
$d\sigma/d\cos\theta$ for the $\gamma p \to \eta p$ reaction.
The curves represent the model results calculated for 
various $W_{\gamma p}$
using $\Lambda_{VNN} = 1.4$~GeV 
and $\Lambda_{V,\,{\rm mon}} = 1.2$~GeV.
The Crystal Ball data in panel (a) 
are from \cite{McNicoll:2010qk}.
The CLAS data on the panel (b) 
are taken from \cite{Williams:2009yj}.
In panels (c), (d) and (e) 
the new CLAS data from \cite{Hu:2020ecf} are presented;
see the black solid circles ($\bullet$)
(these data points were scanned from Fig.~17 of \cite{Hu:2020ecf}).
In panel (c), data from the CBELSA/TAPS experiment \cite{Crede:2009zzb} are shown for comparison.}
\end{figure}

Figure~\ref{fig:A2} shows a comparison of the theoretical results for $d\sigma/dt$
with various experimental data, as described in the figure caption.
Here, one deals with the kinematic region suited for
the reggeized-vector-meson exchange mechanism.
It should be noted that the differential distribution 
at $W_{\gamma p} = 2.38$~GeV 
peaks for $\cos\theta = 0.88$ corresponding to
$-t = 0.23$~GeV$^{2}$,
while at $W_{\gamma p} = 2.98$~GeV 
it peaks for $\cos\theta = 0.94$ corresponding to
$-t = 0.20$~GeV$^{2}$ (see the right panel of Fig.~\ref{fig:A2}).
As a result of these comparisons, one can estimate the form factor
in the $VV \eta$ vertex, 
which describes the $t$ dependencies of the $V$ exchange.
One finds the cutoff parameters
$\Lambda_{VNN} = 1.4$~GeV and $\Lambda_{V,\,{\rm mon}} = 1.3$~GeV
with the same values for $\rho^{0}$ and $\omega$. 
\begin{figure}[!ht]  
\center
\includegraphics[width=6.8cm]{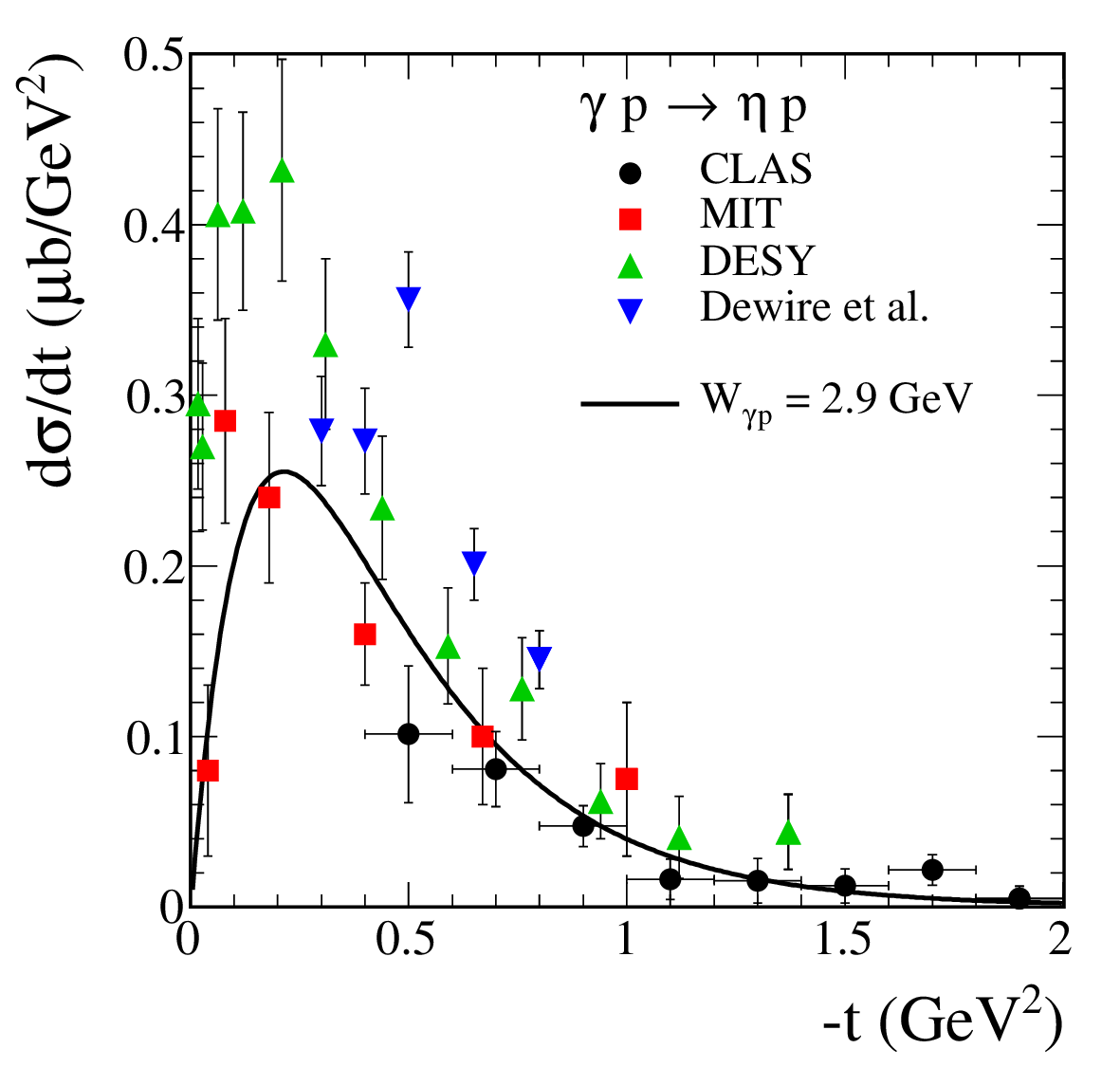}
\includegraphics[width=6.8cm]{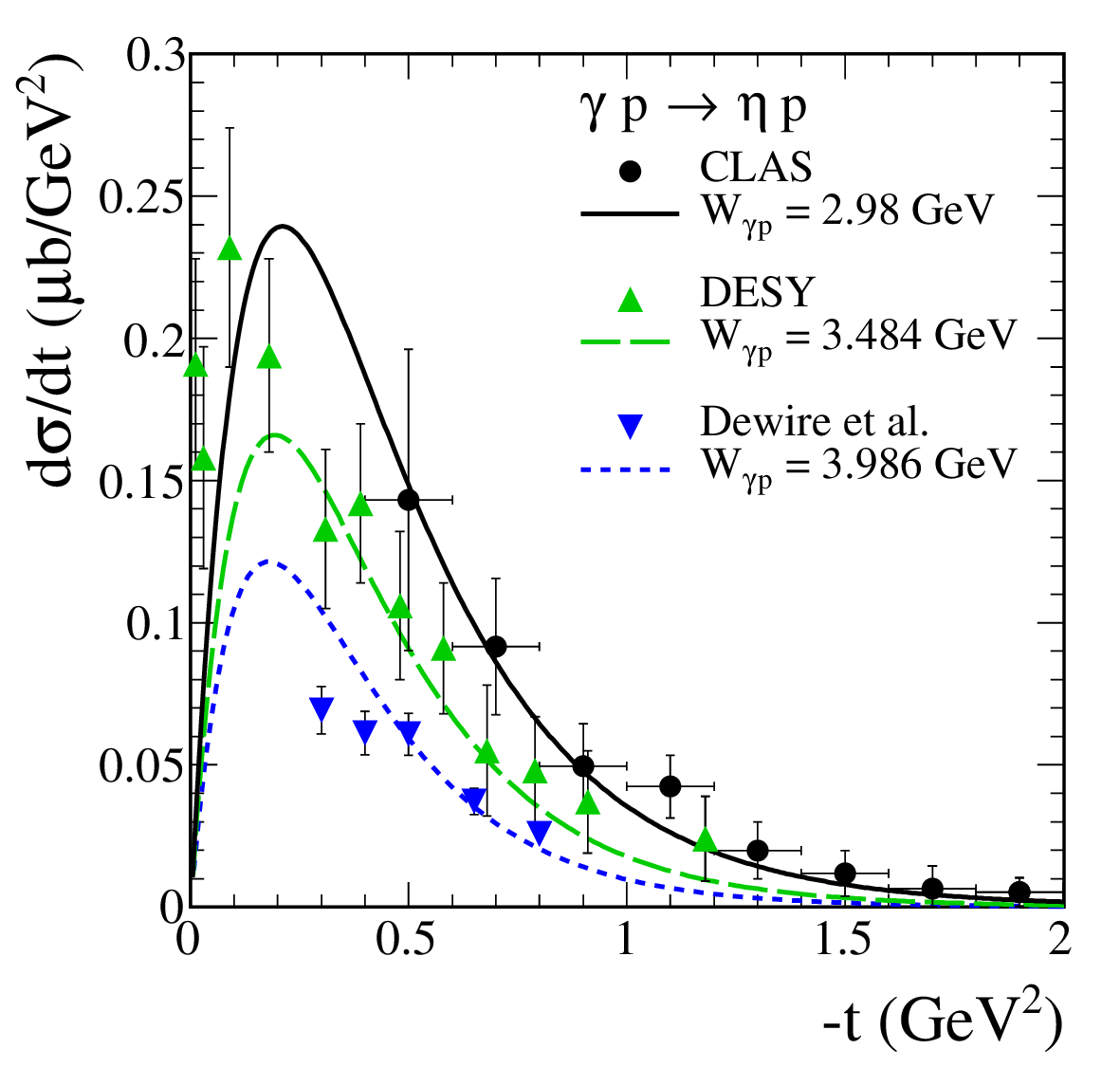}
  \caption{\label{fig:A2}
  \small
The differential cross sections $d\sigma/dt$
for the reaction $\gamma p \to \eta p$.
The curves represent the model results calculated for 
various $W_{\gamma p}$
using $\Lambda_{VNN} = 1.4$~GeV and $\Lambda_{V,\,{\rm mon}} = 1.2$~GeV.
Shown are data from
CLAS \cite{Hu:2020ecf} ($\bullet$)  
(for $W_{\gamma p} = 2.88-2.92$ and 2.98~GeV
in the left and right panels, respectively),
MIT \cite{Bellenger:1968zz} ({\color{red} $\blacksquare$}) 
(for $W_{\gamma p} = 2.694-3.084$~GeV), 
DESY \cite{Braunschweig:1970jb} ({\color{green}$\blacktriangle$}) 
(for $W_{\gamma p} = 2.895$ and 3.484~GeV), 
and Dewire \textit{et al.} \cite{Dewire:1972kk} 
({\color{blue}$\blacktriangledown$}) 
(for $W_{\gamma p} = 2.895$ and 3.986~GeV).
The new CLAS data ($\bullet$)
were scanned from Fig.~18 of \cite{Hu:2020ecf}.}
\end{figure}

\end{widetext}

\section{The $N(1535) \to N \gamma$ and $N(1535) \to N \pi^+ \pi^-$ decays and the $\rho N N(1535)$ coupling constant}
\label{sec:appendixC}

In this Appendix,
the $\rho N N(1535)$ coupling constant
is determined in two ways:
one using the $N(1535) \to N \rho^{0} \to N \gamma$ decay,
and the other using the
$N(1535) \to N \rho^{0} \to N \pi^+ \pi^-$ decay.

In the first method, the radiative decay
of the $N(1535)$ is considered as follows:
\begin{eqnarray}
N(1535)(k_{N^*},\lambda_{N^*}) \to
N(k_{N},\lambda_{N})  + \gamma(k_{\gamma}, \epsilon^{(\gamma)}) \,,
\label{D1}
\end{eqnarray}
where $k_{N}$, $k_{N^{*}}$ and $\lambda_{N}, \lambda_{N^{*}} = \pm 1/2$ 
denote the four-momenta and helicities of the nucleon and resonance,
and $k_{\gamma}$ and $\epsilon^{(\gamma)}$ denote
the four-momentum and polarization vector of photon
with helicities $\lambda_{\gamma} = \pm 1$, respectively.

The amplitude for the reaction (\ref{D1}) is given by

\begin{widetext}

\begin{align}
{\cal M}^{(N^{*} \to N \gamma)}
&= (-i) 
\bar{u}(k_{N},\lambda_{N})
i\Gamma_{\mu}^{(\rho N N^{*})}(k_{N},k_{N^{*}})
u(k_{N^{*}},\lambda_{N^{*}}) \,
i\Delta^{(\rho)\,\mu \nu}(k_{\gamma}) \,
i\Gamma^{(\rho \to \gamma)}_{\nu \kappa}(k_{\gamma})\,  
(\epsilon^{(\gamma)\,\kappa}(\lambda_{\gamma}))^{*}
\nonumber \\
&= \frac{e}{\gamma_{\rho}}
\bar{u}(k_{N},\lambda_{N})
\Gamma_{\mu}^{(\rho N N^{*})}(k_{N},k_{N^{*}})
u(k_{N^{*}},\lambda_{N^{*}}) \,
(\epsilon^{(\gamma)\,\mu}(\lambda_{\gamma}))^{*}\,,
\label{D3}
\end{align}
\end{widetext}
using the VMD ansatz
for the coupling of the $\rho^{0}$ meson to the photon; 
see (3.23)--(3.25) of \cite{Ewerz:2013kda}.
The vertex for the $\rho N N(1535)$ coupling can be derived from an effective coupling Lagrangian.
However, different forms are commonly used in the literature.
The gauge invariant Lagrangian from (A6) of \cite{Nakayama:2008tg} is
\begin{align}
{\cal L}_{\rho N N^{*}_{1/2^{\mp}}}
=& -\frac{1}{2 m_{N}}
\bar{N}^{*} 
\begin{pmatrix}
\gamma_{5}\\
1
\end{pmatrix}
\Big[
g_{\rho N N^{*}}^{(\text{V})}
\Big(
\frac{\gamma_{\mu} \partial^{2}}{m_{N^{*}} + m_{N}}
-i \partial_{\mu}
\Big)\nonumber \\
&-g_{\rho N N^{*}}^{(\text{T})}
\sigma_{\mu \nu}
\partial^{\nu} 
\Big] (\btau \bPhi_{\rho}^{\mu}) 
N + {\rm h.c.} \,,
\label{D3_1}
\end{align}
see Refs.~\cite{Nakayama:2002mu,Nakayama:2003jn,Xie:2008ts},
where the vector- and tensor-type couplings were discussed.
It is worth mentioning that a pure vector type of the form 
$\propto \gamma_{5} [ \gamma_{\mu} \partial^{2} - (m_{N^{*}} + m_{N}) \partial_{\mu} ]$
was derived in \cite{Riska:2000gd}
and another variant of vector-type coupling in (23) of \cite{Pena:2000gb}.
A pure tensor-type coupling 
was used, e.g., in \cite{Shyam:2007iz,Kaptari:2007ss}.
Note that the vector-type term 
of (\ref{D3_1})
vanishes for the real photon in (\ref{D1}), (\ref{D3}) 
in the limit $k_{\gamma}^{2} = 0$.
The consideration here is limited to the
pure tensor-type coupling of (\ref{D3_1}) with
$g_{\rho N N^{*}} \equiv g_{\rho N N^{*}}^{(\text{T})}$.

%
%

In practical calculations, one introduces
in the $\Gamma^{(\rho N N^{*})}_{\mu}$ vertex
[derived from an effective coupling Lagrangian
(\ref{D3_1})]
the form factor
\begin{eqnarray}
F_{\rho N N^{*}}(k_{\rho}^{2}, k_{N}^{2}, k_{N^{*}}^{2}) 
= \tilde{F}_{V}(k_{\rho}^{2}) 
F_{B}(k_{N}^{2})
F_{B}(k_{N^{*}}^{2}) \,,
\label{D4}
\end{eqnarray}
with the assumption 
that $\tilde{F}_{\rho}(k_{\rho}^{2})$ 
is parametrized as
\begin{eqnarray}
\tilde{F}_{\rho}(k_{\rho}^{2})
= \frac{\tilde{\Lambda}_{\rho}^{4}}{\tilde{\Lambda}_{\rho}^{4} + (k_{\rho}^{2} - m_{\rho}^{2})^{2}}\,,
\label{D5}
\end{eqnarray}
where $\tilde{\Lambda}_{\rho}$ is cutoff parameter.
For the on shell case
$F_{\rho N N^{*}}(m_{\rho}^{2}, m_{N}^{2}, m_{N^{*}}^{2}) = 1$.
Then, for (\ref{D1}) one has
\begin{align}
F_{\rho N N^{*}}(k_{\rho}^{2}, k_{N}^{2}, k_{N^{*}}^{2}) 
&= F_{\rho N N^{*}}(0, m_{N}^{2}, m_{N^{*}}^{2}) \nonumber \\
&= 
\tilde{F}_{\rho}(0) 
F_{B}(m_{N}^{2}) 
F_{B}(m_{N^{*}}^{2}) 
\nonumber \\
&= 
\tilde{F}_{\rho}(0)\,.
\label{D4_aux}
\end{align}

In the first method, the radiative decay width is,
\begin{align}
\Gamma(N^{*} \to N \gamma)_{\rm tensor}=
\frac{e^{2}}{\gamma_{\rho}^{2}}
\frac{g_{\rho N N^{*}}}{16 \pi}
\frac{k(m_{N}^{2}-m_{N^{*}}^{2})^{2}}{m_{N}^{2}m_{N^{*}}^{2}}
(\tilde{F}_{\rho}(0))^{2} \,. 
\label{D2}
\end{align}
%
Here $k$ is the photon momentum in the $N^{*}$ rest frame,
$k = (m_{N^{*}}^{2} - m_{N}^{2})/2 m_{N^{*}}$.
Then, the absolute value of coupling constant 
$|g_{\rho N N^{*}}|$,
for $i = 2$ or 3,
can be adjusted to the experimental radiative decay width:
\begin{eqnarray}
\Gamma(N^{*} \to N \gamma) = 
\frac{k^{2}}{\pi} \frac{m_{N}}{m_{N^{*}}} 
|A_{1/2}|^{2}\,,
\label{D8}
\end{eqnarray}
where $A_{1/2}$ represents 
the transverse helicity-1/2 amplitude for the proton,
$A_{1/2}^{p} = 0.105 \pm 0.015$~GeV$^{-1/2}$ 
\cite{ParticleDataGroup:2024cfk}.
Note, that in Table~III of \cite{Hunt:2018wqz} 
the value $A_{1/2}^{p} = 0.107 \pm 0.003$~GeV$^{-1/2}$
is given.

One can compare the results with those obtained using the experimental branching ratio for helicity 1/2 from the PDG compilation,
${\cal B}(N(1535) \to p \gamma) = 0.15 - 0.30 \,\%$ 
\cite{ParticleDataGroup:2024cfk}.
The value of the coupling constant $|g_{\rho N N^{*}}|$ found in this way is smaller by a factor of 1.2 than that based on (\ref{D8}).

In the second method, one considers the decay of $N(1535)$ into a nucleon and two pions via an intermediate $\rho$ meson, taking into account its mass distribution:
\begin{eqnarray}
&&
N(1535)(k_{N^{*}},\lambda_{N^{*}}) 
\nonumber \\
&&
\quad \to N(k_{N},\lambda_{N}) + 
[\rho^{0}(k_{\rho}) \to \pi^{+}(k_{1}) + \pi^{-}(k_{2})] \,. \qquad \;
\label{D20}
\end{eqnarray}
The amplitude for the reaction (\ref{D20}),
denoted by ${\cal M}^{(N^{*} \to N \pi^{+} \pi^{-})}$,
is obtained from (\ref{D3})
by making the replacement
\begin{eqnarray}
&&i\Delta^{(\rho)\,\mu \nu}(k_{\gamma}) \,
i\Gamma^{(\rho \to \gamma)}_{\nu \kappa}(k_{\gamma})\,  
(\epsilon^{(\gamma)\,\kappa}(\lambda_{\gamma}))^{*} \nonumber \\
&&\quad \to 
i\Delta^{(\rho)\,\mu \nu}(k_{\rho}) \,
i\Gamma^{(\rho \pi \pi)}_{\nu}(k_{1},k_{2}) \nonumber \\
&&\quad \quad \;= 
\frac{g_{\rho \pi \pi}}{2} 
\left(-g^{\mu \nu} 
+ \frac{k_{\rho}^{\mu} k_{\rho}^{\nu}}{k_{\rho}^{2}} \right)
\Delta^{(\rho)}_{T}(k_{\rho}^{2})\,
(k_{1}-k_{2})^{\nu} \,, 
\label{D6} \nonumber \\
\end{eqnarray}
and taking 
\begin{eqnarray}
F_{\rho N N^{*}}(k_{\rho}^{2}, k_{N}^{2}, k_{N^{*}}^{2}) 
= F_{\rho N N^{*}}(k_{\rho}^{2}, m_{N}^{2}, m_{N^{*}}^{2})
= \tilde{F}_{\rho}(k_{\rho}^{2})
\label{D7} \nonumber \\
\end{eqnarray}
with $k_{\rho}^{2} = (k_{1}+k_{2})^{2}$.
The $\rho^{0}$~propagator function 
and the $\rho^{0} \pi^{+} \pi^{-}$
coupling in (\ref{D6})
are taken from (4.1)--(4.6)
and (3.35), (3.36) of \cite{Ewerz:2013kda}, respectively.

For the decay process (\ref{D20}) the phase-space integration
can be evaluated numerically using
the \textsc{Decay} Monte Carlo generator \cite{Kycia:2020mgf}.
The value of $g_{\rho N N^{*}}$,
can be obtained from 
the experimental partial decay width 
$\Gamma(N(1535) \to N \rho \to N \pi \pi)$.
Unfortunately, the branching ratio of the
$N(1535)$ state into the $N \rho$ channel 
appears to be not well known in the literature.
In \cite{Hunt:2018wqz} the following values were reported:
${\cal B}(N(1535) \to N \rho) = 14 \pm 2 \, \%$
and ${\cal B}(N(1535) \to N \rho)_{\rm{D-wave}} < 0.3\, \%$.
However, notably smaller value 
for the decay into S wave was found 
by the HADES Collaboration
\cite{HADES:2020kce}
${\cal B}(N(1535) \to N \rho) = 2.7 \pm 0.6\, \%$ 
and $0.5 \pm 0.5\, \%$ for the decay into D wave.
In the calculations, the value from \cite{HADES:2020kce} for the S wave was used.

\begin{figure}[!ht]
\includegraphics[width=6.8cm]{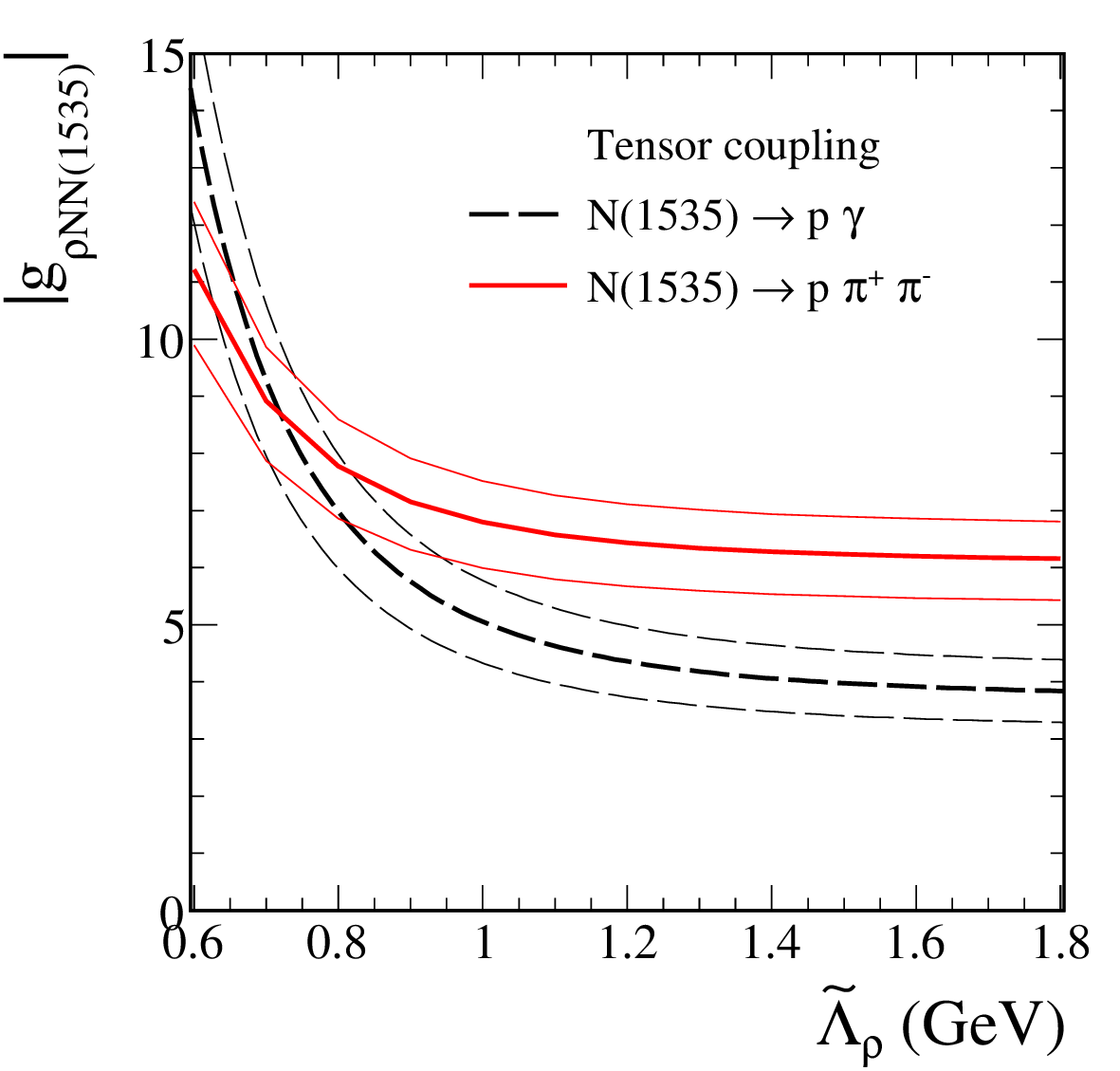}
\caption{
Coupling constant constrained
from the two decays $N(1535)^{+} \to p \gamma$
and $N(1535)^{+} \to p \pi^+ \pi^-$ 
as a function of the cutoff parameter 
$\tilde{\Lambda}_{\rho}$
in $\tilde{F}_{\rho}$ (\ref{D5}).}
\label{fig:app_D}
\end{figure}
The results are shown in Fig.~\ref{fig:app_D}.
From the radiative and two-pion decays of $N(1535)$ resonance, one estimates
the coupling constant
$g_{\rho N N^{*}} \equiv g_{\rho N N^{*}}^{(\text{T})}$ 
occurring in the vertex
$\Gamma_{\mu}^{(\rho N N^{*})}$ (\ref{D3_1})
versus the cutoff parameter $\tilde{\Lambda}_{\rho}$
in $\tilde{F}_{\rho}$ (\ref{D5}).
The values of coupling constants extracted form
the available experimental data on the two decays 
are large. 
This implies that the $\rho$-exchange contribution
may play an important role for the $pp \to pp \eta$ process
and should not be ignored.
The thin upper and lower lines in Fig.~\ref{fig:app_D} represent 
the results that comes from uncertainties in the input data used
($A_{1/2}^{p} = 0.105 \pm 0.015$~GeV$^{-1/2}$ \cite{ParticleDataGroup:2024cfk},
${\cal B}(N(1535) \to N \rho) = 2.7 \pm 0.6\, \%$ \cite{HADES:2020kce}).

\begin{widetext}

\begin{figure}[!ht]
\includegraphics[width = 0.33\textwidth]{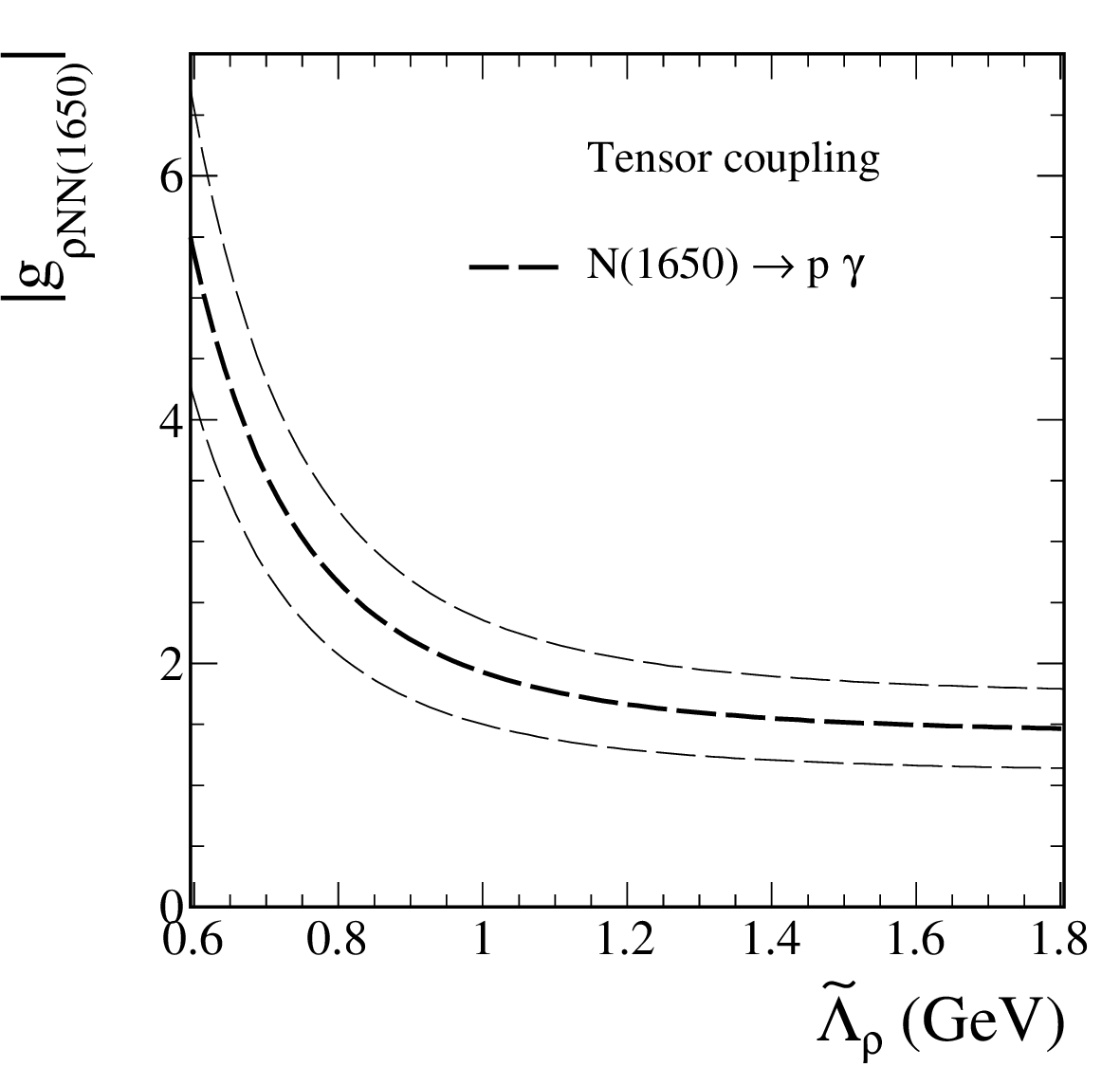}
\includegraphics[width = 0.33\textwidth]{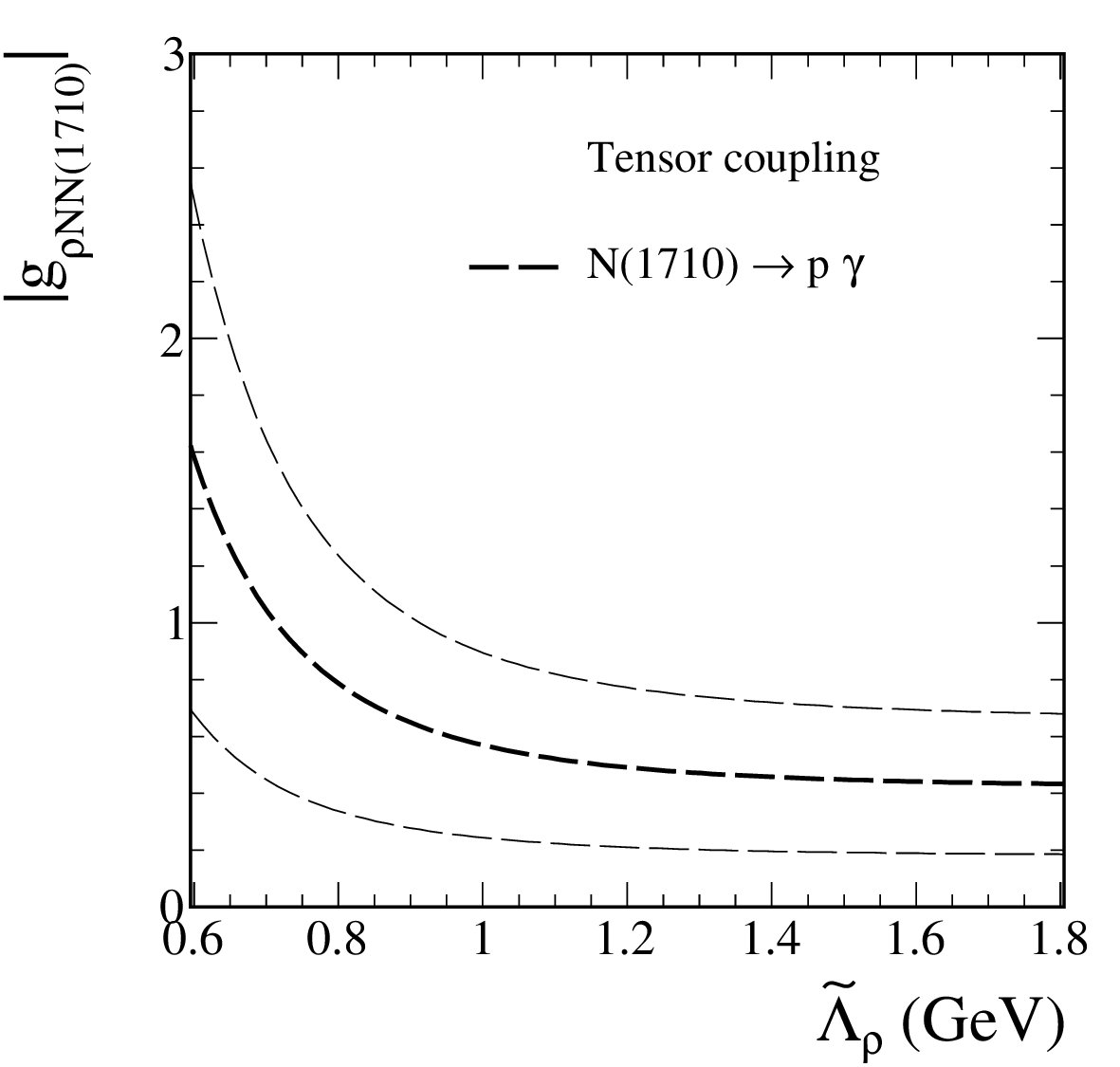}
\includegraphics[width = 0.33\textwidth]{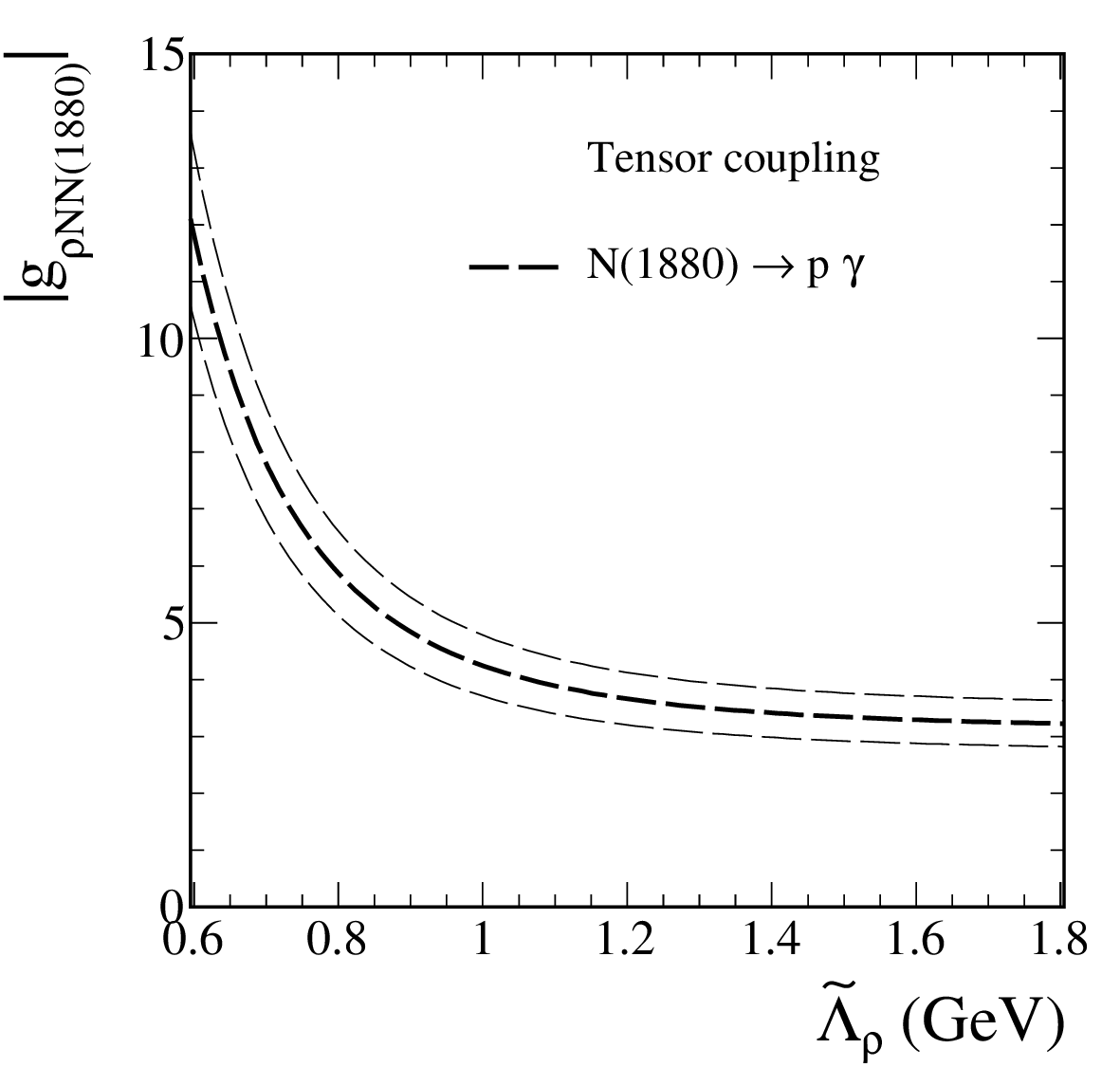}
\includegraphics[width = 0.33\textwidth]{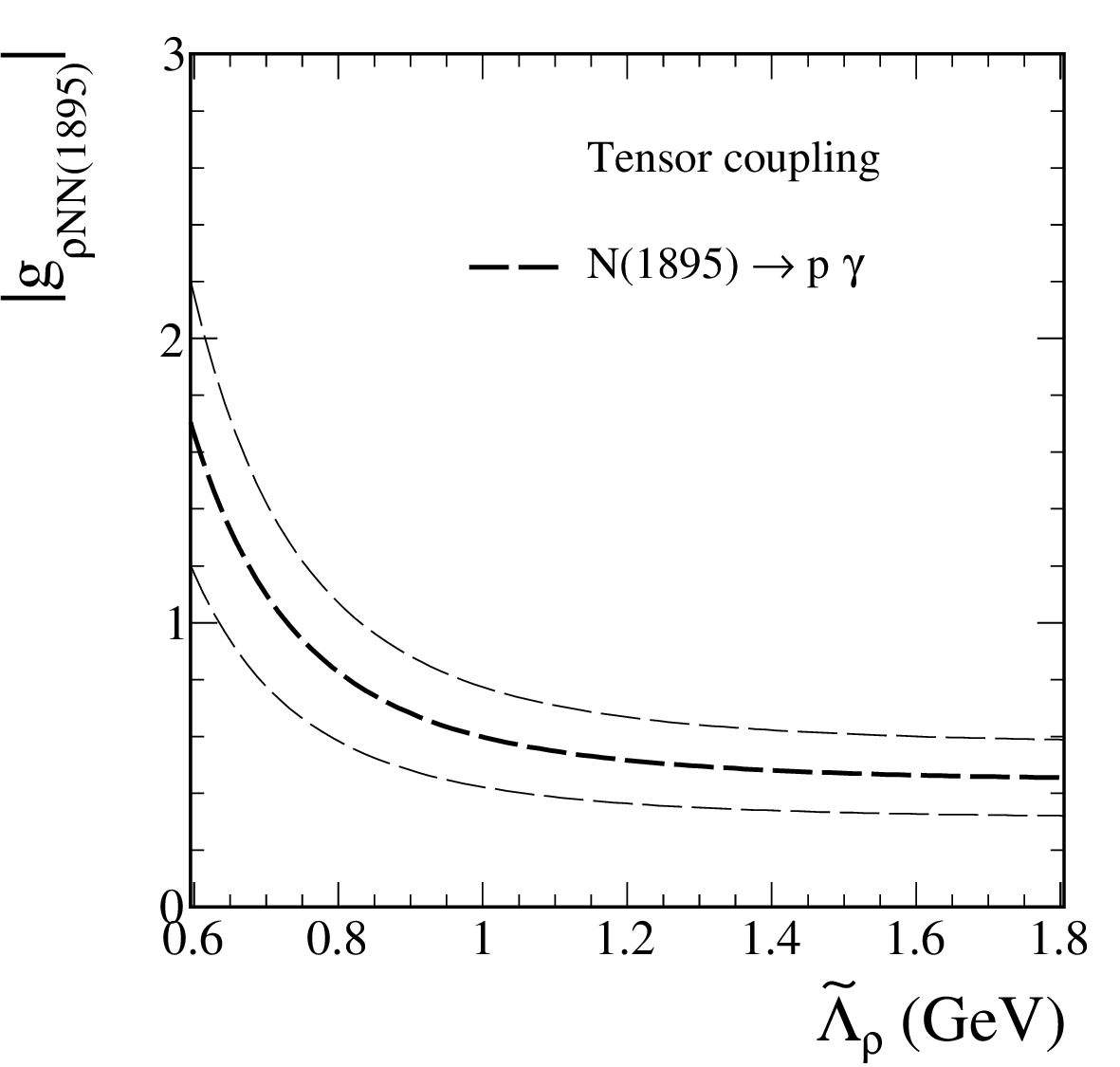}
\includegraphics[width = 0.33\textwidth]{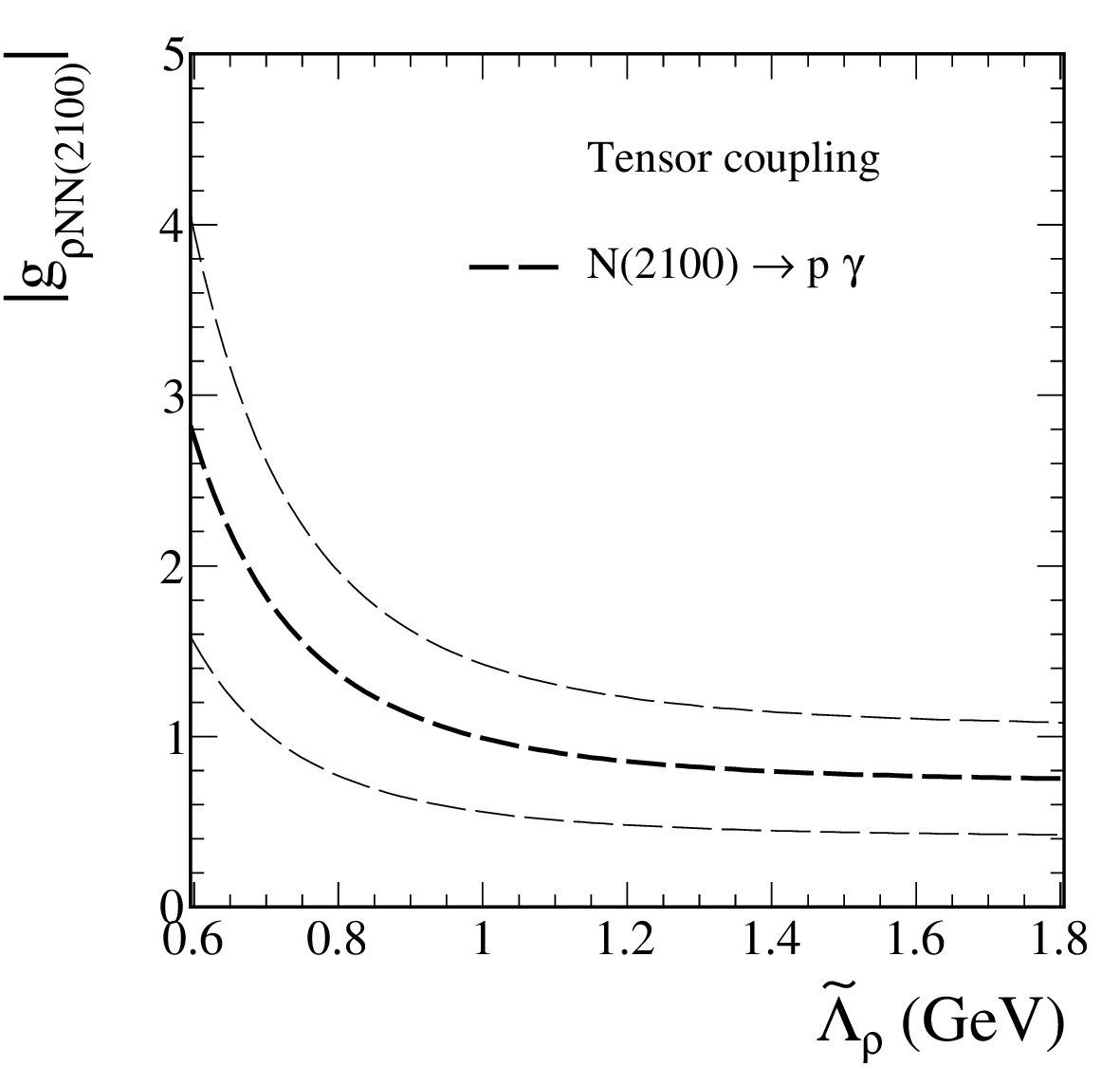}
\caption{
Coupling constant constrained
from the decays $N(1650)^{+}, N(1710)^{+}, N(1880)^{+}, N(1895)^{+}, N(2100)^{+} \to p \gamma$
as a function of $\tilde{\Lambda}_{\rho}$.
The numerical results correspond to the tensor coupling case.}
\label{fig:app_D2}
\end{figure}

\end{widetext}

Finally, Fig.~\ref{fig:app_D2} shows
the numerical values of coupling constants 
$\rho N N(1650)$, 
$\rho N N(1710)$, 
$\rho N N(1880)$,
$\rho N N(1895)$,
and
$\rho N N(2100)$, 
assuming the tensor-type coupling only,
versus $\tilde{\Lambda}_{\rho}$ estimated from (\ref{D2}).
For $N(1650)$ one takes
$A_{1/2}^{p} = 0.045 \pm 0.010$~GeV$^{-1/2}$ 
from \cite{ParticleDataGroup:2024cfk},
for $N(1710)$ one takes
$A_{1/2}^{p} = 0.014 \pm 0.008$~GeV$^{-1/2}$ 
\cite{Hunt:2018wqz},
for $N(1880)$ one takes
$A_{1/2}^{p} = 0.119 \pm 0.015$~GeV$^{-1/2}$ 
\cite{Hunt:2018wqz},
for $N(1895)$ one takes
$A_{1/2}^{p} = 0.017 \pm 0.005$~GeV$^{-1/2}$ 
\cite{Hunt:2018wqz},
and for $N(2100)$ one takes
$A_{1/2}^{p} = 0.032 \pm 0.014$~GeV$^{-1/2}$
\cite{Hunt:2018wqz}.
In the calculations of photon- and hadron-induced reactions the following values were used:
$g_{\rho N N(1535)} = 5.0$ or 4.5 
(respectively, for model 1, 2, or 3),
$g_{\rho N N(1650)} = 1.5$,
$g_{\rho N N(1710)} = 0.4$,
$g_{\rho N N(1880)} = 1.5$,
$g_{\rho N N(1895)} = 0.5$,
$g_{\rho N N(2100)} = 0.5$,
and a universal value 
of $\tilde{\Lambda}_{\rho} = 1.2$~GeV for all resonances.

\bibliography{refs}

\end{document}